\documentclass{emulateapj}
\usepackage{natbib}
\newcommand{\tx}[1]{\textrm{#1}}

\newcommand{\kms}{km~$\tx{s}^{-1}$}

\newenvironment{inlinefigure}{
\def\@captype{figure}
\noindent\begin{minipage}{0.999\linewidth}\begin{center}}
{\end{center}\end{minipage}\smallskip}

\slugcomment{AJ, in accepted}
\shorttitle{Photometric Search for Transients in Galaxy Clusters}
\shortauthors{Sand et al.}

\begin{document}
\title{A Photometric Search for Transients in Galaxy Clusters \altaffilmark{1}}

\author{David J. Sand,$\!$\altaffilmark{2} Dennis Zaritsky, St\'{e}phane
Herbert-Fort, Suresh Sivanandam, and Douglas Clowe\altaffilmark{3}} \affil{Steward Observatory,
Astronomy, Tucson, AZ 91125} \email{dsand@as.arizona.edu}

\begin{abstract}
We have begun a program to search for supernovae and other transients
in the fields of galaxy clusters with the 2.3m Bok Telescope on Kitt
Peak.  We present our automated photometric methods for data
reduction, efficiency characterization, and initial spectroscopy.
With this program, we aim to ultimately identify $\sim$25-35
cluster SN Ia ($\sim$10 of which will be intracluster, hostless
events) and constrain the SN Ia rate associated with old, passive
stellar populations. With these measurements we will constrain the
relative contribution of hostless and hosted SN Ia to the metal
enrichment of the intracluster medium.  In the current work, we have
identified a central excess of transient events within $1.25 r_{200}$
in our cluster fields after statistically subtracting out the
'background' transient rate taken from an off-cluster CCD chip.  Based
on the published rate of SN Ia for cluster populations we estimate
that $\sim$20 percent of the excess cluster transients are due to
cluster SN Ia, a comparable fraction to core collapse (CC) supernovae
and the remaining are likely to be active galactic nuclei.
Interestingly, we have identified three intracluster SN candidates,
all of which lay beyond $R>r_{200}$.  These events, if truly
associated with the cluster, indicate a large deficit of intracluster
(IC) SN at smaller radii, and may be associated with the IC stars of
infalling groups or indicate that the intracluster light (ICL) in the
cluster outskirts is actively forming stars which contribute CC SN or
prompt SN Ia.

\end{abstract}
\altaffiltext{1}{Observations reported here were obtained at the MMT Observatory, a joint facility of the University of Arizona and the Smithsonian Institution.}
\altaffiltext{2}{Chandra Fellow}
\altaffiltext{3}{Current address: Ohio University, Dept. of Physics \& Astronomy, Clippinger Labs 251B, Athens, OH 45701}

\section{Introduction}

Because little can escape the deep gravitational potential wells of
galaxy clusters, they are among the only sites where one can
confidently construct constrained models of chemical enrichment.  The
intracluster medium (ICM) contains most of the baryons in clusters
\citep[e.g.,][]{Gonzalez07} and is metal enriched, $ \langle Fe/H
\rangle \sim 0.3 Z_{\odot}$ \citep[e.g.,][]{Balestra07}.  However,
none of the proposed mechanisms to transport the metals produced by
supernovae from the galaxies in which they lie to the ICM have proven
adequate --- unless large gas outflows from cluster members are
invoked \citep[][see also discussion in \citealt{Dupke00}]{Renzini97}.
We are left to conclude either that galaxies somehow eject most of
their metals, some basic part of the chemical evolution model, such as
the SNe rate or yield, is incorrect, or an enrichment channel has been
overlooked.

One such channel that has recently been proposed is enrichment by
intracluster stars (ICS) whose SNe pollute the ICM in situ
\citep{Sasaki01,Zaritsky04,Domainko04,Lin04b,Sivanandam07}.  Because
the ICS comprise $\sim$30\% of the total cluster stellar budget
\citep{Gonzalez05,Gonzalez07}, a comparable fraction of the ICM metals
should come from this population even if galactic winds are highly
efficient.  A quantitative attempt to determine whether the dominant
source of metals is the ICS or the cluster galaxies has been unable to
distinguish between the two sources because of uncertainties in
cluster SNe rates and SN Ia progenitors \citep{Sivanandam07}.

Despite the widespread use of SNe Ia's as a probe of the cosmic
expansion history, the progenitor systems of SNe Ia remain
unidentified \citep[e.g.,][]{Hillebrandt00}.  Understanding the
origins and basic physics behind SNe Ia is essential, given that their
use as standard candles rests on the assumption that their basic
properties are constant as a function of cosmic time and that their
luminosities can be calibrated with an unchanging empirical relation
\citep[see recent review by][]{Filippenko95}.  One interpretation of
the shortfall of metals in the chemical evolutionary models is that
there exists a population of SNe that we are missing. Because we do
not understand the origin of Ia's, this possibility must be
considered.

Independent recent results indeed suggest that SN Ia come from two
different stellar populations
\citep{Mannucci05,Scannapieco05,Sullivan06}.  One type is associated
with star forming regions, which tend to be overluminous, have a rate
proportional to the star formation rate, and appear within 1 Gyr of
the formation of the stellar population. The other type is associated
with old stellar populations, have a rate that is proportional to the
total stellar mass of the system, and only appear many Gyrs after the
formation of the stellar population.  A SN search that probes both
young and old stellar populations must use the star formation history
of the survey volume to disentangle the two SNe Ia populations
\citep[e.g.,][]{Sullivan06} and measure the SN rate for each
population. This necessary modeling leads to larger uncertainties in
the inferred rates.

Old, elliptical galaxies only give rise to SN Ia associated with the
long delay time population \citep{Sullivan06}, which makes galaxy
clusters the natural environment in which to constrain the rate of
this type of SNe Ia.  By pursuing a SN survey in low redshift galaxy
clusters the modeling problem is averted because old stellar
populations dominate. The star formation history of these galaxies can
be approximated to high accuracy as a star formation burst at high
redshift \citep[e.g.,][]{Holden05}.  Furthermore, the profusion of old
ellipticals in clusters also makes it possible to measure the rate of
these SNe even though the rate is quite low.

There have been several modern SN surveys that have focused on the
galaxy cluster environment, beginning with that of \citet{Norgaard89}.
The first systematic survey, the Mount Stromlo Abell Cluster Supernova
Search \citep{Reiss98}, ran for 3.5 years and discovered 52 SNe, 25 of
which were associated with the cluster targets.  The survey focused on
Abell clusters at $0.02 < z < 0.08$, and of the 25 cluster SN Ia, 14
were confidently determined to be SN Ia.  This survey did not attempt
a SN Ia rate measurement.  \citet{Galyam02} searched {\it Hubble Space
Telescope} archival images of nine galaxy clusters that had more than
one imaging epoch for SNe.  Despite finding only two probable cluster
SN, they were able to put the first real limits on the SNe rate in
clusters at both low and high redshift.  This same group has
undertaken a longer term program in low redshift ($0.06 < z < 0.19$)
galaxy clusters, as part of the Wise Observatory Optical Transient
Search \citep[WOOTS;][]{GalYam07a}.  Their results include the
discovery of two IC SN Ia \citep{Galyam03} and a measurement of the SN
Ia rate in galaxy clusters based on six SN Ia \citep{Sharon07}.  Very
recently, \citet{Mannucci07b} presented a measurement of both the SN
Ia and CC SN rate in local galaxy clusters, measuring the cluster SN
Ia rate to be $\sim$3 times higher than that in field E/S0 galaxies.
Work has also begun at higher redshift \citep{Sharonhigh}.

Motivated by all of the above, we began a long-term photometric
monitoring campaign of $\sim$60 X-ray selected galaxy clusters at $0.1
< z < 0.2$ using the 2.3 m Bok telescope on Kitt Peak with its 90Prime
1-degree field imager \citep{90prime}.  We aim to measure the
old-population, SN Ia rate, both in cluster galaxies and among the
ICS. The data are also being used to study other transient phenomena.
Here, we present the observational design of the survey, transient
detection techniques and efficiency calculations, initial spectroscopy
and first science results.  In \S~\ref{sec:surveydesign}, we describe
the cluster sample, survey design and scientific goals.  In
\S~\ref{sec:search} we discuss our automated image reduction pipeline
and transient detection technique, followed by our calculated
transient detection efficiency in \S~\ref{sec:efficiency}.  We present
our current transient catalog in \S~\ref{sec:varcat}, along with a
sample of transients that do not appear to lie within galaxies, which
are generally referred to as hostless.  Our initial spectroscopy is
presented in \S~\ref{sec:spec}.  We present evidence for a central
overdensity of transient sources associated with our cluster sample in
\S~\ref{sec:overdense} and discuss the nature of these sources in
\S~\ref{sec:whatrexcess}.  We devote \S~\ref{sec:ICspec} to a
discussion of the intriguing implications of the three IC SN
candidates at $R>r_{200}$.  Finally, in \S~\ref{sec:future} we outline
our future plans.  Where necessary, we adopt a flat cosmology with
$H_{0}$=70 \kms Mpc$^{-1}$, $\Omega_{m}$=0.3, and
$\Omega_{\Lambda}$=0.7.

\section{Survey Design and Goals}\label{sec:surveydesign}

The basic observational goal of this survey is to confidently identify
SNe Ia within the galaxy clusters in our sample at $0.1 < z < 0.2$,
whether they lie among the ICS (hostless) or within cluster galaxies
(hosted).  Given our observational strategy and depth, we will
eventually correct the observed SN rate to the true rate, measure
whether there are significant differences among the hosted and
hostless events, and ultimately improve the model of the metal
enrichment history of the ICM.

Various considerations play a role in the survey design, including the
available field of view, image depth, survey cadence, and sample size.
From the depth we achieve with modest exposure times (total 1600 sec,
see below), and a desire to have control fields near the clusters, we
decided to constrain the sample to redshifts between 0.1 and 0.2.  As
described in more detail below, we roughly center our clusters on CCD
chip 1 of 90Prime, the prime focus wide field imager on the 2.3m Bok
telescope on Kitt Peak.  We use the other three chips to measure the
background transient detection rate, given that there is a $\sim$500
arcsecond gap between them.  The $\sim$30' x 30' field of chip 1
corresponds to a radius of $\sim$1.6 Mpc at z=0.1 and $\sim$3.0 Mpc at
z=0.2 and allows us to probe out beyond $r_{200}$ in all of our
clusters.

The peak magnitude of type Ia supernovae is $M_{B}=-19.25$, with a
small intrinsic dispersion of $0.2-0.3$ mag \citep{Phillips93}.  For
the redshift range of our cluster sample, this corresponds roughly to
$19 < g < 20$ mag.  Type Ia SNe fade by $\sim$ 1 mag in 15 rest frame
days.  By taking four 400 second $g$ band exposures with small dithers
we easily find SN candidates down to $g\sim21.5$ during good observing
conditions (see \S~\ref{sec:efficiency}), allowing us to observe SNe
Ia's well before and after peak brightness.  We will present our
formal effective visibility time, the time during which a given survey
image is sensitive to SNe, in future work as we determine the SNe Ia
rate in our cluster sample. We present preliminary calculations in
\S~\ref{sec:howmanysnia}.

For an exposure time of 4$\times$400 seconds per cluster field and a
generous allocation of $\sim$4 nights per month on the 2.3m Bok
telescope (excluding the southern Arizona monsoon season), we can
survey about 60 galaxy clusters monthly.  In addition to spectroscopic
followup of a fraction of our cluster SNe candidates, we are also
experimenting with multi-band imaging of each cluster field at each
epoch to exploit recent progress in single-epoch phototyping of
supernovae candidates \citep{Poznanski06}, but do not describe the
latter here.  We also employ other methods for statistical
determinations of the rate of cluster transients (see
\S~\ref{sec:future}).

As will be discussed in \S~\ref{sec:howmanysnia}, we expect to find
$\sim$2-3 cluster SN Ia per month of 90Prime imaging employing the
above strategy.  If the program runs for two observing seasons, each
spanning six months, we will discover $\sim$25-35 cluster SN Ia, with
between $\sim$5 and 10 of these being hostless, IC events.  This
survey, when completed, will have 4-6 times the number of cluster SN
Ia than previous focused searches which have sought to measure the
rate in clusters.  Overcoming previous surveys' small number
statistics is essential for pinning down the rate and understanding
the long-delay time SN Ia population.

Several secondary science goals can also be pursued with these data.
For example, active galactic nuclei (AGN) and quasi-stellar objects
(QSOs) constitute a large fraction of our transient objects (see
\S~\ref{sec:varcat} and \ref{sec:spec}).  Identification of AGN/QSOs
through their variability is highly complementary to other AGN
selection techniques \citep{Sarajedini03,Sarajedini06,Totani05}, and
may be useful in cluster fields where there is a known overdensity of
AGN \citep{Ruderman05,Martini07}.  Another example is the use of our
4$\times$400 second image sequence to study short time scale
variability \citep[e.g.,][]{Becker04}.  A final example is the study
of rare, mysterious transient events in cluster cores such as that
identified in Abell 267 which at the time could not be fit by any
known supernova template \citep{Stern04}.

\subsection{Sample Selection}\label{sec:sample}

We chose our cluster sample from among $0.1 < z < 0.2$ X-ray selected
galaxy clusters visible from northern hemisphere telescopes with a
preference for galaxy clusters with known strong lensing features (to
increase our chances, albeit low, of discovering strongly lensed
transient phenomena).  Clusters were preferably chosen to have a high
X-ray luminosity, although not in any systematic way.
The cluster sample, along with the observing epochs included here, is
detailed in Table~\ref{table:clustertable}.  The X-ray luminosities
are those reported directly from the BAX online cluster data base,
which attempts to homogenize luminosities based on flux measurements
and uses a $H_{0}$=50 \kms Mpc$^{-1}$ and $\Omega_{m}$=1.0 cosmology.
We refer the reader to the BAX web site for
details\footnote{http://bax.ast.obs-mip.fr/}.

\section{Photometric Transient Search}\label{sec:search}

The observations were conducted with the prime focus imager, 90Prime
\citep{90prime}, on the 2.3m Bok telescope on Kitt Peak.  The focal
plane array consists of a mosaic of four Lockheed 4k $\times$ 4k blue
sensitive CCDs (the peak QE is near 4500 \AA).  With the pixel scale
of $0\farcs45$ per pixel, this detector array gives a field of view of
about one square degree.  The chip gaps are large, approximately 500
arcseconds on the sky.  For our purposes, these gaps are actually an
advantage because it places the other chips farther away from the
cluster.
We center our cluster fields on Chip 1 (cosmetically the best out of
the four), giving coverage out to at least $r_{200}$ for all
targets. The other three chips provide control fields.  Here, we
present transients from chip 3, in addition to chip 1, to quantify the
background transient rate.  A charge trap in chip 3 decreases the
effective field of view by approximately one-eighth, but otherwise the
chip is relatively free of defects.

The images considered here were obtained between January and June
2006, during four observing runs (our pilot program), and one night in
December 2006.  The single December night was followed by one night of
spectroscopy, which we present in \S~\ref{sec:spec}.  Several other
runs beyond the pilot program, which implemented this imaging plus
spectroscopy strategy, were spoiled either due to weather or
instrument problems, leaving only the one useable night in December
2006 where everything worked. There are five clusters (noted in
Table~\ref{table:clustertable}) whose only additional epoch of imaging
were taken during the period of the Fall 2006 campaign when the
90Prime imager was operating with anomalously high readout noise.
Since these images are of lower quality, they are not generally
included in this study, but are used only as the reference image for
differencing of the good December data in these five clusters.

\subsection{Basic Image Calibration}{\label{sec:imagecal}}

The image processing consists of various steps that we describe here.
We do the initial image reductions using a nearly automated {\sc IRAF}
script that relies heavily on the {\it mscred} mosaic data reduction
system.  This script corrects for cross talk, trims and subtracts the
overscan region, applies an additional bias subtraction to remove
structure seen in the bias exposures, flat-fields the data and rejects
cosmic rays in each individual exposure using the {\sc LACOSMIC} task
\citep{vandokkum01}.  Regions around saturated and nearly-saturated
objects are heavily masked because residuals near poorly subtracted
bright objects can be one of the largest sources of false detections
in our transient detection pipeline.  We calculate exposure weight
maps using a combination of a normalized twilight flat field and the
cosmic ray rejection map.  We edit each image header to include the
refined world coordinate system solution obtained using the {\it
msccmatch} task and the USNO-A2 catalog. The final astrometric
solution has a typical rms scatter of $\sim$0\farcs2-0\farcs3.

To combine the individual exposures, we rescale and resample each onto
the same astrometric grid using SExtractor \citep{sexbib} and {\sc
SWarp}\footnote{version 2.15.7; http://terapix.iap.fr/soft/swarp}.  A
final median combined cluster image is made for each observing epoch
using the individual exposure weight maps and {\sc SWarp}.  We chose
to median combine the frames to remove fast-moving objects, although
slowly moving objects, those that move by less than the typical seeing
at the Bok telescope during the course of a 4 x 400 image sequence
(see Figure~\ref{fig:psf}), remain.

We photometrically calibrate the final images (both the combined and
individual exposures) using non-saturated point sources in common with
the Sloan Digital Sky Survey (SDSS;
\citet{sdsscite})\footnote{http://www.sdss.org} when there is overlap.
Of the fifty-five clusters in this study, thirty have SDSS coverage.
Otherwise, we bootstrap calibrations using the photometric zeropoints
of SDSS clusters on clear nights.  Every cluster was observed on at
least one photometric night (loosely defined to be a night in which
the Sloan derived zeropoint did not vary by more than 0.1 mag
throughout the night).  We use SExtractor to generate object catalogs
of individual and combined images, which also provides a measure of
the average image point spread function (PSF) across the field.  See
Figure~\ref{fig:psf} for the distribution of image FWHMs and
Figure~\ref{fig:zpthist} for the distribution of relative zeropoints
during our imaging runs. Both of these quantities play an important
role in determining our transient detection efficiency
(\S~\ref{sec:efficiency}), and will be discussed briefly here.

When measuring the image PSF and zeropoints by comparison with SDSS,
we select only the high S/N objects with SExtractor.  For
concreteness, an object must have a minimum of fifteen pixels (the
{\sc detect\_minarea} SExtractor parameter) above a detection
threshold of $20-\sigma$ ({\sc detect\_thresh} parameter) in order to
be considered for the PSF measurement and to be used in the
calculation of the image zeropoint.  For the automatic PSF
measurement, stars that met these criteria were further flagged to
have a flux between 2$\times10^5$ and 1$\times10^6$ counts.  These
flux limits were set empirically to minimize outliers due to galaxy
contamination and near-saturation objects.  A final PSF measurement
(confirmed empirically for a subset of objects) was found by taking
the output SExtractor parameter {\sc fwhm\_image} for these objects
and computing their mean FWHM using the IDL function {\sc
resistant\_mean}, which clips based on the median and the median
absolute deviation.  As can be seen in Figure~\ref{fig:psf}, the FWHM
is not good, although this has been factored into our survey planning
and has not kept us from the survey depth and transient detection
efficiency desired (see \S~\ref{sec:efficiency}).

For the zeropoint calculation, we take the above high S/N SExtracted
objects and WCS match with point sources from the SDSS archive
(matches are made to within five 90Prime pixels -- $\sim$2.5 arcsec).
The mean image zeropoint was calculated directly using the IDL
function {\sc resistant\_mean} (typically hundreds of SDSS point
sources and 90Prime counterparts were matched).  On those nights
determined to be photometric, a linear fit was performed on the
SDSS-derived zeropoint as a function of 90Prime imaging airmass to
derive an atmospheric extinction correction for those clusters without
counterpart SDSS imaging.  Using this, cluster fields without SDSS
imaging were given a zeropoint based on the image airmass and a bright
star catalog was created to be used as a photometric catalog for all
other imaging epochs of that cluster.

\subsection{Difference imaging and transient detection}{\label{sec:trandetect}}

We implement an automated pipeline to PSF-match and difference an
individual epoch image from a reference image, identify candidate
variable objects, and post image cut-outs that are reviewed by a human
using a combination of Perl and IDL scripts whose core utilizes the
ISIS software package \citep{Alard00}.

In detail, the sequence is as follows.  When a new image is obtained,
the code identifies a reference image by searching for the
corresponding archived image with the tightest PSF.  The new image is
registered, resampled onto the reference image grid, PSF matched, and
subtracted. Residual objects are classified as candidate
transients. If the newly acquired image happens to be the best to
date, it is used as the reference image and the second best image is
subtracted.  The image differencing is performed with the {\it
mrj\_phot} program within the ISIS package and we refer the reader to
\citet{Alard00} for details.  We found the procedure works well when
we split each 4k $\times$ 4k CCD into 16 subregions for subtraction
and the PSF kernel convolution is allowed to vary over each subregion.
We set our upper flux limit for detection and image rescaling well
below the saturation level (see \S~\ref{sec:efficiency}).  We utilize
{\it detect.csh} and {\it find.csh} from within ISIS to find high
variance peaks in the differenced images and produce a first list of
candidate variables.  We then trim the list by removing objects near
bad columns ($<25$ pixels), near masked saturated stars ($<25$
pixels), nearby significant negative values in the differenced image
($<7-\sigma$ below a local sky value calculated with an annulus of
50-70 pixels; most likely due to a slight registration misalignment
that produces dipoles in the difference image), and near another
transient ($<25$ pixels). The latter are likely either residuals from
the poor subtraction of a large object or astrometric variables.  We
quantify the likelihood of the rejection of real variables in
\S~\ref{sec:efficiency}.  These cuts tend to reduce the number of
candidates from several hundred to less than several tens in a given
image.  Despite our best efforts, some obviously spurious transient
candidates remain, and so human review is necessary.

The principal long time scale variable candidates in this project are
expected to be variable stars, active galactic nuclei (AGN), novae and
supernovae.  Since the primary goal of this work is to search for SNe
in the clusters, we tried to screen out known AGN and variable stars
by crosschecking with NED through a batch query.  Nonetheless,
previously unknown AGN and variable stars are a major source of
contamination in this work, but will lead to exciting auxiliary
science.

\section{Transient Detection Efficiency} {\label{sec:efficiency}}

To understand the recovery efficiency of transient events in our
pipeline as a function of seeing and transient brightness, we add
point sources to individual images and attempt to recover them with
our transient detection pipeline with the unmodified reference image.
These test point sources are randomly placed in images with a
uniformly distributed $g$ magnitude between $17 < m_g < 24$.

A typical efficiency run for a given image consists of 50 separate
realizations, each of which incorporates 30 random amplitude, randomly
placed point sources. The final efficiency calculation for that image
is then based on the recovery statistics of 1500 fake point
sources. Although we have not devoted the computer resources to
measuring our detection efficiency in each of our survey images, we
have sampled the parameter space (PSF and transparency) covered by our
data.

We present typical results as a function of these two parameters in
Figures~\ref{fig:totsceff} and \ref{fig:totzptsceff}.  The recovery
rate is $\sim$80-90\% over a magnitude range of $g\sim 18-21.5$ during
clear sky conditions and the full range of seeing conditions.  When we
do not recover the transient it is sometimes due to its proximity to a
masked portion of the image.  Occasionally, a catastrophic image
subtraction error in one of the sixteen subtraction subregions occurs.
Normally this manifests itself in the form of a slight misregistration
of the subregion, which produces dipole residuals in the differenced
image, and no viable transient candidates.  Occasionally if a
subregion has a very bright star in its sector, it will corrupt the
whole subregion, leaving a poor subtraction.  An examination of when
this occurs did not reveal any trend towards the outer edges of the
field of view, and so this is unlikely to be the cause of the central
excess of transients that we find in \S~\ref{sec:overdense}.  In the
absence of these relatively rare subtraction errors in a subregion, we
recover over 90\% of our fake transients with $18.5 < g < 21.5$.  In
addition to recovering the vast majority of the transients, we also
generally measure the correct magnitude. In Figure~\ref{fig:magdiff},
we plot the difference between input magnitude and that recovered as a
function of input transient magnitude.  At the faintest magnitudes
where we detect transients ($g\sim 22.5$), the scatter is $\sim$0.2
mag, with no bias.  The transient detection efficiency in chip 3 (the
off cluster chip) is consistent with that of chip 1 to within the
noise of the inidividual magnitude bins, and for the rest of this work
we consider them identical.

One additional experiment was performed to gauge the relative
detection efficiency for transients in galaxy cores versus those in
regions with no apparent galaxy host.  We placed artificial SN with
random $g$ magnitudes between 21.0 and 21.5 (the faint end out to
which we reliably detect transients) at varying annuli from the BCG
center in the same sample of clusters that we used to determine the
detection efficiency as a function of image PSF.  To be explicit, 50
artificial transients per bin were placed at annuli between 0 and 1
PSF, 1 and 2 PSFs, 2 and 5 PSFs, 5 and 10 PSFs, and continuing out to
a maximum radius of five hundred pixels from the center of the cluster
BCG.  Between 95\% and 100\% of the artificial transients were
recovered, with no trend seen as a function of radius.  Those few
artificial transients that were not found by our detection pipeline
were not found due to proximity with a masked saturated star or other
image defect, as determined by visual inspection.  For this reason we
do not distinguish detection efficiencies for objects found in the
cores of galaxies versus those with no host galaxy.  A similar result
was obtained by \citet{Neill06} for the Supernova Legacy Survey,
although the WOOTS cluster search did find a significantly decreased
detection efficiency in the cores of galaxy hosts versus hostless
events \citep{Galyam03}.

We will use these calculated detection efficiencies in
\S~\ref{sec:whatrexcess} to estimate the approximate number of cluster
SN Ia we should observe and to compare with the observations, although
as we mentioned previously this treatment does not properly account
for the varying maximum brightness and light curve of SN Ia.

\section{Identified variable objects}\label{sec:varcat}

We present our transient catalogs for the cluster-centered chip 1 in
Table~\ref{table:vartable} and the off-cluster chip 3 in
Table~\ref{table:vartable3}. These tables contain the position, epoch
of discovery, the apparent $g$ magnitude of the transient, the
projected distance from the transient to the brightest cluster galaxy,
and any additional comments.  Such comments may include, for example,
if the transient was detected in more than one epoch as a variable
(repeat detection; RD), if there is a correspondence to a NED source,
if the transient is an IC, hostless SN candidate (see
\S~\ref{sec:iccands}), or if we were able to get a spectrum of the
object (see \S~\ref{sec:spec}).  To get an idea of the magnitude
distribution of our discovered transients, we plot magnitude
histograms in Figure~\ref{fig:maghist}. The dashed histogram shows the
transient magnitude histogram after removal of sources that were
detected as variables at multiple observing epochs, and the dotted
line shows the transient magnitude histogram after removal of all
sources known to not be at the cluster redshift (whether found through
NED or our own spectroscopy).

Of the variable objects associated with NED sources, most are QSOs.
In the cluster-centered chip 1, 20 transients were associated with
known QSOs (9 in the off-cluster chip 3).  In addition to these, three
transients in the cluster-centered chip 1 are associated with cluster
members that are known AGN or QSOs from SDSS.  Two additional
transients in the off-cluster chip 3 are SDSS AGN at the cluster
redshift, but are $\sim$5.2 and 7.7 Mpc from the presumed cluster
center.  These AGN may be associated with infalling groups at the
outer edges of the cluster, with implications for the origin of AGNs
in cluster environments \citep[e.g.,][]{Martini07}.  We will discuss
these cluster AGN further in \S~\ref{sec:whatrexcess}.  Fourteen
transients in chip 1 (four in chip 3) were 'repeat detections';
several of these are known QSOs, but given our visibility time (see
\S~\ref{sec:howmanysnia}), some may be SN detected at different parts
of their light curve. Three other transients in chip 1 are associated
with galaxies at the cluster redshift, none in chip 3.

\subsection{Transients without Hosts}\label{sec:iccands}

One of the primary goals of this survey is to identify IC supernovae.
When a potential IC event is identified, we combine all of
the imaging epochs in which the transient was not detected into a
single deep image to search for a possible faint host galaxy.  The
images are combined as a weighted average (more appropriate for
reaching faint limiting magnitudes than a median combination).  We run
SExtractor on these deep, combined images with a $3-\sigma$ detection
threshold and minimum area of three pixels (MAGAUTO magnitudes are
computed).  We consider a transient to be an IC candidate if its
Sextractor extraction radius does not overlap any object in the deep
object catalog.

A synopsis of our IC SN candidates is presented in
Table~\ref{table:ICsumm}.  We list the position, magnitude, projected
distance from the BCG ($R_{BCG}$), that projected distance in terms of
$r_{200}$, and our $g$-band magnitude limits (both apparent,
$m_{lim,g}$, and absolute, $M_{lim,g}$, assuming that the object
is at the cluster redshift).  To calculate the limiting magnitude for
the faint host detection, we define the point source detection limit
identically to that of \citet{Becker04} and the NOAO Archive
definition of photometric depth, as
\begin{equation}\label{eqn:maglimit}
m_{l} = m_{0} - 2.5 \log (1.2 W \sigma_{sky} n) 
\end{equation}

\noindent where $m_{0}$ is the image zero point, $W$ is the seeing of
the image, $\sigma_{sky}$ is the sky dispersion around the mode, and
$n=3$ represents the $3-\sigma$ detection limit.  Point source
limiting magnitudes are given since dwarf galaxies at the cluster
redshift (with our typical seeing) should be unresolved
\citep[e.g.~dwarf galaxies in the Coma cluster have a typical size of
$\sim$1 arcsec;][]{Komiyama02}.  To put the absolute $g$ band
detection limits in perspective, one can adopt the Virgo Cluster
luminosity function as presented by \citet{Trentham01}.  Assuming this
luminosity function is representative of the clusters in our sample,
then $\sim$0.2\% of the galaxy light in the cluster would come from
dwarf galaxies below $M_{g} = -14$.

Apparently hostless transients may have a variety of explanations.
First, these could be true IC SNe.  Second, they may simply lie within
hosts that are below the detection threshold of our current imaging,
despite the detection limits described above.  As the survey revisits
our cluster fields more and more, the constraints will tighten.
Third, they may be flaring stars or novae in our galaxy, with
precursors too faint given our current detection limits.  Fourth, they
may be CC supernovae associated with star formation in
IC space, as suggested by recent observations of $H\alpha$ in a tidal
tail in the cluster A3627 \citep{Sun07}.  Finally, they may be highly
variable higher-redshift objects, such as AGN, that we only detect in
high-state.

To resolve this confusion, we are making spectroscopic identification
of all of our IC SN candidates a high priority as the survey
progresses.  Due to the limited amount of spectroscopic time we have
had so far, we were able only to attempt confirmation of the one IC
candidate in Abell 516 (see \S~\ref{sec:spec}). Unfortunately, the
resulting spectrum is inconclusive due to poor signal to noise. Its
magnitude at discovery was $g$=23.83, well below our nominal limit.
We will discuss the number of IC SN candidates found relative to the
total number of SN Ia expected in our survey up till now in
\S~\ref{sec:howmanysnia} and we speculate further in
\S~\ref{sec:ICspec} under the assumption that the three IC SN
candidates found in chip 1 are indeed part of the cluster and not some
foreground or background population.

\section{Initial Followup Spectroscopy}\label{sec:spec}

Spectroscopic identification of a significant subsample of our cluster
SNe candidates representing the complete range in brightness and
environment is critical to our understanding of the relative
populations of foreground SNe, cluster SN Ia, uncataloged AGN, and
CC SNe associated with the cluster.

We were allocated five nights for followup spectroscopy during the
final trimester of 2006 to implement our full cluster supernovae
program.  Unfortunately, as discussed in \S~\ref{sec:search}, the
90Prime noise levels were greatly elevated during most of the
observing term, except for the final imaging night when a grounding
problem was fixed and the instrument returned to nominal.  Hence, we
only have one night of imaging followed by one night of Blue Channel
Spectrograph followup at the MMT in this initial post-pilot portion of
the program.  We present these data here as a demonstration of our
ability to rapidly identify and confirm transient sources.

For the spectroscopy, we use the 300 line grating with a central
wavelength of 5800 \AA~(corresponding to a wavelength range of 3200 -
8400 \AA) for our Blue Channel Spectrograph \citep{BCS} observations
on the night of December 19-20 2006.  The 1\farcs0 slit was used
throughout the night. We use IRAF to do the standard CCD processing,
one dimensional spectrum extraction, and wavelength and flux
calibration.  No attempt was made to remove potential host-galaxy
contamination from any spectrum.  We intend to revisit this issue in
future work, especially for the hosted events.

\begin{figure*}
\begin{center}

\mbox{\epsfysize=18.0cm \epsfbox{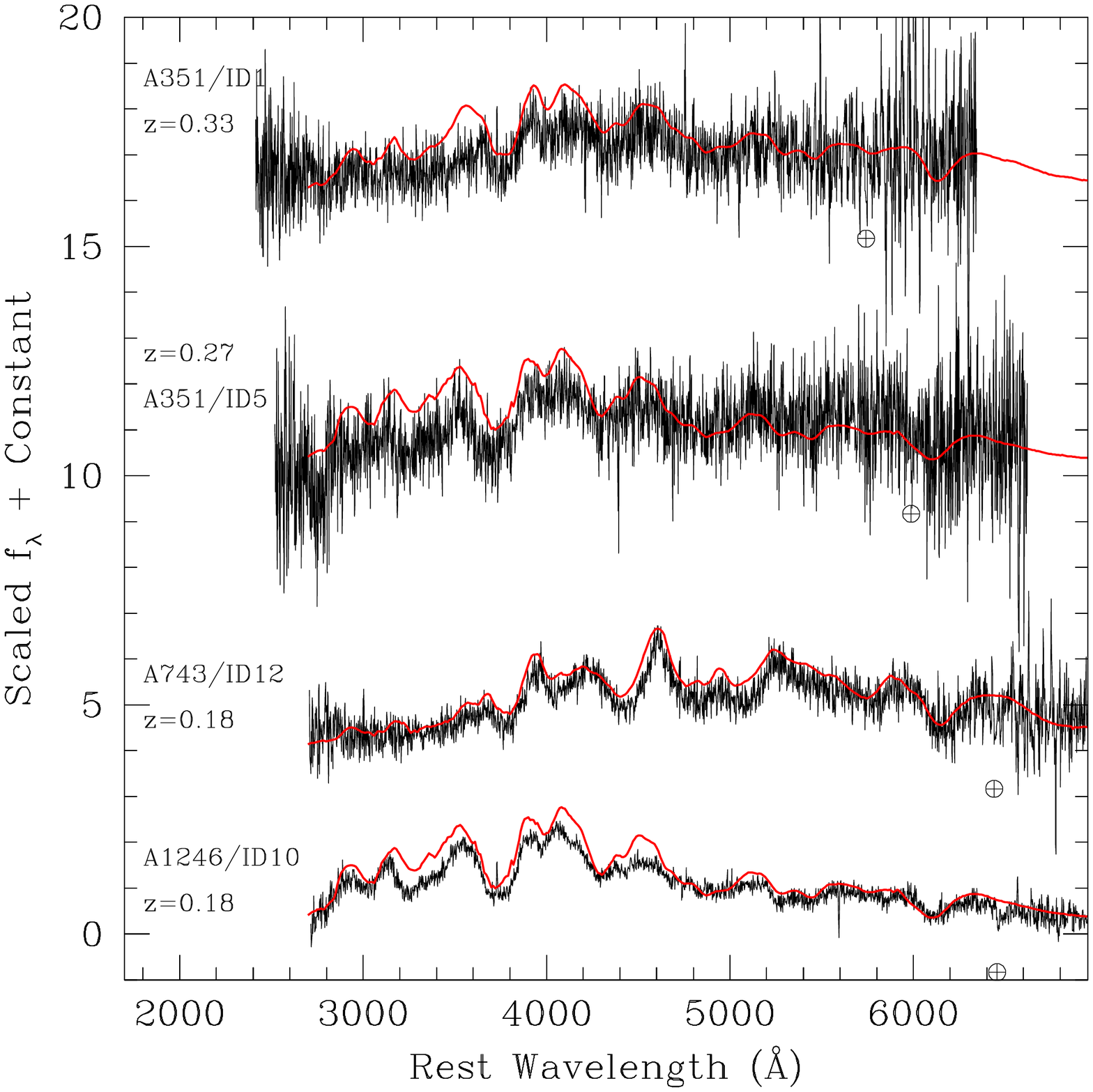}}

\caption{Rest wavelength spectra of SNe Ia.  Overplotted are SNe Ia
template spectra from Peter Nugent \citep{Nugent02}.
\label{fig:sn1a}}
\end{center}
\end{figure*}

\begin{figure*}
\begin{center}

\mbox{\epsfysize=18.0cm \epsfbox{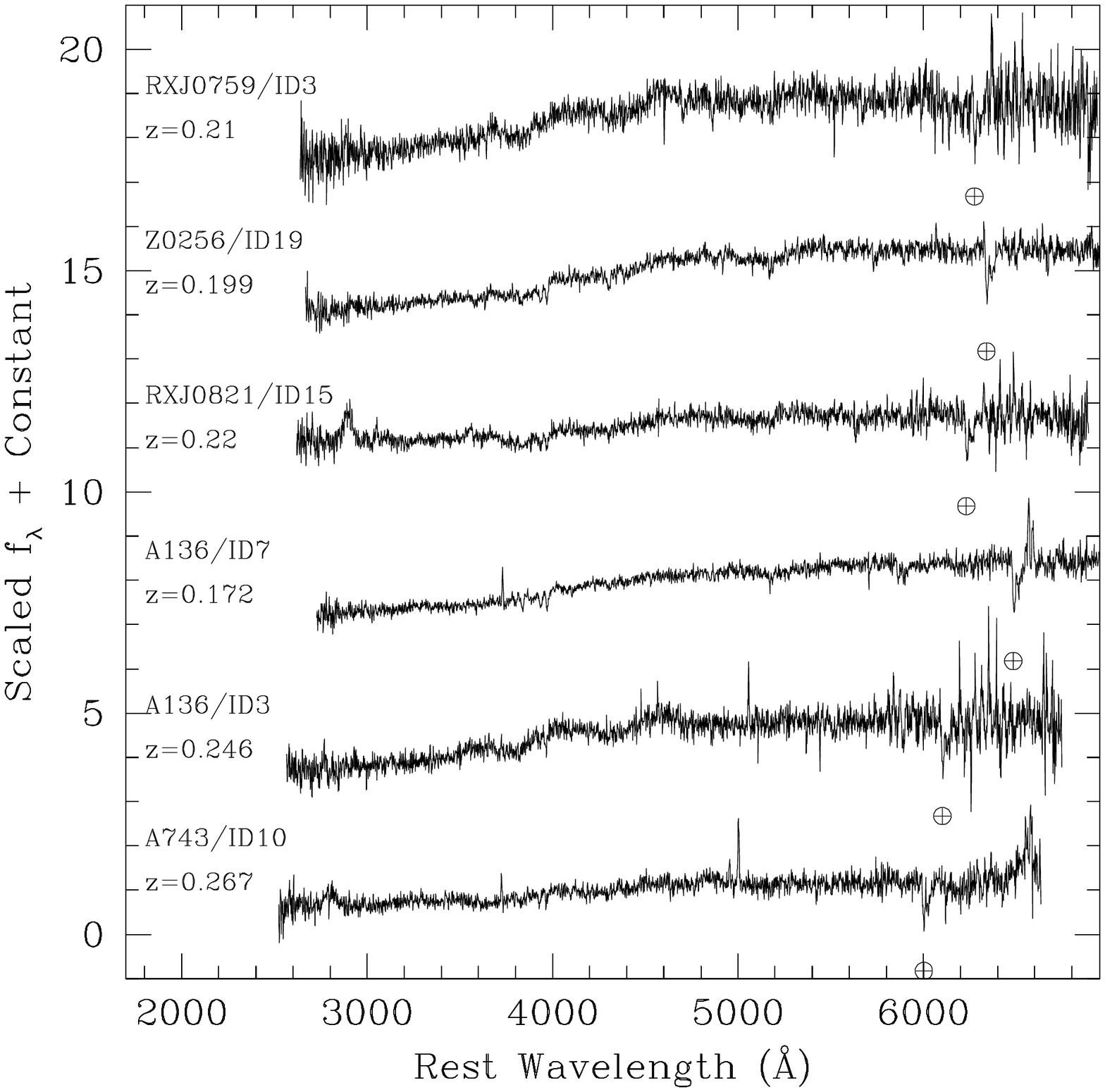}}

\caption{Spectra of six low redshift ($z<0.3$) galaxies.  For a
discussion, see \S~\ref{sec:lowzspecs}.
\label{fig:speclozgals}}
\end{center}
\end{figure*}

\begin{figure*}
\begin{center}

\mbox{\epsfysize=18.0cm \epsfbox{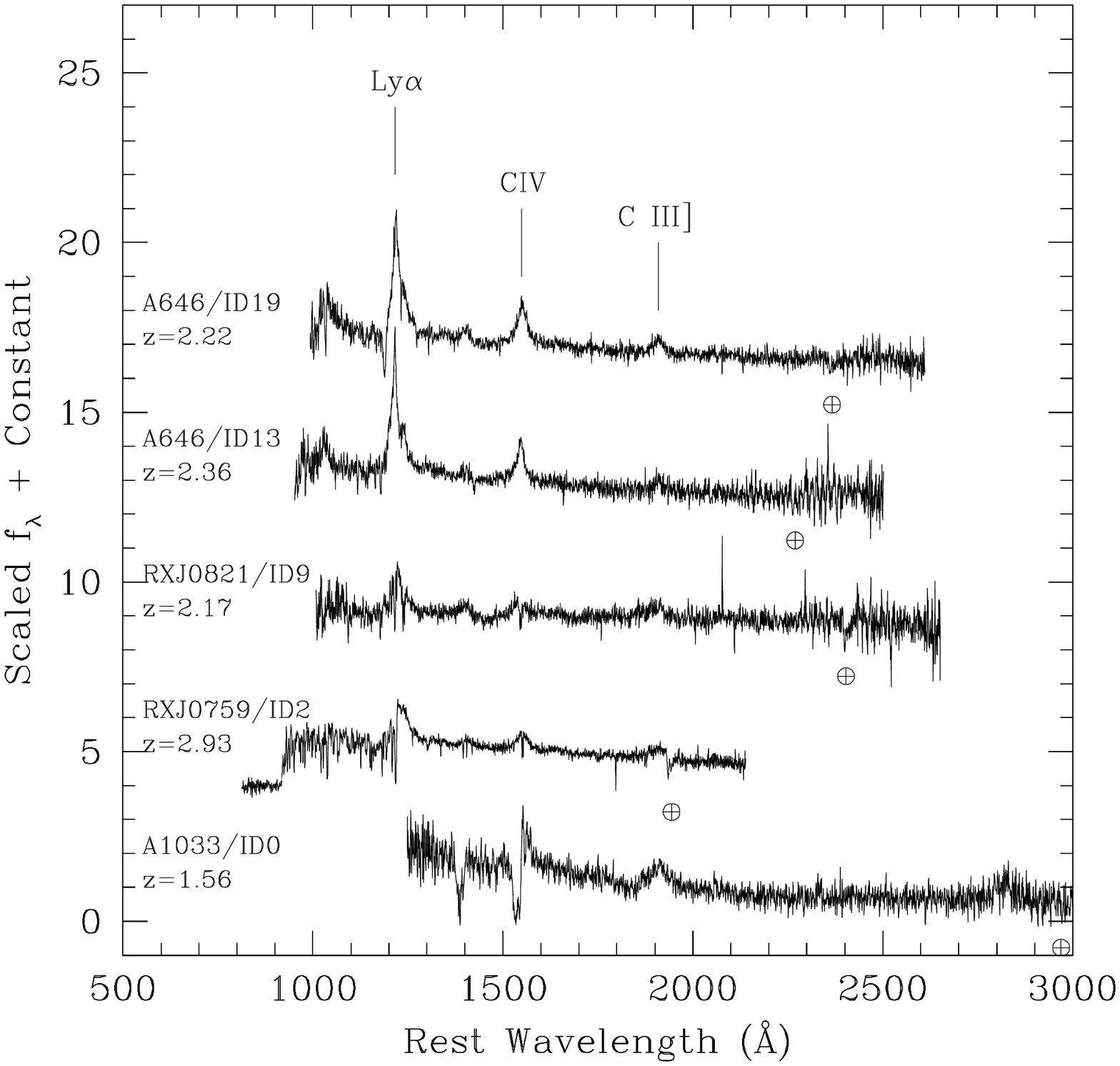}}

\caption{Spectra of five high redshift QSOs.  The location of
the A-band atmospheric absorption feature is indicated with a telluric
symbol.  No attempt was made to remove this feature.
\label{fig:hizagn}}
\end{center}
\end{figure*}

\begin{figure*}
\begin{center}

\mbox{\epsfysize=18.0cm \epsfbox{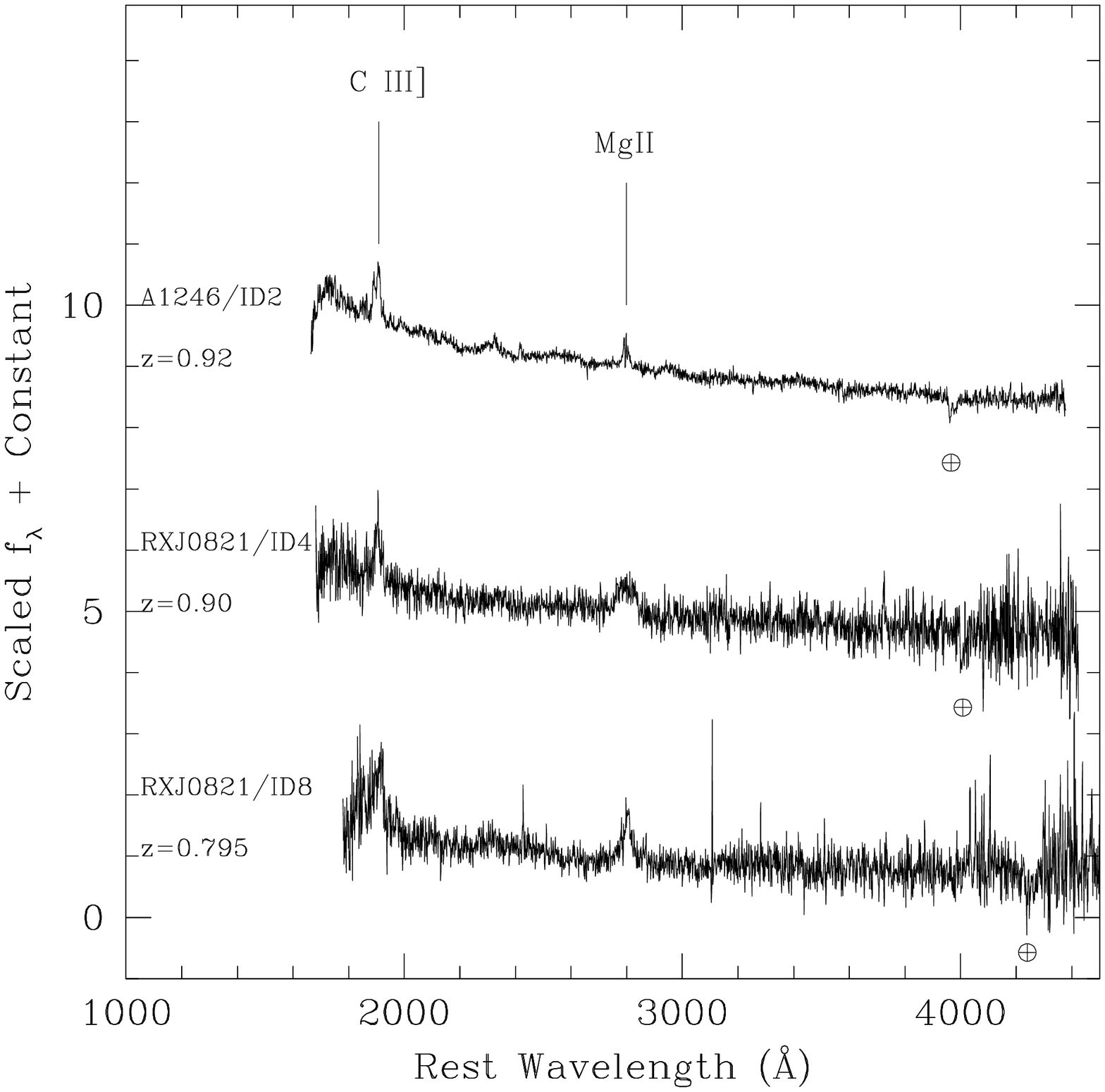}}

\caption{Spectra of five intermediate redshift QSOs.  The location of
the A-band atmospheric absorption feature is indicated with a telluric
symbol.  No attempt was made to remove this feature.
\label{fig:lozagn}}
\end{center}
\end{figure*}

\begin{figure*}
\begin{center}

\mbox{\epsfysize=18.0cm \epsfbox{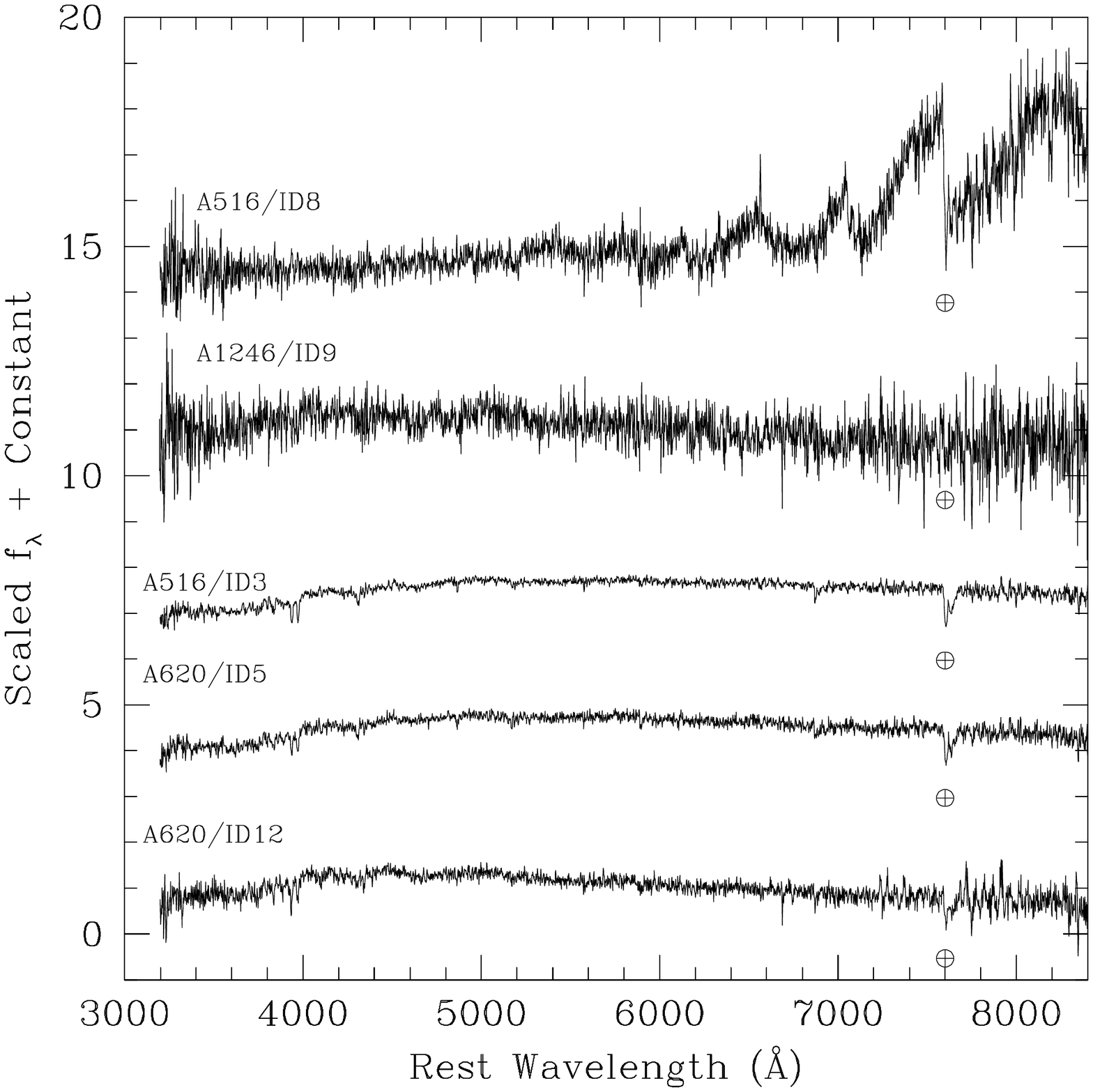}}

\caption{Spectra of five variable stars.  The location of the A-band
atmospheric absorption feature is indicated with a telluric symbol.
No attempt was made to remove this feature.
\label{fig:specstars}}
\end{center}
\end{figure*}

\begin{figure*}
\begin{center}

\mbox{\epsfysize=18.0cm \epsfbox{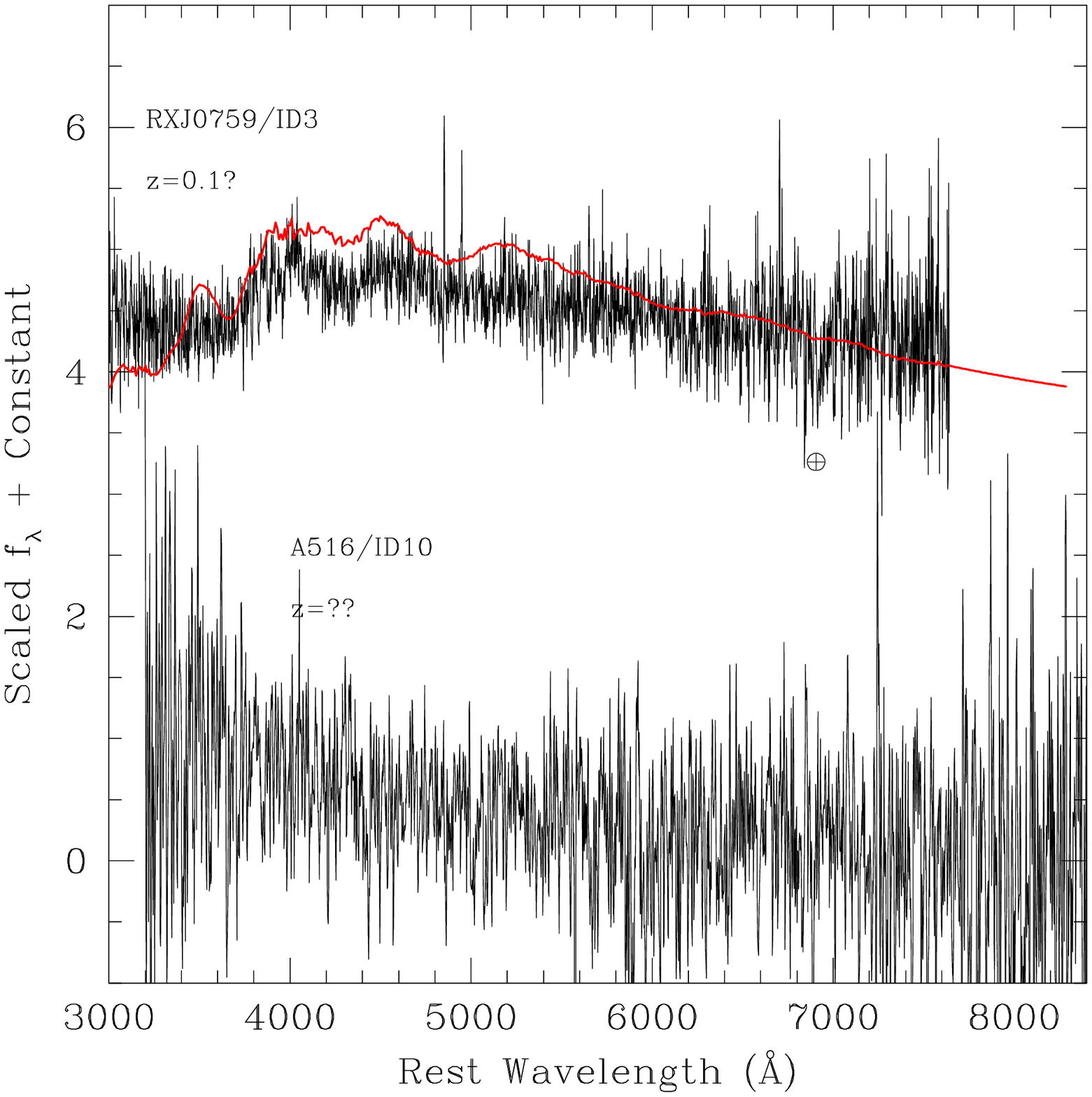}}

\caption{Spectra of two unknown objects, although we suspect that the
top spectrum may be a SNI b/c roughly eight days past explosion at a
redshift of $z=0.1$, with some obvious galaxy contamination.  We
overplot a template SN spectrum with these characteristics using the
template of \citet{Nugent02}.  Even though Abell 516/ID10 was very
faint at discovery, we devoted some spectroscopic resources to it
since it was an apparent IC event.  The location of the
A-band atmospheric absorption feature is indicated with a telluric
symbol.  No attempt was made to remove this feature.
\label{fig:unknown}}
\end{center}
\end{figure*}

A summary of our initial spectroscopic observations is presented in
Table~\ref{table:bcs}.  Target selection is difficult to quantify. We
chose targets based on their position on the sky at the given time
during the night, with preference given to spectroscopically observing
multiple targets in the same cluster field in order to minimize slew
and acquisition times.  We gave targets higher priority if they were
near the cluster center ($R<r_{200}$) and slightly offset from their
host galaxy (note that all four of our confirmed SN Ia's are within
$r_{200}$, but two of the four show no apparent offset from their
unresolved galaxy host). The IC SN candidate Abell 516/ID11 (whose
identification was unsuccessful) was also given highest priority.
Otherwise, individual targets were selected more or less at random.
Given our seeing ($\sim$2-2.5 arcsec), it was difficult to
include host galaxy morphology as a criteria for spectroscopic
selection.  A seeing of 2 arcseconds corresponds to $\sim$3.7-6.6 kpc
throughout our cluster redshift range.  To determine accurate, broad
morphological classification, however, requires resolutions of $\sim$1
kpc or better \citep{Lotz04}.  All spectra are presented in
Figures~\ref{fig:sn1a}, \ref{fig:speclozgals}, \ref{fig:hizagn},
\ref{fig:lozagn}, \ref{fig:specstars}, and \ref{fig:unknown}.  Of the
50 unique transient sources (that is, transients that were not
found to be variable in other epochs) discovered in our December data
(Table~\ref{table:vartable}), we were able to observe 25
spectroscopically, although two of these transients were not
confidently confirmed spectroscopically.  We now briefly present the
four broad categories of transients we found and highlight individual
interesting objects.  We conclude this section with a
discussion.

\subsection{Supernova Ia}

We confirmed four type Ia supernova \citep{Sand07} in our initial
spectroscopic followup (Figure~\ref{fig:sn1a}), one of which is
clearly associated with the cluster in our sample (A1246/ID10).  The
other three are background.  We match our spectra to spectral
templates taken from \citet{Nugent02}.  No formal fit was done, but
matching by eye was sufficient to identify SNe Ia and estimate the
time since peak luminosity.

\subsection{Variable Stars}

Five of the transients targeted for spectroscopy proved to be 
stars (Figure~\ref{fig:specstars}).

\subsection{Quasars}

We spectroscopically identified several previously unknown quasars
selected from their optical variability, with a redshift range between
$z\sim$0.8 and 2.9, and we present their spectra in
Figures~\ref{fig:hizagn} and \ref{fig:lozagn}.  As expected, the
transient position is always coincident with the center of the galaxy
host position.

\subsection{Low-z Galaxies and AGN}\label{sec:lowzspecs}

Our spectroscopic sample includes six low redshift ($z<0.3$) galaxies
(Figure~\ref{fig:speclozgals}). The host galaxy for transient
Z0256/ID19 is a cluster member, and since there are no star
formation signatures this galaxy may have hosted a SN Ia. The
cluster redshift for Abell 136 ($z=0.157$) is based on one galaxy
redshift \citep{Struble91} and so we tentatively consider A136/ID7 to
also be a cluster member.  Since this host galaxy has [O II] in
emission, indicating star formation, the nature of any hidden SN is
more ambiguous since both CC and SN Ia occur in star forming regions.
Both of these transients are coincident with the galaxy host core.
Future followup spectroscopy of these host galaxies will allow us to
subtract out the host contamination and possibly identify underlying
cluster SNe in these systems.

Two other sources show signatures of AGN activity.  Broad MgII in
A743/ID10 indicates that this is an AGN.  We measure the redshift of
RXJ821/ID15 to be at $z=0.22$ from the Ca H and K absorption lines.
At this redshift, the broad emission line near 2800\AA~(assuming
$z=0.22$) is not at the correct wavelength to be MgII.  There is a
point source $\sim$1.7 arcseconds from the galaxy which is the source
of the observed variability (we centered our slit on this point source
and not the intervening galaxy, although contamination was
inevitable).  We tentatively conclude that this single broad emission
line is from a quasar at unknown redshift and that this is the source
of the observed variability, rather than the $z=0.22$ foreground
galaxy that is contaminating our spectrum.  Similar flares from
background AGN superimposed on foreground 'host' galaxies have
contaminated SN surveys in the past \citep{Gal-Yam05}.  Finally, it
should be noted that just because a host galaxy shows AGN activity
does not mean that this is the source of the transient; a SNe is still
possible in these systems.

\subsection{Unknown}\label{sec:unknownspec}

We have two spectra that we could not confidently identify
(Figure~\ref{fig:unknown}).  The spectrum of A516/ID10 corresponds to
a very faint, but IC SN candidate.  This is a blue,
relatively featureless spectrum.  The other transient is offset from
the host galaxy center with a spectrum that is also relatively blue,
but has several broad features.  We tentatively identify this spectrum
as that of a SN Ib/c roughly eight days post explosion. The
correspondence can be judged relative to the overplotted template
spectrum \citep{Nugent02}.  However, due to significant host galaxy
contamination and relatively low signal to noise, we categorize this
identification as tentative.

\subsection{Discussion of spectroscopic results}

It is worth comparing our spectroscopic results with those
obtained in similar general SNe and cluster SNe searches.  We discuss
three subjects in particular -- the AGN frequency, the relative number
of background/foreground SNe to cluster SNe, and the field ratio of SN
Ia to CC SN. Throughout this discussion, it must be kept in
mind that we are dealing with small number statistics based on one
night's worth of spectroscopy.


AGN are a perennial contaminant in SNe surveys, some of which
preferentially avoid SNe candidates in galaxy cores for this specific
reason \citep[e.g.~Section 3 of][]{Matheson05} or have many epochs of
imaging with which to identify intermittent AGN flaring.  Given our
poor seeing, many of our SNe candidates appear centered on galaxies
(including unresolved galaxies); indeed, two of the four SNe
spectroscopically discovered were coincident with point sources 
in the reference frame.  This suggests that with our seeing
conditions, we should not avoid SN candidates for followup which are
in the nuclear regions of their host.  It is difficult for us to
compare our fraction of AGN to SNe with other surveys since very few
publish all of their spectra or have much more involved SNe candidate
identification algorithms.

Of the four confirmed SNe Ia, one is in a cluster and three are
background.  While these numbers are subject to incompleteness, very
small number statistics, and survey field mismatch, it is worth
comparing them to the other cluster SNe surveys.  First, the Mount
Stromlo Abell Cluster Supernova Search discovered 48 SNe, half of
which were hosted by cluster members \citep{Germany04}.  Similarly,
the WOOTS cluster SNe survey found that seven out of their twelve
confirmed SNe were in the targeted clusters.  While it appears that
these other cluster SNe surveys found a slightly higher cluster to
field SNe ratio, given our small sample size we are in agreement with
this previous work.

Finally, the ratio of field SNe Ia to CC SNe (three to one
if the ambiguous spectrum from \S~\ref{sec:unknownspec} is a field SN
Ib/c, although our consistency does not change regardless) is
consistent with the spectroscopically complete field results of
\citet{GalYam07a}, who found that four out of five of their field SN
were SN Ia, although this is a difficult comparison to make since our
survey is not spectroscopically complete.  As in the flux-limited
WOOTS survey, the effective volume probed for the more luminous SNe Ia
is large enough that it outweighs the fact that CC SN are
more common per unit volume.

From the above, it is clear that future work must focus on
intelligently selecting transient targets for spectroscopic followup.
For this purpose, we are obtaining multi-band, multi-epoch data for
all of our transients as this survey continues in order to more
accurately determine a phototype prior to spectroscopic followup
\citep[e.g.~][]{Poznanski06}.

\section{Is there an overdensity of transients associated with the clusters?}\label{sec:overdense}

While spectroscopic confirmation is the most direct method of
confirming an association between transients and the cluster, a
statistical excess within the cluster is another method of confirming
a general correspondence.  This approach may not, however, enable a
full distinction between certain classes of transient. For example,
while variable stars will not correlate with the cluster position,
both SNe and AGN will.

To examine the statistical properties of the candidate transients, we
place all of the clusters on a similar physical scale by taking their
published X-ray luminosities, as shown in
Table~\ref{table:clustertable}, and the $L_{X}$-$M_{200}$ relation as
found by \citet{Reiprich02} to calculate $M_{200}$ for each cluster.
After converting $M_{200}$ to our adopted $\Lambda$CDM cosmology, we
then calculate $r_{200}$ using $M_{200}$ = $200(4/3 \pi
r_{200}^{3}\rho_{crit})$, where $\rho_{crit}$ is evaluated at the
cluster redshift.

We plot the transient projected density in the cluster-centered CCD
chip 1 as a function of scaled radius, $R/r_{200}$, in
Figure~\ref{fig:radexcess}.  Error bars are based on Poisson
statistics.
We measure the 'background' or 'field' projected density of transients
using the off cluster CCD chip 3 events (Table~\ref{table:vartable3}).
This background transient rate, $0.45\pm0.05$ per $A_{200}$ ($A_{200}$
is the area enclosed within $R_{200}$), is plotted as a horizontal
band in Figure~\ref{fig:radexcess}.  Beyond a scaled radius of 1.25,
the cluster transient projected density matches the background.

The number of transients per $A_{200}$ within $R/r_{200}=1$ is
$0.78\pm0.08$.  The statistical significance of the excess in this
inner region relative to the background rate is 3.7 $\sigma$. In terms
of the 107 unique candidates we have (within $r_{200}$), these numbers
suggest that we have roughly 45 cluster-associated transients within
$R/r_{200} = 1$.

\section{Cluster Transient Populations}\label{sec:whatrexcess}

There are three plausible sources for the central excess of cluster
transients: cluster SN Ia, cluster CC SN, and cluster AGN.
We now discuss the relative contribution of each of these classes.

\subsection{Cluster SN-Ia}\label{sec:howmanysnia}

To estimate the expected number of cluster SN Ia in our sample we
utilize the standard SN Ia rate per unit stellar mass equation as
calculated by a survey:
\begin{equation}\label{eqn:sn1arate}
R_{SN Ia}{=}\frac{N}{\sum_{i}\Delta t_{i} M_{\odot,i}}
\end{equation}

\noindent where $N$ is the number of SN Ia discovered, $\Delta t_{i}$
is the visibility time of the $i$-th observation, and $M_{\odot,i}$ is
the stellar mass of the cluster being observed in observation $i$.
The sum is over all individual observation epochs.  By solving for
$N$, given known visibility times, cluster stellar masses, and an
adopted SN Ia rate from the literature we can approximately determine
the fraction of our excess cluster transients that are SN Ia.

For the SN Ia rate, we adopt the rate measured by \citet{Sharon07},
$0.098$ SNuM, for several reasons.  First, their sample is the most
analogous to the current sample -- they measured the SN Ia rate in
galaxy clusters at $0.06 < z < 0.19$.  Second, this rate is a
relatively large one in comparison to other SN Ia rate measurements of
E/S0 galaxy populations (although it is consistent with other measures
within the uncertainties) and so provides a conservative upper limit
on the the number of SN Ia expected in the current survey.  Note
that if we adopted the SN Ia rate in local galaxy clusters found by
\citet{Mannucci07b} -- which is consistent within the uncertainties of
Sharon et al. -- our final SN Ia number would be reduced by two-thirds.
For our purposes of estimating the expected number of SN Ia at the
factor of two level, the adopted rate is not crucial.

To calculate the stellar masses of our galaxy clusters, we utilize the
$L_{K,200}-M_{200}$ relation found by \citet{Lin04} and our $M_{200}$
values calculated from the $L_{X}$-$M_{200}$ relation utilized in
\S~\ref{sec:overdense}.  Adopting $M/L_{*,K}=0.78$ \citep[see Figure 3
of][]{Lin03}, we assign a stellar mass to each cluster.  This
calculation relies on aspects constructed from different cluster
samples and techniques and so should be treated cautiously.

Finally, we need a measure of the effective visibility time of an
observation, $\Delta t$, or the time during which a SN Ia in the cluster
would be visible given the detection limits and transient detection
efficiency of that image.
We postpone a detailed calculation of the visibility time till future
work where we will measure the SN Ia rate in low redshift clusters
directly.  For the moment, an accurate estimate of the visibility time
will allow us to estimate the number of SN Ia we should have observed
thus far.  With this goal in mind, we use a highly simplified model
for our detection efficiency based on our calculations in
\S~\ref{sec:efficiency},

\begin{equation}\label{eqn:deteff_par}
\eta(m)= \left\{ \begin{array}{ll}
  0,&m\le18.5 \\
0.8,&18.5 < m < 21.5 \\
0,&m > 21.5 \\
\end{array} \right.
\end{equation}

\noindent Although this parameterization is only approximate,
it has the advantage of being simple to implement.

The final step is to determine the time that a SN Ia spends between
$18.5 < g < 21.5$ as a function of cluster redshift.  We adopt a
non-stretched peak absolute B-magnitude for SN Ia of $M_{B}=-19.25$
\citep{Sullivan06}.  In this estimate, we neglect the uncertainty in
this value ($\sim$0.15 mag) and the correlation between light curve
width and peak magnitude \citep[SN Ia with a brighter peak magnitude
have a longer rise and fall time around maximum brightness;
parameterized by the 'stretch' relation, e.g,][]{perlmutter99}. The
latter leads to an upper limit because the SN Ia originating from old
stellar populations, such as those in clusters, tend to have a smaller
stretch factor than typical \citep{Sullivan06}.  To obtain $g$ band
light curves for SN Ia at a given cluster redshift, we begin with the
multiepoch SN Ia spectral and photometric templates of
\citet{Nugent02}. Using the $B-$band filter transmission curves, we
normalize the \citet{Nugent02} spectra to produce the correct $B$
magnitude at a given point on the light curve, redshift the spectra to
the cluster redshift, and then synthesize the $g$ magnitude using the
$g-$band filter transmission curve.  This method is analogous to that
presented by \citet{Sharon07}, and we show sample light curves at
$z=0.1,0.2$ in Figure~\ref{fig:samplecurve}.  As a check, we also
considered host galaxy extinction by taking $E(B-V)=0.2$ \citep[the
1-$\sigma$ dispersion value used by][]{Neill06} and rerunning our
analysis.  Including this level of host extinction would cause
$\sim$30\% fewer cluster SN Ia to be indentified by our survey, which
for the purposes of the following discussion we neglect.  In any case,
the inclusion of host galaxy extinction into our calculation would
increase the number of unaccounted for cluster transients.

Putting all of these ingredients together, we calculate that $N$ in
Eqn~\ref{eqn:sn1arate} is $\sim$10.  Although this number is a coarse
estimate for reasons outlined above, the uncertainty in this number is
tens of percent rather than a factor of two.  If we say that $\sim$10 of
the excess cluster transients are cluster SN Ia, then the other
$\sim$30-35 excess events must be either CC SN or cluster
AGN. Importantly, the vast majority of the candidates are not SN Ia's.

Because of the dominance of contamination, it is imperative to
classify the transients. Spectroscopy is clearly the preferred way to
do this and from the results presented so far we can test whether the
above estimate is reasonable. If we divide the number of expected
cluster SN Ia (ten) by the number of imaging nights presented in this
study (thirteen) to conclude that our detection rate is $\sim$0.8
cluster SN Ia per imaging night (the true detection rate of cluster SN
Ia is slightly higher because of partial night losses due to weather
and instrument problems, but this simple calculation serves our
purpose). This value is consistent with the one cluster SN Ia
spectroscopically confirmed.
It is difficult at this time to push further on this comparison
because the cluster SN Ia (Abell 1246/ID10) was preferentially
observed due to its proximity to the cluster center and clear offset
from the host galaxy.



A separate consistency check involves comparing the number of expected
SN Ia with the number of IC SN candidates (although it should be noted
that all of our IC SN candidates are at $R>r_{200}$ while the number
of expected cluster SN Ia was calculated for $R<r_{200}$).  Assuming
that all three of the IC SN candidates found in the cluster-centered
chip 1 are actual IC SN Ia, then $\sim$3/13 would be the resulting
hostless to total SN Ia fraction, implying a $\sim$20\% IC stellar
mass fraction.  This number is in broad agreement with direct
measurements of the IC stellar population
\citep{Feldmeier04,Gonzalez05,Zibetti05,Krick07}.  We discuss the IC
SN candidates further in \S~\ref{sec:ICspec}.


\subsection{Core Collapse SN}

Core collapse SN progenitors are massive, necessarily young stars,
which is consistent with the negligibly small CC supernova rate in
E/S0 type galaxies \citep[e.g.,][]{Mannucci05}.  Therefore one might
not expect CC SN to contribute significantly to our survey.  In
general, CC SN are fainter than their SN Ia counterparts, leading to
shorter visibility times.  In detail this is not always the case
since individual CC SNe can be quite bright or have a very long
plateau phase \citep{Hamuy03}, which would prolong the visibility
time.  However, it so happens that, as we describe below, the CC SN
rate in our survey is comparable to that of Ia's.

We begin the exploration of this class by examining results from
previous surveys. The Mount Stromlo Abell cluster supernova survey
\citep{Germany04,Reiss98}, which ran for 3.5 years, searched for SN in
galaxy clusters at a redshift of $0.04 < z < 0.2$.  Of the 25 cluster
SN discovered, 14 were SN Ia (identified spectroscopically or by their
light curve), while the remaining 11 are likely to be CC events
(because the Mount Stromlo Abell cluster supernova survey obtained
light curves for all their candidate SN, it is not likely that these
events were AGN or variable stars).
Alternatively, taking the 136 SNe sample of \citet{Cappellaro99}, and
selecting the cluster SNe, \citet{Mannucci07b} found 44 cluster SNe, 16
of which are CC.  These two results suggest that somewhere in the
neighborhood of 40\% of the detected cluster SNe may be CC.

Empirically, no cluster SNe search has uncovered more CC SNe than
SNe Ia \citep{Germany04,GalYam07a,Mannucci07b}.  As such, we
conservatively suggest that even if $\sim$10 of our excess cluster
transients are due to CC SN, we still have an unexplained population
of $\sim$20-25 cluster transients.

\subsection{Cluster AGN}

Active galactic nuclei and quasars exhibit optical variability on a
variety of time scales, from days to years
\citep{Webb00,Totani05,Klesman07}, with somewhere between 30-50\% of
all AGN displaying variability during the course of these surveys.
Indeed, many of the transients found in this survey with known
counterparts are QSOs (see \S~\ref{sec:efficiency}).  We now test
whether the remaining detected central transients can be AGN.

Multiwavelength studies of clusters find that roughly 5\% of cluster
members with $M_{R} < -20$ are AGN \citep{Martini07}.  These objects
are more centrally concentrated than cluster galaxies of the same
luminosity, although \citet{Martini07} only probe the area covered by
the ACIS-I chips on board {\it Chandra}, which is always $R< r_{200}$
for their sample.
Adopting a 5\% AGN fraction and applying the \citet{Popesso07}
$N_{200}$ vs $M_{200}$ scaling relation (for $M_{r} < -20$) we
calculate that our sample contains $\sim$200 cluster AGN.  Of these,
$\sim30-50$\% are variable implying that as many as 60 to 100 cluster
AGN could have been detected.

There are various reasons why this number exceeds the actual number we
are trying to account for, $\sim 20-25$.  First, our temporal coverage
is sparse and many of our clusters only have two imaging epochs so
far.  As the survey continues we expect to detect more of the cluster
AGN population.
Also, our estimate for the number of cluster AGNs relies heavily on
galaxy cluster scaling relations based on different samples and
techniques, which can plausibly introduce an error of a factor of
$\sim$2.  Although this doesn't introduce a bias in one direction, it
does result in making the difference between the prediction and
observation less statistically significant.  We conclude that it is
plausible that the remainder of the statistical transient excess in
clusters is due to AGN.  Indeed, if we have over-estimated the
number of cluster SNIa and CC SNe in the previous two sections (we
have tried to get them correct at the factor of $\sim$2 level), it is
also plausible that cluster AGN can make up the difference and account
for the cluster transient excess.

\section{The Intracluster Supernovae Candidates}\label{sec:ICspec}

As has been discussed, the number of IC SN candidates in the
cluster-centered chip 1 is in rough agreement with the expected
$\sim$10 cluster SN Ia in the survey thus far (see
\S~\ref{sec:iccands} and \ref{sec:howmanysnia}).  Even if the
simplest explanation for the three IC SN candidates (all at
$R>r_{200}$) is that they are background events whose host was not
detectable in our stacked images, it is interesting for the sake of
argument to explore the implications of the three IC SN
candidates being associated with the clusters.  Note that the
following discussion is not significantly altered if one takes the
'background' rate of hostless events to be that measured by chip 3,
from which we would conclude that only two out of three of our IC SN
candidates are actually associated with the cluster.

A comparison between our IC SN candidates and ICL studies in the
literature points to an untapped regime in IC studies.  First the
three IC SN candidates all lie at $R>r_{200}$ (or $R>1500$ kpc) from
the BCG.  This is an unexplored radial regime among observational ICL
studies, given that the study with the highest radial measurement is
that of \citet{Zibetti05} who stacked 683 SDSS clusters and was able
to probe the average ICL out to $\sim$900 kpc.  A simple reference
point from \citet{Gonzalez07} can be used to show how unusual three IC
SN at $R>r_{200}$ really is.  From their sample of 24 clusters with
direct low surface brightness measurements of the ICL,
\citet{Gonzalez07} find that 80\% of the BCG+ICL light (of which the
ICL dominates) is located within 300 kpc of the cluster center,
although extrapolation is necessary to extend this result out to
$R>r_{200}$.  Using this as a baseline, finding three IC SN at
$R>r_{200}$ would imply that we would expect at least 12 IC SN within
the central 300 kpc of our clusters!  The fact that none were found
suggests that either the three IC SN are in fact not associated with
the cluster, there is an excess of heretofore undetected ICL beyond
$r_{200}$, or that the stellar population in the cluster outskirts is
more conducive to SN events.  The answer could plausibly be a
combination of the second and third possibilities just mentioned.  For
instance, infalling groups at the cluster outskirts would presumably
have their own ICL (unaccounted for in studies at smaller radii), and
it may be IC SN in these infalling groups that we are detecting.  If
these IC stars (whether or not they are associated with infalling
groups) are more recently stripped than those in the central regions
of the cluster (or if a sprinkle of star formation has been induced in
the process of the stars becoming unbound), then these outskirt IC SN
could be either CC events (which is compatible with their
single epoch magnitude) or prompt SN Ia seen to be associated with
star forming regions.  In any case, the possibility of IC SN
in the outskirts of clusters is extremely intriguing and warrants a
more coordinated effort.  To investigate this potential large cluster
radius IC SN population further requires continued cluster SN searches
utilizing wide-field imagers capable of imaging out to the cluster
outskirts.

It should also be noted that given that this initial phase of the
survey expected to find $\sim$10 cluster SN, some IC SN candidates
within $r_{200}$ would be expected given that the ICL fraction
typically seen in the literature is $\sim$20\%.  The fact that we
found no candidates within $r_{200}$ is surprising, and a larger data
set is needed to investigate if the true cluster SN Ia rate is lower
than expected from previous cluster studies.  For instance, if we take
the early galaxy type SN Ia rate measurement from \citet{Mannucci05}
(with a SN Ia rate of 0.038 SNuM) as our cluster SN Ia rate rather
than of \citet{Sharon07}, then we would have expected only $\sim$4
cluster SN Ia in the survey thus far and plausibly no IC SN Ia.
Further study is needed to resolve this situation.


\section{Summary and Future Work}\label{sec:future}

We have presented initial results from our program to search for and
identify SNe Ia in a sample of X-ray selected clusters at $0.1<z<0.2$.
Of particular interest are hostless SN whose progenitors are IC stars.
This survey will measure the SN Ia rate in galaxy clusters and the
mean ICL fraction by noting the relative number of hosted to hostless
events.  Ultimately, we aim to constrain the metal enrichment of
the ICM and the role that IC SNe play.

The data presented in this paper include our intitial
photometry-only campaign followed by our first spectroscopic results.
We use the 30 $\times$ 30 arcminute field of view of Chip 1 on the
90Prime imager as our cluster-centered field, which allows us to probe
beyond $r_{200}$ in all of our clusters.  We also use one of the
off-cluster CCDs as a probe of the background rate of transients.

The initial results presented in the paper include -- 

\begin{enumerate} 

\item Using our automated pipeline -- which reduces, difference
images, and posts candidate transients to the web for human inspection
-- we have identified 218 unique transient sources in the
cluster-centered Chip 1 (107 of these were within $r_{200}$) and 102
on the off-cluster Chip 3.

\item Of the transient sources we have identified, four were
apparently hostless events (three of the four were in the
cluster-centered chip 1).  Continued imaging of these fields will put
tighter constraints on possible faint hosts, while our current limits
reach a magnitude limit of at least $g\sim$25 (corresponding to a
galaxy of $M_g > -14.3$ in all cases).  Interestingly, the three IC
candidates on the cluster-centered chip were all at $R > r_{200}$,
indicating that they are either not genuine cluster events or that
some mechanism must be invoked to cause an excess of IC SN in the
cluster outskirts.  In any case, further study is needed.


\item Initial spectroscopic results from one night of MMT Blue Channel
Spectrograph data shows that we can identify and followup cluster SN
Ia and other interesting variables.  Of the four confirmed SNe Ia, one
is in a cluster. Additionally, we tentatively identify a SNe Ib/c that
could be in a cluster. Spectroscopic followup will become a major
focus as the survey continues.

\item Plotting the number of transients as a function of a scaled
radius, $R/r_{200}$, reveals an excess of transients within $1.25
r_{200}$ in our clusters.  Of the $\sim$40-50 excess transients within
$r_{200}$, we conservatively estimate that $\sim$10 of these are
cluster SN Ia, $\sim$10 are CC SN of some form, and the rest are
cluster AGN.

\end{enumerate}

This survey is in its beginning stages and there are several avenues
of future research.  The main focus will continue to be on finding,
and spectroscopically identifying, hostless and hosted cluster SN Ia.
Inevitably, this will allow for other time domain and cluster science.
For instance, our variability survey offers an opportunity to study
cluster AGN out to a much larger radius than is feasible with most
multi-object spectrographs and X-ray instrumentation
\cite[e.g.,][]{Martini07}.  Using the full data set (all four CCDs) to
search for all transients on long, $\sim$monthly, time scales along
with fast optical transients (objects that vary on $\sim$1000 sec time
scales) and faint eclipsing systems are among our future plans.

We are in the process of obtaining deep auxiliary r-band images of all
of our cluster fields.  The resulting red seqence identification will
confidently identify old, early-type galaxies.  Any transients
associated with these galaxies are highly likely to be SN Ia, since
CC SN rarely occur in old stellar populations.  This will
also be a unique, $0.1 < z < 0.2$ data set for studying the cluster
red sequence and the luminous red galaxy halo occupation distribution.

Measuring the SN rate in low redshift galaxy clusters (both from
hosted and hostless systems) will provide a snapshot of the processes
that contribute to the metal enrichment of the ICM.  In combination
with ongoing cluster SN research at higher redshift, we aim to
constrain the SN Ia progenitor associated with old stellar
populations, and ultimately understand the metal enrichment history of
the ICM.

\acknowledgments

We would like to acknowledge the constructive comments of Avishay
Gal-Yam.  We thank Christy Tremonti for help in identifying some of
the spectra presented in this work and Tom Matheson for advice on the
survey and for help typing SNe.  DJS would also like to thank Richard
Cool and Ed Olszewski for advice on using 90Prime and reducing the
data.  DJS acknowledges support provided by NASA through Chandra
Postdoctoral Fellowship grant number PF5-60041.  DZ acknowledges that
this research was supported in part by the National Science Foundation
under Grant No. PHY99-07949 during his visit to KITP, a Guggenheim
fellowship, generous support from the NYU Physics department and
Center for Cosmology and Particle Physics during his sabbatical there,
NASA LTSA grant NNG05GE82G, and NSF grant AST-0307482.  This research
has made use of the X-Rays Clusters Database (BAX) which is operated
by the Laboratoire d'Astrophysique de Tarbes-Toulouse (LATT), under
contract with the Centre National d'Etudes Spatiales (CNES).  This
research has made use of the NASA/IPAC Extragalactic Database (NED)
which is operated by the Jet Propulsion Laboratory, California
Institute of Technology, under contract with the National Aeronautics
and Space Administration.

\bibliographystyle{apj}
\bibliography{apj-jour,mybib}

\begin{thebibliography}{72}
\expandafter\ifx\csname natexlab\endcsname\relax\def\natexlab#1{#1}\fi

\bibitem[{{Adelman-McCarthy} {et~al.}(2006){Adelman-McCarthy}, {Ag{\"u}eros},
  {Allam}, {Anderson}, {Anderson}, {Annis}, {Bahcall}, {Baldry}, {Barentine},
  {Berlind}, {Bernardi}, {Blanton}, {Boroski}, {Brewington}, {Brinchmann},
  {Brinkmann}, {Brunner}, {Budav{\'a}ri}, {Carey}, {Carr}, {Castander},
  {Connolly}, {Csabai}, {Czarapata}, {Dalcanton}, {Doi}, {Dong}, {Eisenstein},
  {Evans}, {Fan}, {Finkbeiner}, {Friedman}, {Frieman}, {Fukugita}, {Gillespie},
  {Glazebrook}, {Gray}, {Grebel}, {Gunn}, {Gurbani}, {de Haas}, {Hall},
  {Harris}, {Harvanek}, {Hawley}, {Hayes}, {Hendry}, {Hennessy}, {Hindsley},
  {Hirata}, {Hogan}, {Hogg}, {Holmgren}, {Holtzman}, {Ichikawa}, {Ivezi{\'c}},
  {Jester}, {Johnston}, {Jorgensen}, {Juri{\'c}}, {Kent}, {Kleinman}, {Knapp},
  {Kniazev}, {Kron}, {Krzesinski}, {Kuropatkin}, {Lamb}, {Lampeitl}, {Lee},
  {Leger}, {Lin}, {Long}, {Loveday}, {Lupton}, {Margon},
  {Mart{\'{\i}}nez-Delgado}, {Mandelbaum}, {Matsubara}, {McGehee}, {McKay},
  {Meiksin}, {Munn}, {Nakajima}, {Nash}, {Neilsen}, {Newberg}, {Newman},
  {Nichol}, {Nicinski}, {Nieto-Santisteban}, {Nitta}, {O'Mullane}, {Okamura},
  {Owen}, {Padmanabhan}, {Pauls}, {Peoples}, {Pier}, {Pope}, {Pourbaix},
  {Quinn}, {Richards}, {Richmond}, {Rockosi}, {Schlegel}, {Schneider},
  {Schroeder}, {Scranton}, {Seljak}, {Sheldon}, {Shimasaku}, {Smith}, {Smol{\v
  c}i{\'c}}, {Snedden}, {Stoughton}, {Strauss}, {SubbaRao}, {Szalay},
  {Szapudi}, {Szkody}, {Tegmark}, {Thakar}, {Tucker}, {Uomoto}, {Vanden Berk},
  {Vandenberg}, {Vogeley}, {Voges}, {Vogt}, {Walkowicz}, {Weinberg}, {West},
  {White}, {Xu}, {Yanny}, {Yocum}, {York}, {Zehavi}, {Zibetti}, \&
  {Zucker}}]{sdsscite}
{Adelman-McCarthy}, J.~K., {Ag{\"u}eros}, M.~A., {Allam}, S.~S., {Anderson},
  K.~S.~J., {Anderson}, S.~F., {Annis}, J., {Bahcall}, N.~A., {Baldry}, I.~K.,
  {Barentine}, J.~C., {Berlind}, A., {Bernardi}, M., {Blanton}, M.~R.,
  {Boroski}, W.~N., {Brewington}, H.~J., {Brinchmann}, J., {Brinkmann}, J.,
  {Brunner}, R.~J., {Budav{\'a}ri}, T., {Carey}, L.~N., {Carr}, M.~A.,
  {Castander}, F.~J., {Connolly}, A.~J., {Csabai}, I., {Czarapata}, P.~C.,
  {Dalcanton}, J.~J., {Doi}, M., {Dong}, F., {Eisenstein}, D.~J., {Evans},
  M.~L., {Fan}, X., {Finkbeiner}, D.~P., {Friedman}, S.~D., {Frieman}, J.~A.,
  {Fukugita}, M., {Gillespie}, B., {Glazebrook}, K., {Gray}, J., {Grebel},
  E.~K., {Gunn}, J.~E., {Gurbani}, V.~K., {de Haas}, E., {Hall}, P.~B.,
  {Harris}, F.~H., {Harvanek}, M., {Hawley}, S.~L., {Hayes}, J., {Hendry},
  J.~S., {Hennessy}, G.~S., {Hindsley}, R.~B., {Hirata}, C.~M., {Hogan}, C.~J.,
  {Hogg}, D.~W., {Holmgren}, D.~J., {Holtzman}, J.~A., {Ichikawa}, S.-i.,
  {Ivezi{\'c}}, {\v Z}., {Jester}, S., {Johnston}, D.~E., {Jorgensen}, A.~M.,
  {Juri{\'c}}, M., {Kent}, S.~M., {Kleinman}, S.~J., {Knapp}, G.~R., {Kniazev},
  A.~Y., {Kron}, R.~G., {Krzesinski}, J., {Kuropatkin}, N., {Lamb}, D.~Q.,
  {Lampeitl}, H., {Lee}, B.~C., {Leger}, R.~F., {Lin}, H., {Long}, D.~C.,
  {Loveday}, J., {Lupton}, R.~H., {Margon}, B., {Mart{\'{\i}}nez-Delgado}, D.,
  {Mandelbaum}, R., {Matsubara}, T., {McGehee}, P.~M., {McKay}, T.~A.,
  {Meiksin}, A., {Munn}, J.~A., {Nakajima}, R., {Nash}, T., {Neilsen}, Jr.,
  E.~H., {Newberg}, H.~J., {Newman}, P.~R., {Nichol}, R.~C., {Nicinski}, T.,
  {Nieto-Santisteban}, M., {Nitta}, A., {O'Mullane}, W., {Okamura}, S., {Owen},
  R., {Padmanabhan}, N., {Pauls}, G., {Peoples}, J.~J., {Pier}, J.~R., {Pope},
  A.~C., {Pourbaix}, D., {Quinn}, T.~R., {Richards}, G.~T., {Richmond}, M.~W.,
  {Rockosi}, C.~M., {Schlegel}, D.~J., {Schneider}, D.~P., {Schroeder}, J.,
  {Scranton}, R., {Seljak}, U., {Sheldon}, E., {Shimasaku}, K., {Smith}, J.~A.,
  {Smol{\v c}i{\'c}}, V., {Snedden}, S.~A., {Stoughton}, C., {Strauss}, M.~A.,
  {SubbaRao}, M., {Szalay}, A.~S., {Szapudi}, I., {Szkody}, P., {Tegmark}, M.,
  {Thakar}, A.~R., {Tucker}, D.~L., {Uomoto}, A., {Vanden Berk}, D.~E.,
  {Vandenberg}, J., {Vogeley}, M.~S., {Voges}, W., {Vogt}, N.~P., {Walkowicz},
  L.~M., {Weinberg}, D.~H., {West}, A.~A., {White}, S.~D.~M., {Xu}, Y.,
  {Yanny}, B., {Yocum}, D.~R., {York}, D.~G., {Zehavi}, I., {Zibetti}, S., \&
  {Zucker}, D.~B. 2006, \apjs, 162, 38

\bibitem[{{Alard}(2000)}]{Alard00}
{Alard}, C. 2000, \aaps, 144, 363

\bibitem[{{Balestra} {et~al.}(2007){Balestra}, {Tozzi}, {Ettori}, {Rosati},
  {Borgani}, {Mainieri}, {Norman}, \& {Viola}}]{Balestra07}
{Balestra}, I., {Tozzi}, P., {Ettori}, S., {Rosati}, P., {Borgani}, S.,
  {Mainieri}, V., {Norman}, C., \& {Viola}, M. 2007, \aap, 462, 429

\bibitem[{{Balogh} {et~al.}(2002){Balogh}, {Couch}, {Smail}, {Bower}, \&
  {Glazebrook}}]{Balogh02}
{Balogh}, M.~L., {Couch}, W.~J., {Smail}, I., {Bower}, R.~G., \& {Glazebrook},
  K. 2002, \mnras, 335, 10

\bibitem[{{Becker} {et~al.}(2004){Becker}, {Wittman}, {Boeshaar},
  {Clocchiatti}, {Dell'Antonio}, {Frail}, {Halpern}, {Margoniner}, {Norman},
  {Tyson}, \& {Schommer}}]{Becker04}
{Becker}, A.~C., {Wittman}, D.~M., {Boeshaar}, P.~C., {Clocchiatti}, A.,
  {Dell'Antonio}, I.~P., {Frail}, D.~A., {Halpern}, J., {Margoniner}, V.~E.,
  {Norman}, D., {Tyson}, J.~A., \& {Schommer}, R.~A. 2004, \apj, 611, 418

\bibitem[{{Bertin} \& {Arnouts}(1996)}]{sexbib}
{Bertin}, E. \& {Arnouts}, S. 1996, \aaps, 117, 393

\bibitem[{{Cappellaro} {et~al.}(1999){Cappellaro}, {Evans}, \&
  {Turatto}}]{Cappellaro99}
{Cappellaro}, E., {Evans}, R., \& {Turatto}, M. 1999, \aap, 351, 459

\bibitem[{{Croom} {et~al.}(2004){Croom}, {Smith}, {Boyle}, {Shanks}, {Miller},
  {Outram}, \& {Loaring}}]{Croom04}
{Croom}, S.~M., {Smith}, R.~J., {Boyle}, B.~J., {Shanks}, T., {Miller}, L.,
  {Outram}, P.~J., \& {Loaring}, N.~S. 2004, \mnras, 349, 1397

\bibitem[{{Domainko} {et~al.}(2004){Domainko}, {Gitti}, {Schindler}, \&
  {Kapferer}}]{Domainko04}
{Domainko}, W., {Gitti}, M., {Schindler}, S., \& {Kapferer}, W. 2004, \aap,
  425, L21

\bibitem[{{Dupke} \& {White}(2000)}]{Dupke00}
{Dupke}, R.~A. \& {White}, III, R.~E. 2000, \apj, 537, 123

\bibitem[{{Feldmeier} {et~al.}(2004){Feldmeier}, {Mihos}, {Morrison},
  {Harding}, {Kaib}, \& {Dubinski}}]{Feldmeier04}
{Feldmeier}, J.~J., {Mihos}, J.~C., {Morrison}, H.~L., {Harding}, P., {Kaib},
  N., \& {Dubinski}, J. 2004, \apj, 609, 617

\bibitem[{{Filippenko}(2005)}]{Filippenko95}
{Filippenko}, A.~V. 2005, in Astrophysics and Space Science Library, Vol. 332,
  White dwarfs: cosmological and galactic probes, ed. E.~M. {Sion},
  S.~{Vennes}, \& H.~L. {Shipman}, 97--133

\bibitem[{{Gal-Yam}(2005)}]{Gal-Yam05}
{Gal-Yam}, A. 2005, The Astronomer's Telegram, 586, 1

\bibitem[{{Gal-Yam} {et~al.}(2003){Gal-Yam}, {Maoz}, {Guhathakurta}, \&
  {Filippenko}}]{Galyam03}
{Gal-Yam}, A., {Maoz}, D., {Guhathakurta}, P., \& {Filippenko}, A.~V. 2003,
  \aj, 125, 1087

\bibitem[{{Gal-Yam} {et~al.}(2007){Gal-Yam}, {Maoz}, {Guhathakurta}, \&
  {Filippenko}}]{GalYam07a}
---. 2007, ArXiv e-prints, 711

\bibitem[{{Gal-Yam} {et~al.}(2002){Gal-Yam}, {Maoz}, \& {Sharon}}]{Galyam02}
{Gal-Yam}, A., {Maoz}, D., \& {Sharon}, K. 2002, \mnras, 332, 37

\bibitem[{{Gandhi} {et~al.}(2004){Gandhi}, {Crawford}, {Fabian}, \&
  {Johnstone}}]{Gandhi04}
{Gandhi}, P., {Crawford}, C.~S., {Fabian}, A.~C., \& {Johnstone}, R.~M. 2004,
  \mnras, 348, 529

\bibitem[{{Germany} {et~al.}(2004){Germany}, {Reiss}, {Schmidt}, {Stubbs}, \&
  {Suntzeff}}]{Germany04}
{Germany}, L.~M., {Reiss}, D.~J., {Schmidt}, B.~P., {Stubbs}, C.~W., \&
  {Suntzeff}, N.~B. 2004, \aap, 415, 863

\bibitem[{{Gonzalez} {et~al.}(2005){Gonzalez}, {Zabludoff}, \&
  {Zaritsky}}]{Gonzalez05}
{Gonzalez}, A.~H., {Zabludoff}, A.~I., \& {Zaritsky}, D. 2005, \apj, 618, 195

\bibitem[{{Gonzalez} {et~al.}(2007){Gonzalez}, {Zaritsky}, \&
  {Zabludoff}}]{Gonzalez07}
{Gonzalez}, A.~H., {Zaritsky}, D., \& {Zabludoff}, A.~I. 2007, ArXiv e-prints,
  705

\bibitem[{{Hamuy}(2003)}]{Hamuy03}
{Hamuy}, M. 2003, \apj, 582, 905

\bibitem[{{Hao} {et~al.}(2005){Hao}, {Strauss}, {Tremonti}, {Schlegel},
  {Heckman}, {Kauffmann}, {Blanton}, {Fan}, {Gunn}, {Hall}, {Ivezi{\'c}},
  {Knapp}, {Krolik}, {Lupton}, {Richards}, {Schneider}, {Strateva}, {Zakamska},
  {Brinkmann}, {Brunner}, \& {Szokoly}}]{Hao05}
{Hao}, L., {Strauss}, M.~A., {Tremonti}, C.~A., {Schlegel}, D.~J., {Heckman},
  T.~M., {Kauffmann}, G., {Blanton}, M.~R., {Fan}, X., {Gunn}, J.~E., {Hall},
  P.~B., {Ivezi{\'c}}, {\v Z}., {Knapp}, G.~R., {Krolik}, J.~H., {Lupton},
  R.~H., {Richards}, G.~T., {Schneider}, D.~P., {Strateva}, I.~V., {Zakamska},
  N.~L., {Brinkmann}, J., {Brunner}, R.~J., \& {Szokoly}, G.~P. 2005, \aj, 129,
  1783

\bibitem[{{Hewett} {et~al.}(1995){Hewett}, {Foltz}, \& {Chaffee}}]{Hewett95}
{Hewett}, P.~C., {Foltz}, C.~B., \& {Chaffee}, F.~H. 1995, \aj, 109, 1498

\bibitem[{{Hillebrandt} \& {Niemeyer}(2000)}]{Hillebrandt00}
{Hillebrandt}, W. \& {Niemeyer}, J.~C. 2000, \araa, 38, 191

\bibitem[{{Holden} {et~al.}(2005){Holden}, {van der Wel}, {Franx},
  {Illingworth}, {Blakeslee}, {van Dokkum}, {Ford}, {Magee}, {Postman}, {Rix},
  \& {Rosati}}]{Holden05}
{Holden}, B.~P., {van der Wel}, A., {Franx}, M., {Illingworth}, G.~D.,
  {Blakeslee}, J.~P., {van Dokkum}, P., {Ford}, H., {Magee}, D., {Postman}, M.,
  {Rix}, H.-W., \& {Rosati}, P. 2005, \apjl, 620, L83

\bibitem[{{Klesman} \& {Sarajedini}(2007)}]{Klesman07}
{Klesman}, A. \& {Sarajedini}, V. 2007, \apj, 665, 225

\bibitem[{{Komiyama} {et~al.}(2002){Komiyama}, {Sekiguchi}, {Kashikawa},
  {Yagi}, {Doi}, {Iye}, {Okamura}, {Shimasaku}, {Yasuda}, {Mobasher}, {Carter},
  {Bridges}, \& {Poggianti}}]{Komiyama02}
{Komiyama}, Y., {Sekiguchi}, M., {Kashikawa}, N., {Yagi}, M., {Doi}, M., {Iye},
  M., {Okamura}, S., {Shimasaku}, K., {Yasuda}, N., {Mobasher}, B., {Carter},
  D., {Bridges}, T.~J., \& {Poggianti}, B.~M. 2002, \apjs, 138, 265

\bibitem[{{Krick} \& {Bernstein}(2007)}]{Krick07}
{Krick}, J.~E. \& {Bernstein}, R.~A. 2007, \aj, 134, 466

\bibitem[{{Lin} \& {Mohr}(2004)}]{Lin04b}
{Lin}, Y.-T. \& {Mohr}, J.~J. 2004, \apj, 617, 879

\bibitem[{{Lin} {et~al.}(2003){Lin}, {Mohr}, \& {Stanford}}]{Lin03}
{Lin}, Y.-T., {Mohr}, J.~J., \& {Stanford}, S.~A. 2003, \apj, 591, 749

\bibitem[{{Lin} {et~al.}(2004){Lin}, {Mohr}, \& {Stanford}}]{Lin04}
---. 2004, \apj, 610, 745

\bibitem[{{Lotz} {et~al.}(2004){Lotz}, {Primack}, \& {Madau}}]{Lotz04}
{Lotz}, J.~M., {Primack}, J., \& {Madau}, P. 2004, \aj, 128, 163

\bibitem[{{Mannucci} {et~al.}(2005){Mannucci}, {Della Valle}, {Panagia},
  {Cappellaro}, {Cresci}, {Maiolino}, {Petrosian}, \& {Turatto}}]{Mannucci05}
{Mannucci}, F., {Della Valle}, M., {Panagia}, N., {Cappellaro}, E., {Cresci},
  G., {Maiolino}, R., {Petrosian}, A., \& {Turatto}, M. 2005, \aap, 433, 807

\bibitem[{{Mannucci} {et~al.}(2007){Mannucci}, {Maoz}, {Sharon}, {Botticella},
  {Della Valle}, {Gal-Yam}, \& {Panagia}}]{Mannucci07b}
{Mannucci}, F., {Maoz}, D., {Sharon}, K., {Botticella}, M.~T., {Della Valle},
  M., {Gal-Yam}, A., \& {Panagia}, N. 2007, ArXiv e-prints, 710

\bibitem[{{Martini} {et~al.}(2007){Martini}, {Mulchaey}, \&
  {Kelson}}]{Martini07}
{Martini}, P., {Mulchaey}, J.~S., \& {Kelson}, D.~D. 2007, ArXiv e-prints, 704

\bibitem[{{Matheson} {et~al.}(2005){Matheson}, {Blondin}, {Foley}, {Chornock},
  {Filippenko}, {Leibundgut}, {Smith}, {Sollerman}, {Spyromilio}, {Kirshner},
  {Clocchiatti}, {Aguilera}, {Barris}, {Becker}, {Challis}, {Covarrubias},
  {Garnavich}, {Hicken}, {Jha}, {Krisciunas}, {Li}, {Miceli}, {Miknaitis},
  {Prieto}, {Rest}, {Riess}, {Salvo}, {Schmidt}, {Stubbs}, {Suntzeff}, \&
  {Tonry}}]{Matheson05}
{Matheson}, T., {Blondin}, S., {Foley}, R.~J., {Chornock}, R., {Filippenko},
  A.~V., {Leibundgut}, B., {Smith}, R.~C., {Sollerman}, J., {Spyromilio}, J.,
  {Kirshner}, R.~P., {Clocchiatti}, A., {Aguilera}, C., {Barris}, B., {Becker},
  A.~C., {Challis}, P., {Covarrubias}, R., {Garnavich}, P., {Hicken}, M.,
  {Jha}, S., {Krisciunas}, K., {Li}, W., {Miceli}, A., {Miknaitis}, G.,
  {Prieto}, J.~L., {Rest}, A., {Riess}, A.~G., {Salvo}, M.~E., {Schmidt},
  B.~P., {Stubbs}, C.~W., {Suntzeff}, N.~B., \& {Tonry}, J.~L. 2005, \aj, 129,
  2352

\bibitem[{{McMahon} {et~al.}(2002){McMahon}, {White}, {Helfand}, \&
  {Becker}}]{McMahon02}
{McMahon}, R.~G., {White}, R.~L., {Helfand}, D.~J., \& {Becker}, R.~H. 2002,
  \apjs, 143, 1

\bibitem[{{Neill} {et~al.}(2006){Neill}, {Sullivan}, {Balam}, {Pritchet},
  {Howell}, {Perrett}, {Astier}, {Aubourg}, {Basa}, {Carlberg}, {Conley},
  {Fabbro}, {Fouchez}, {Guy}, {Hook}, {Pain}, {Palanque-Delabrouille},
  {Regnault}, {Rich}, {Taillet}, {Aldering}, {Antilogus}, {Arsenijevic},
  {Balland}, {Baumont}, {Bronder}, {Ellis}, {Filiol}, {Gon{\c c}alves},
  {Hardin}, {Kowalski}, {Lidman}, {Lusset}, {Mouchet}, {Mourao}, {Perlmutter},
  {Ripoche}, {Schlegel}, \& {Tao}}]{Neill06}
{Neill}, J.~D., {Sullivan}, M., {Balam}, D., {Pritchet}, C.~J., {Howell},
  D.~A., {Perrett}, K., {Astier}, P., {Aubourg}, E., {Basa}, S., {Carlberg},
  R.~G., {Conley}, A., {Fabbro}, S., {Fouchez}, D., {Guy}, J., {Hook}, I.,
  {Pain}, R., {Palanque-Delabrouille}, N., {Regnault}, N., {Rich}, J.,
  {Taillet}, R., {Aldering}, G., {Antilogus}, P., {Arsenijevic}, V., {Balland},
  C., {Baumont}, S., {Bronder}, J., {Ellis}, R.~S., {Filiol}, M., {Gon{\c
  c}alves}, A.~C., {Hardin}, D., {Kowalski}, M., {Lidman}, C., {Lusset}, V.,
  {Mouchet}, M., {Mourao}, A., {Perlmutter}, S., {Ripoche}, P., {Schlegel}, D.,
  \& {Tao}, C. 2006, \aj, 132, 1126

\bibitem[{{Norgaard-Nielsen} {et~al.}(1989){Norgaard-Nielsen}, {Hansen},
  {Jorgensen}, {Aragon Salamanca}, \& {Ellis}}]{Norgaard89}
{Norgaard-Nielsen}, H.~U., {Hansen}, L., {Jorgensen}, H.~E., {Aragon
  Salamanca}, A., \& {Ellis}, R.~S. 1989, \nat, 339, 523

\bibitem[{{Nugent} {et~al.}(2002){Nugent}, {Kim}, \& {Perlmutter}}]{Nugent02}
{Nugent}, P., {Kim}, A., \& {Perlmutter}, S. 2002, \pasp, 114, 803

\bibitem[{{Oegerle} {et~al.}(1991){Oegerle}, {Fitchett}, {Hill}, \&
  {Hintzen}}]{Oegerle91}
{Oegerle}, W.~R., {Fitchett}, M.~J., {Hill}, J.~M., \& {Hintzen}, P. 1991,
  \apj, 376, 46

\bibitem[{{Perlmutter} {et~al.}(1999){Perlmutter}, {Aldering}, {Goldhaber},
  {Knop}, {Nugent}, {Castro}, {Deustua}, {Fabbro}, {Goobar}, {Groom}, {Hook},
  {Kim}, {Kim}, {Lee}, {Nunes}, {Pain}, {Pennypacker}, {Quimby}, {Lidman},
  {Ellis}, {Irwin}, {McMahon}, {Ruiz-Lapuente}, {Walton}, {Schaefer}, {Boyle},
  {Filippenko}, {Matheson}, {Fruchter}, {Panagia}, {Newberg}, {Couch}, \& {The
  Supernova Cosmology Project}}]{perlmutter99}
{Perlmutter}, S., {Aldering}, G., {Goldhaber}, G., {Knop}, R.~A., {Nugent}, P.,
  {Castro}, P.~G., {Deustua}, S., {Fabbro}, S., {Goobar}, A., {Groom}, D.~E.,
  {Hook}, I.~M., {Kim}, A.~G., {Kim}, M.~Y., {Lee}, J.~C., {Nunes}, N.~J.,
  {Pain}, R., {Pennypacker}, C.~R., {Quimby}, R., {Lidman}, C., {Ellis}, R.~S.,
  {Irwin}, M., {McMahon}, R.~G., {Ruiz-Lapuente}, P., {Walton}, N., {Schaefer},
  B., {Boyle}, B.~J., {Filippenko}, A.~V., {Matheson}, T., {Fruchter}, A.~S.,
  {Panagia}, N., {Newberg}, H.~J.~M., {Couch}, W.~J., \& {The Supernova
  Cosmology Project}. 1999, \apj, 517, 565

\bibitem[{{Phillips}(1993)}]{Phillips93}
{Phillips}, M.~M. 1993, \apjl, 413, L105

\bibitem[{{Popesso} {et~al.}(2007){Popesso}, {Biviano}, {B{\"o}hringer}, \&
  {Romaniello}}]{Popesso07}
{Popesso}, P., {Biviano}, A., {B{\"o}hringer}, H., \& {Romaniello}, M. 2007,
  \aap, 464, 451

\bibitem[{{Poznanski} {et~al.}(2006){Poznanski}, {Maoz}, \&
  {Gal-Yam}}]{Poznanski06}
{Poznanski}, D., {Maoz}, D., \& {Gal-Yam}, A. 2006, ArXiv Astrophysics e-prints

\bibitem[{{Reiprich} \& {B{\"o}hringer}(2002)}]{Reiprich02}
{Reiprich}, T.~H. \& {B{\"o}hringer}, H. 2002, \apj, 567, 716

\bibitem[{{Reiss} {et~al.}(1998){Reiss}, {Germany}, {Schmidt}, \&
  {Stubbs}}]{Reiss98}
{Reiss}, D.~J., {Germany}, L.~M., {Schmidt}, B.~P., \& {Stubbs}, C.~W. 1998,
  \aj, 115, 26

\bibitem[{{Rengstorf} {et~al.}(2004){Rengstorf}, {Mufson}, {Andrews},
  {Honeycutt}, {Vivas}, {Abad}, {Adams}, {Bailyn}, {Baltay}, {Bongiovanni},
  {Brice{\~n}o}, {Bruzual}, {Coppi}, {Della Prugna}, {Emmet}, {Ferr{\'{\i}}n},
  {Fuenmayor}, {Gebhard}, {Hern{\'a}ndez}, {Magris}, {Musser}, {Naranjo},
  {Oemler}, {Rosenzweig}, {Sabbey}, {S{\'a}nchez}, {S{\'a}nchez}, {Schaefer},
  {Schenner}, {Sinnott}, {Snyder}, {Sofia}, {Stock}, \& {van
  Altena}}]{Rengstorf04}
{Rengstorf}, A.~W., {Mufson}, S.~L., {Andrews}, P., {Honeycutt}, R.~K.,
  {Vivas}, A.~K., {Abad}, C., {Adams}, B., {Bailyn}, C., {Baltay}, C.,
  {Bongiovanni}, A., {Brice{\~n}o}, C., {Bruzual}, G., {Coppi}, P., {Della
  Prugna}, F., {Emmet}, W., {Ferr{\'{\i}}n}, I., {Fuenmayor}, F., {Gebhard},
  M., {Hern{\'a}ndez}, J., {Magris}, G., {Musser}, J., {Naranjo}, O., {Oemler},
  A., {Rosenzweig}, P., {Sabbey}, C.~N., {S{\'a}nchez}, G., {S{\'a}nchez}, G.,
  {Schaefer}, B., {Schenner}, H., {Sinnott}, J., {Snyder}, J.~A., {Sofia}, S.,
  {Stock}, J., \& {van Altena}, W. 2004, \apj, 617, 184

\bibitem[{{Renzini}(1997)}]{Renzini97}
{Renzini}, A. 1997, \apj, 488, 35

\bibitem[{{Richards} {et~al.}(2004){Richards}, {Nichol}, {Gray}, {Brunner},
  {Lupton}, {Vanden Berk}, {Chong}, {Weinstein}, {Schneider}, {Anderson},
  {Munn}, {Harris}, {Strauss}, {Fan}, {Gunn}, {Ivezi{\'c}}, {York},
  {Brinkmann}, \& {Moore}}]{Richards05}
{Richards}, G.~T., {Nichol}, R.~C., {Gray}, A.~G., {Brunner}, R.~J., {Lupton},
  R.~H., {Vanden Berk}, D.~E., {Chong}, S.~S., {Weinstein}, M.~A., {Schneider},
  D.~P., {Anderson}, S.~F., {Munn}, J.~A., {Harris}, H.~C., {Strauss}, M.~A.,
  {Fan}, X., {Gunn}, J.~E., {Ivezi{\'c}}, {\v Z}., {York}, D.~G., {Brinkmann},
  J., \& {Moore}, A.~W. 2004, \apjs, 155, 257

\bibitem[{{Ruderman} \& {Ebeling}(2005)}]{Ruderman05}
{Ruderman}, J.~T. \& {Ebeling}, H. 2005, \apjl, 623, L81

\bibitem[{{Sand} {et~al.}(2007){Sand}, {Zaritsky}, {Herbert-Fort},
  {Sivanandram}, {Clowe}, \& {Matheson}}]{Sand07}
{Sand}, D., {Zaritsky}, D., {Herbert-Fort}, S., {Sivanandram}, S., {Clowe}, D.,
  \& {Matheson}, T. 2007, Central Bureau Electronic Telegrams, 831, 1

\bibitem[{{Sarajedini} {et~al.}(2003){Sarajedini}, {Gilliland}, \&
  {Kasm}}]{Sarajedini03}
{Sarajedini}, V.~L., {Gilliland}, R.~L., \& {Kasm}, C. 2003, \apj, 599, 173

\bibitem[{{Sarajedini} {et~al.}(2006){Sarajedini}, {Koo}, {Phillips},
  {Kobulnicky}, {Gebhardt}, {Willmer}, {Vogt}, {Laird}, {Im}, {Iverson}, \&
  {Mattos}}]{Sarajedini06}
{Sarajedini}, V.~L., {Koo}, D.~C., {Phillips}, A.~C., {Kobulnicky}, H.~A.,
  {Gebhardt}, K., {Willmer}, C.~N.~A., {Vogt}, N.~P., {Laird}, E., {Im}, M.,
  {Iverson}, S., \& {Mattos}, W. 2006, \apjs, 166, 69

\bibitem[{{Sasaki}(2001)}]{Sasaki01}
{Sasaki}, S. 2001, \pasj, 53, 53

\bibitem[{{Scannapieco} \& {Bildsten}(2005)}]{Scannapieco05}
{Scannapieco}, E. \& {Bildsten}, L. 2005, \apjl, 629, L85

\bibitem[{{Schmidt} {et~al.}(1989){Schmidt}, {Weymann}, \& {Foltz}}]{BCS}
{Schmidt}, G.~D., {Weymann}, R.~J., \& {Foltz}, C.~B. 1989, \pasp, 101, 713

\bibitem[{{Sharon} {et~al.}(2007{\natexlab{a}}){Sharon}, {Gal-Yam}, {Maoz},
  {Donahue}, {Ebeling}, {Ellis}, {Filippenko}, {Foley}, {Freedman}, {Kirshner},
  {Kneib}, {Matheson}, {Mulchaey}, {Sarajedini}, \& {Voit}}]{Sharonhigh}
{Sharon}, K., {Gal-Yam}, A., {Maoz}, D., {Donahue}, M., {Ebeling}, H., {Ellis},
  R.~S., {Filippenko}, A.~V., {Foley}, R., {Freedman}, W.~L., {Kirshner},
  R.~P., {Kneib}, J.-P., {Matheson}, T., {Mulchaey}, J.~S., {Sarajedini},
  V.~L., \& {Voit}, M. 2007{\natexlab{a}}, in American Institute of Physics
  Conference Series, Vol. 924, American Institute of Physics Conference Series,
  460--463

\bibitem[{{Sharon} {et~al.}(2007{\natexlab{b}}){Sharon}, {Gal-Yam}, {Maoz},
  {Filippenko}, \& {Guhathakurta}}]{Sharon07}
{Sharon}, K., {Gal-Yam}, A., {Maoz}, D., {Filippenko}, A.~V., \&
  {Guhathakurta}, P. 2007{\natexlab{b}}, \apj, 660, 1165

\bibitem[{{Sivanandam} {et~al.}(2007){Sivanandam}, {Zabludoff}, {Zaritsky},
  {Gonzalez}, \& {Kelson}}]{Sivanandam07}
{Sivanandam}, S., {Zabludoff}, A.~I., {Zaritsky}, D., {Gonzalez}, A.~H., \&
  {Kelson}, D.~D. 2007, {\it in preparation}

\bibitem[{{Stern} {et~al.}(2004){Stern}, {van Dokkum}, {Nugent}, {Sand},
  {Ellis}, {Sullivan}, {Bloom}, {Frail}, {Kneib}, {Koopmans}, \&
  {Treu}}]{Stern04}
{Stern}, D., {van Dokkum}, P.~G., {Nugent}, P., {Sand}, D.~J., {Ellis}, R.~S.,
  {Sullivan}, M., {Bloom}, J.~S., {Frail}, D.~A., {Kneib}, J.-P., {Koopmans},
  L.~V.~E., \& {Treu}, T. 2004, \apj, 612, 690

\bibitem[{{Struble} \& {Rood}(1991)}]{Struble91}
{Struble}, M.~F. \& {Rood}, H.~J. 1991, \apjs, 77, 363

\bibitem[{{Sullivan} {et~al.}(2006){Sullivan}, {Le Borgne}, {Pritchet},
  {Hodsman}, {Neill}, {Howell}, {Carlberg}, {Astier}, {Aubourg}, {Balam},
  {Basa}, {Conley}, {Fabbro}, {Fouchez}, {Guy}, {Hook}, {Pain},
  {Palanque-Delabrouille}, {Perrett}, {Regnault}, {Rich}, {Taillet}, {Baumont},
  {Bronder}, {Ellis}, {Filiol}, {Lusset}, {Perlmutter}, {Ripoche}, \&
  {Tao}}]{Sullivan06}
{Sullivan}, M., {Le Borgne}, D., {Pritchet}, C.~J., {Hodsman}, A., {Neill},
  J.~D., {Howell}, D.~A., {Carlberg}, R.~G., {Astier}, P., {Aubourg}, E.,
  {Balam}, D., {Basa}, S., {Conley}, A., {Fabbro}, S., {Fouchez}, D., {Guy},
  J., {Hook}, I., {Pain}, R., {Palanque-Delabrouille}, N., {Perrett}, K.,
  {Regnault}, N., {Rich}, J., {Taillet}, R., {Baumont}, S., {Bronder}, J.,
  {Ellis}, R.~S., {Filiol}, M., {Lusset}, V., {Perlmutter}, S., {Ripoche}, P.,
  \& {Tao}, C. 2006, \apj, 648, 868

\bibitem[{{Sun} {et~al.}(2007){Sun}, {Donahue}, \& {Voit}}]{Sun07}
{Sun}, M., {Donahue}, M., \& {Voit}, G.~M. 2007, ArXiv e-prints, 706

\bibitem[{{Totani} {et~al.}(2005){Totani}, {Sumi}, {Kosugi}, {Yasuda}, {Doi},
  \& {Oda}}]{Totani05}
{Totani}, T., {Sumi}, T., {Kosugi}, G., {Yasuda}, N., {Doi}, M., \& {Oda}, T.
  2005, \apjl, 621, L9

\bibitem[{{Trentham} \& {Tully}(2002)}]{Trentham01}
{Trentham}, N. \& {Tully}, R.~B. 2002, \mnras, 335, 712

\bibitem[{{van Dokkum}(2001)}]{vandokkum01}
{van Dokkum}, P.~G. 2001, \pasp, 113, 1420

\bibitem[{{Vanden Berk} {et~al.}(2006){Vanden Berk}, {Shen}, {Yip},
  {Schneider}, {Connolly}, {Burton}, {Jester}, {Hall}, {Szalay}, \&
  {Brinkmann}}]{VandenBerk06}
{Vanden Berk}, D.~E., {Shen}, J., {Yip}, C.-W., {Schneider}, D.~P., {Connolly},
  A.~J., {Burton}, R.~E., {Jester}, S., {Hall}, P.~B., {Szalay}, A.~S., \&
  {Brinkmann}, J. 2006, \aj, 131, 84

\bibitem[{{Webb} \& {Malkan}(2000)}]{Webb00}
{Webb}, W. \& {Malkan}, M. 2000, \apj, 540, 652

\bibitem[{{Williams} {et~al.}(2004){Williams}, {Olszewski}, {Lesser}, \&
  {Burge}}]{90prime}
{Williams}, G.~G., {Olszewski}, E., {Lesser}, M.~P., \& {Burge}, J.~H. 2004, in
  Ground-based Instrumentation for Astronomy. Edited by Alan F. M. Moorwood and
  Iye Masanori. Proceedings of the SPIE, Volume 5492, pp. 787-798 (2004)., ed.
  A.~F.~M. {Moorwood} \& M.~{Iye}, 787--798

\bibitem[{{Zaritsky} {et~al.}(2004){Zaritsky}, {Gonzalez}, \&
  {Zabludoff}}]{Zaritsky04}
{Zaritsky}, D., {Gonzalez}, A.~H., \& {Zabludoff}, A.~I. 2004, \apjl, 613, L93

\bibitem[{{Zibetti} {et~al.}(2005){Zibetti}, {White}, {Schneider}, \&
  {Brinkmann}}]{Zibetti05}
{Zibetti}, S., {White}, S.~D.~M., {Schneider}, D.~P., \& {Brinkmann}, J. 2005,
  \mnras, 358, 949

\end{thebibliography}

\clearpage

\begin{inlinefigure}
\begin{center}
\resizebox{\textwidth}{!}{\includegraphics{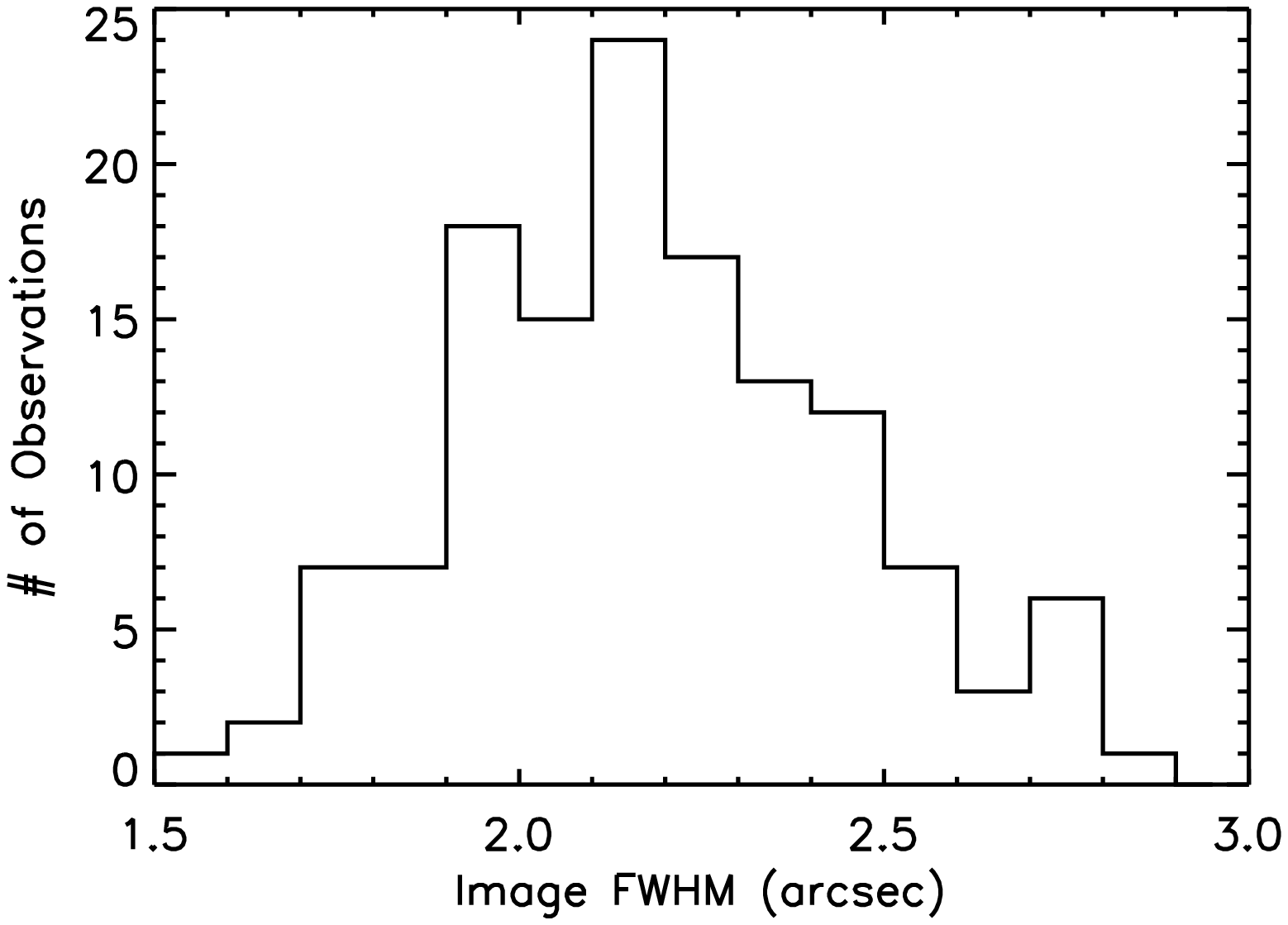}}
\end{center}
\figcaption{Histogram of the distribution of stellar FWHM during
our 90Prime imaging campaign. \label{fig:psf}}
\end{inlinefigure}

\clearpage

\begin{inlinefigure}
\begin{center}
\resizebox{\textwidth}{!}{\includegraphics{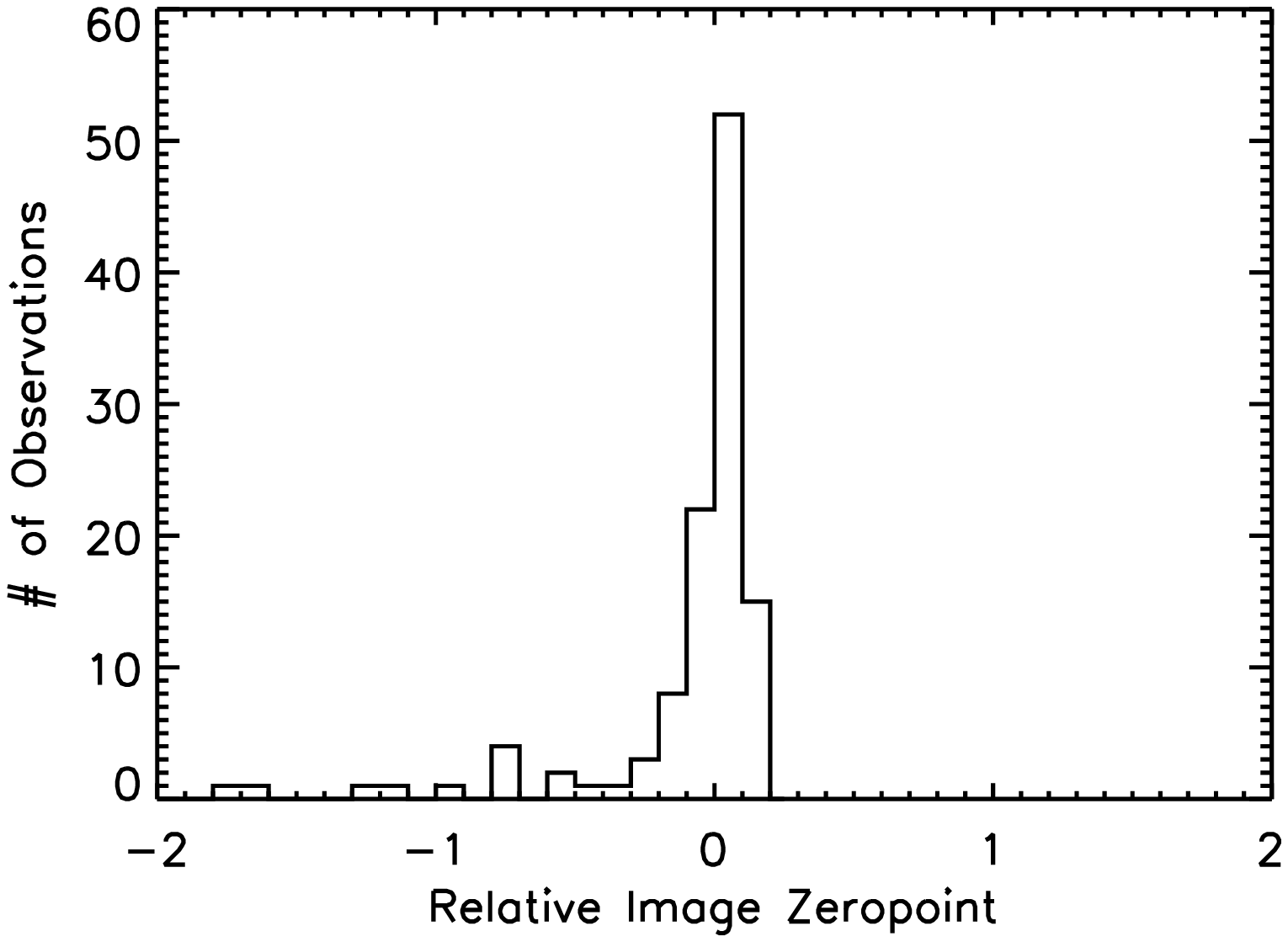}}
\end{center}
\figcaption{Histogram of the distribution of image zeropoints
with respect to the median of the distribution.\label{fig:zpthist}}
\end{inlinefigure}

\clearpage

\begin{inlinefigure}
\begin{center}
\resizebox{\textwidth}{!}{\includegraphics{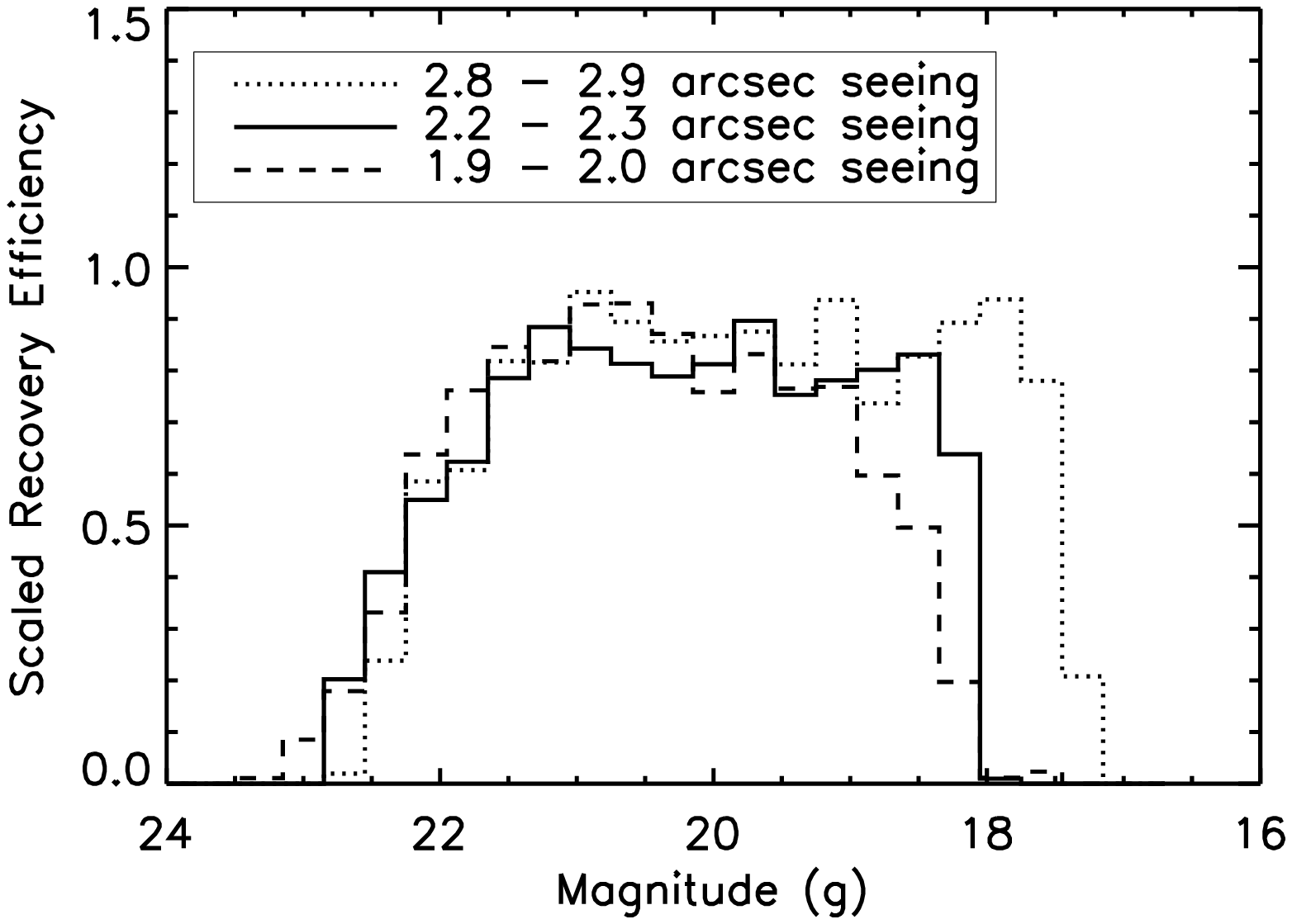}}
\end{center}
\figcaption{Transient point source detection efficiency as a function
of input magnitude for three values of the seeing.  The bright end
cutoff in efficiency is due to our strict rejection of bright
candidates. We are better able to detect
bright transients during poorer seeing conditions simply because the
transients are further from saturation.  The detection
efficiency has been rescaled to account for the area of the chip that
was masked for any reason.  \label{fig:totsceff}}
\end{inlinefigure}

\clearpage

\begin{inlinefigure}
\begin{center}
\resizebox{\textwidth}{!}{\includegraphics{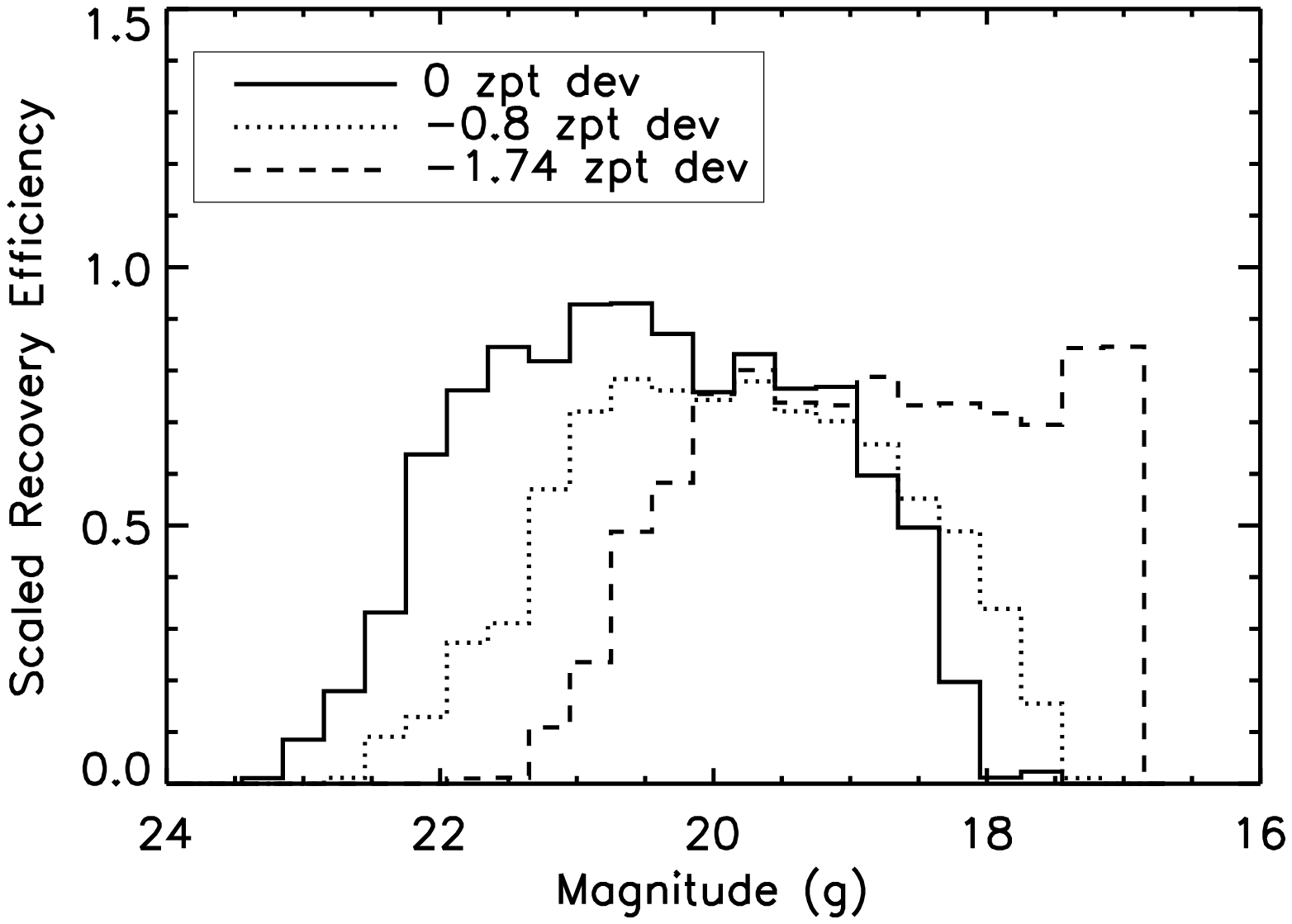}}
\end{center}
\figcaption{Transient point source detection efficiency as a function
of zeropoint deviation from the median.
To achieve high detection
efficiency over the $g\sim$19-21.5 region necessary for this survey,
clear conditions are necessary.  The detection efficiency has been
rescaled to account for the area of the chip that was masked for any
reason.  \label{fig:totzptsceff}}
\end{inlinefigure}

\clearpage

\begin{inlinefigure}
\begin{center}
\resizebox{\textwidth}{!}{\includegraphics{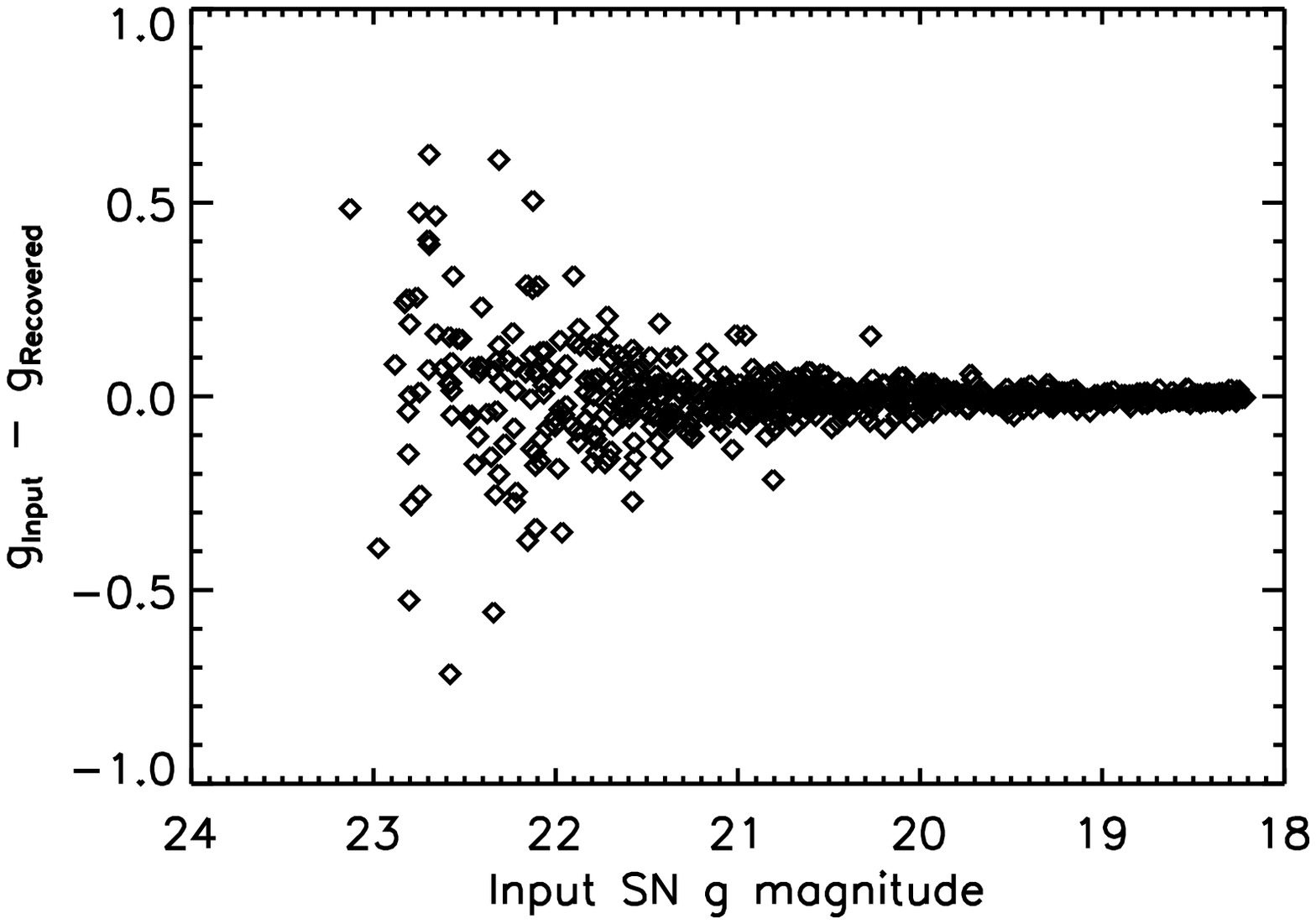}}
 \end{center}
\figcaption{The difference between the magnitude of the artificial transient
placed into an image versus the recovered magnitude 
as a function of input magnitude.  Shown are the
results from a typical run as discussed in
\S~\ref{sec:efficiency}.\label{fig:magdiff}}
\end{inlinefigure}

\clearpage

\begin{inlinefigure}
\begin{center}
\resizebox{\textwidth}{!}{\includegraphics{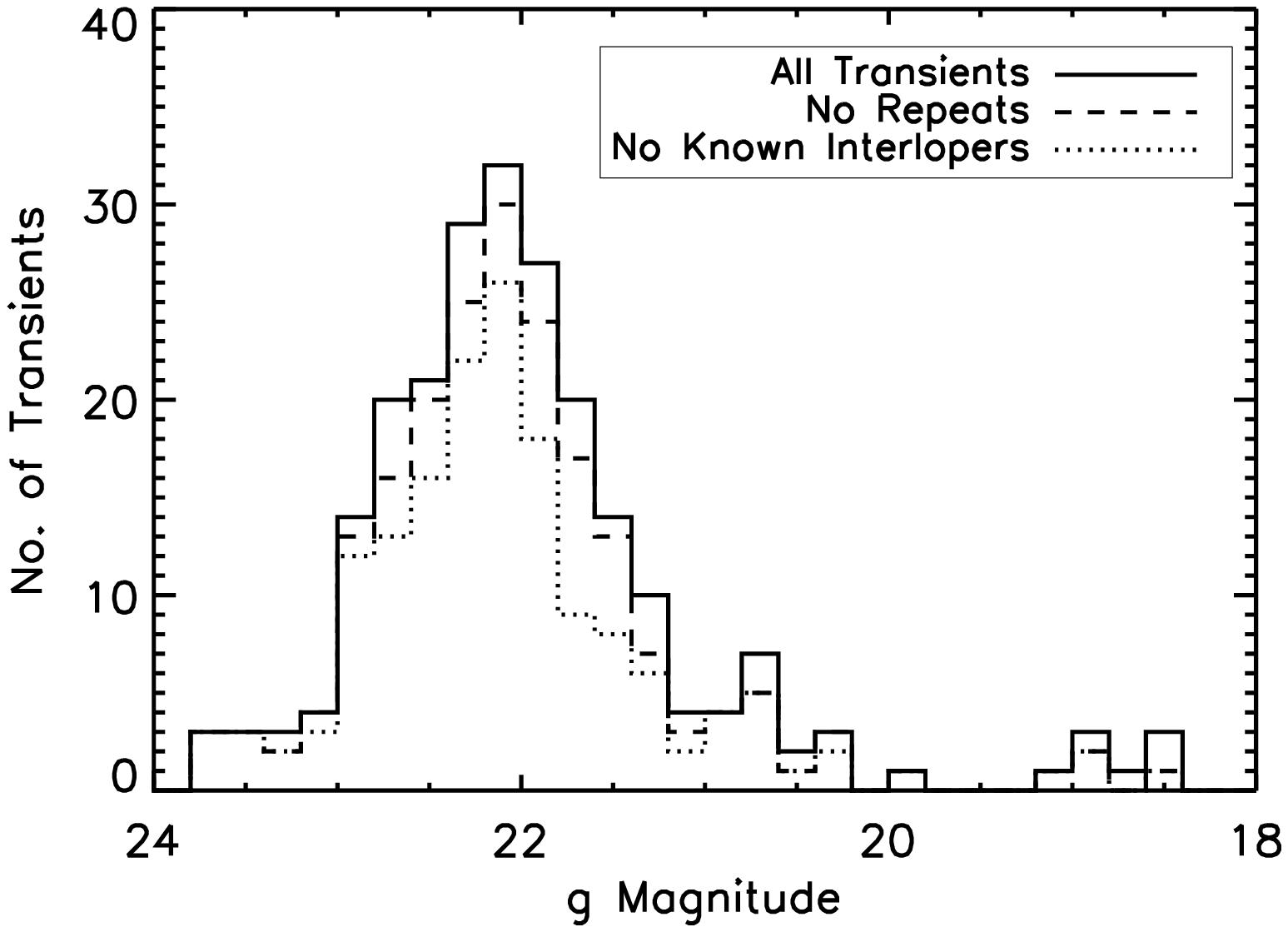}}
\end{center}
\figcaption{Histogram showing the magnitude distribution of identified
transients. The dashed histogram is plotted after the removal of
transients that were detected as variables in multiple observing
epochs.  The dotted histogram is the transient magnitude distribution
after removal of all sources known to not be at the cluster redshift.
\label{fig:maghist}}
\end{inlinefigure}

\clearpage

\clearpage

\clearpage

\clearpage

\clearpage

\clearpage

\clearpage

\begin{inlinefigure}
\begin{center}
\resizebox{\textwidth}{!}{\includegraphics{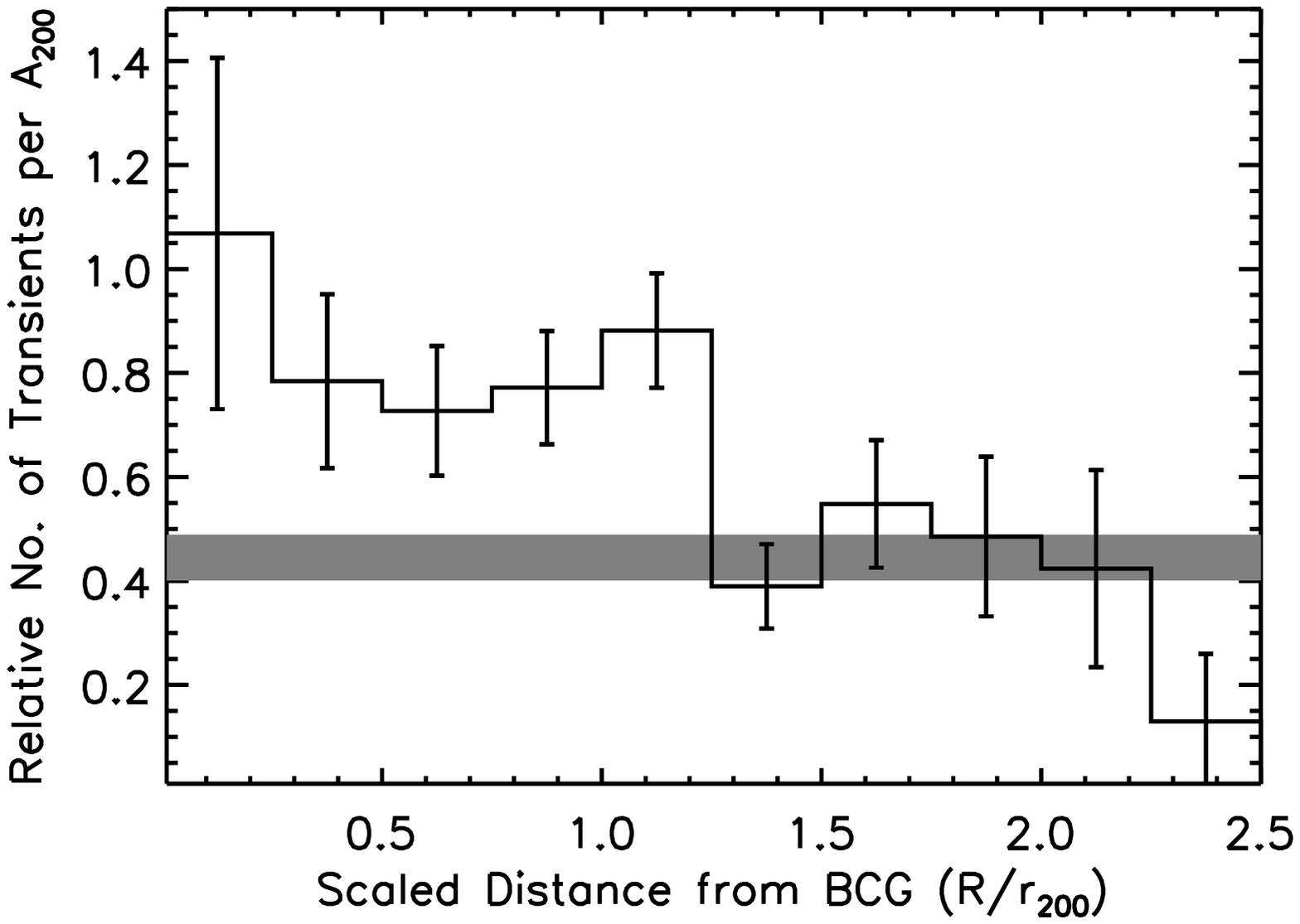}}
\end{center}
\figcaption{Radial profile of the variable sources as a function of
clustercentric radius, scaled by $r_{200}$.  All values have been
renormalized to account for the different areas probed by the radial
bins (including masked regions).  All of the clusters have been probed
out to $r_{200}$.  The 'background' transient rate is represented by
the horizontal bar and was measured from chip 3.  
\label{fig:radexcess}}
\end{inlinefigure}

\clearpage

\begin{inlinefigure}
\begin{center}
\resizebox{\textwidth}{!}{\includegraphics{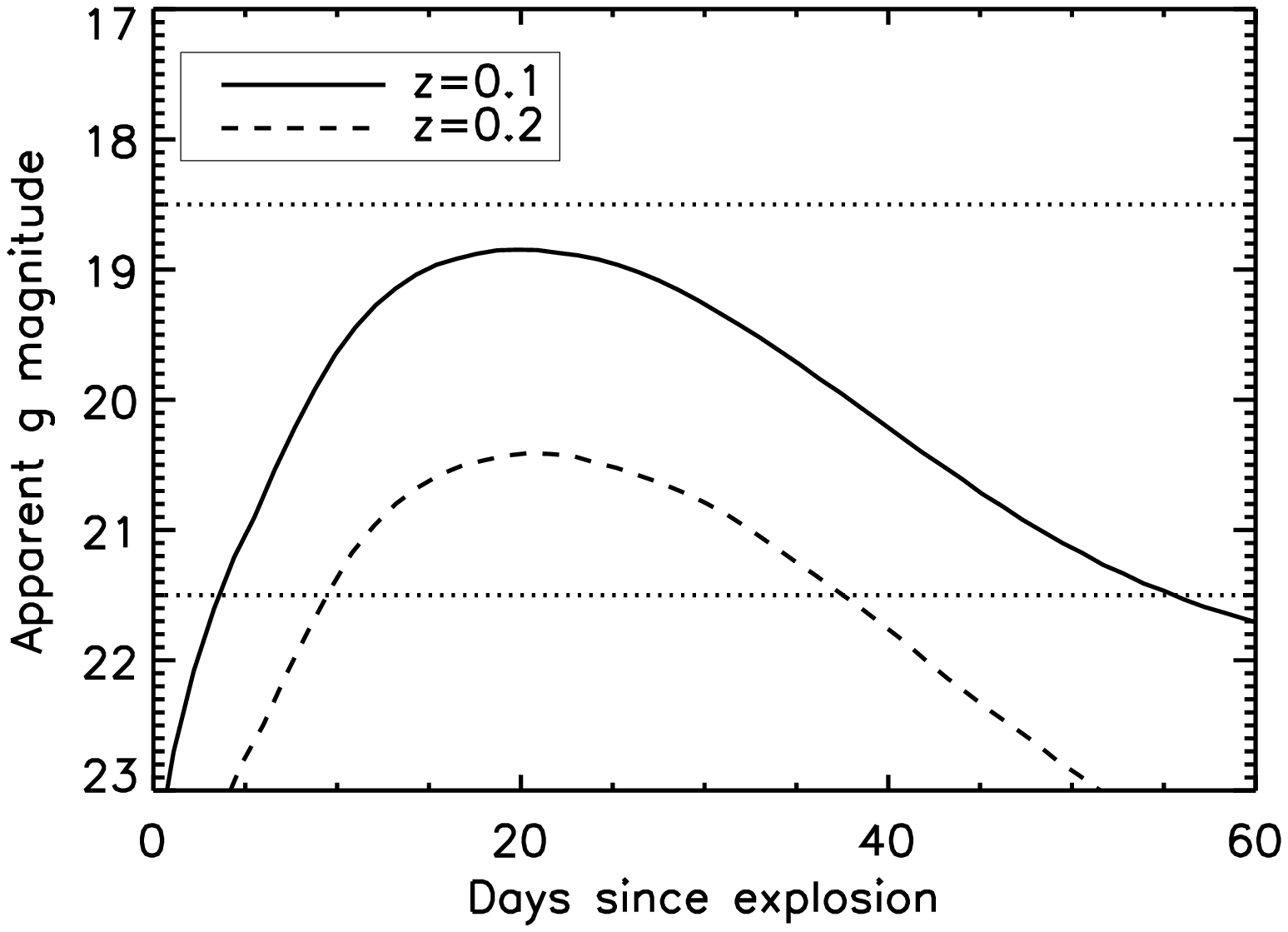}}
 \end{center}
\figcaption{Examples of template $g$ band SN Ia light curves at the
redshift edges of our cluster sample ($0.1 < z <
0.2$).\label{fig:samplecurve}}
\end{inlinefigure}

\clearpage

\begin{deluxetable}{lcccccccc}
\tabletypesize{\small}
\tablecolumns{9}
\tablecaption{List of Cluster Fields \label{table:clustertable}}
\tablehead{
\colhead{Cluster} & \colhead{$z_{clus}$} & \colhead{$\alpha$} & \colhead{$\delta$} & \colhead{$L_{X}$} & \colhead{Epoch 1} & \colhead{Epoch 2} & \colhead{Epoch 3} & \colhead{Epoch 4}\\
\colhead{} & \colhead{} & \colhead{(J2000.0)} & \colhead{(J2000.0)} & \colhead{[$10^{44} {\rm ergs\  s}^{-1}]$} & \colhead{} & \colhead{} & \colhead{} & \colhead{}}
\startdata
A2703&0.114&00:05:30.6&+16:5:11.9&2.72&11/18/06\tablenotemark{$\star$}&12/15/06&&\\
A136&0.157&01:04:14.4&+25:05:03.9&1.32&11/17/06\tablenotemark{$\star$}&12/15/06&&\\
A351&0.105&02:25:42.3&-08:40:21.8&0.21&11/18/06\tablenotemark{$\star$}&12/15/06&&\\
Z0256&0.194&02:59:32.6&+00:13:56.1&0.85&11/17/06\tablenotemark{$\star$}&12/15/06&&\\
A516&0.141&04:50:24.9&-08:49:24.6&0.17&11/18/06\tablenotemark{$\star$}&12/15/06&&\\
A565&0.105&07:06:54.0&+71:40:56.0&1.36&01/28/06&02/24/06&04/22/06&\\
A572&0.104&07:14:18.7&+54:41:03.1&1.38&01/28/06&02/24/06&04/24/06&\\
A580&0.118&07:25:57.2&+41:23:07.7&1.90&01/28/06&02/24/06&04/23/06&\\
RXJ0759&0.104&07:59:41.1&+54:00:25.7&1.97&01/28/06&04/23/06\tablenotemark{$\dagger$}&12/15/06\tablenotemark{$\dagger$}&\\
RXJ0804&0.187&08:04:21.4&+46:47:12.9&2.40&01/28/06&02/24/06&04/23/06&12/15/06\\
A667&0.145&08:28:06.4&+44:46:03.6&4.45&01/28/06&02/24/06&04/23/06&\\
A750&0.163&09:09:12.7&+10:58:29.3&8.74&01/28/06&02/24/06&04/23/06&\\
A853&0.165&09:42:15.0&+15:22:53.3&4.06&01/28/06&02/24/06&04/24/06&\\
A961&0.124&10:16:23.2&+33:38:19.8&3.12&01/28/06&02/24/06&04/23/06&\\
A1201&0.169&11:12:54.6&+13:26:09.8&6.05&01/28/06\tablenotemark{$\dagger$}&04/23/06\tablenotemark{$\dagger$}&&\\
A1413&0.143&11:55:18.2&+23:24:12.4&10.83&01/28/06&02/24/06&04/23/06&\\
A1689&0.183&13:11:29.9&-01:20:33.3&20.95&01/28/06&02/24/06&04/23/06&06/19/06\\
A1361&0.117&11:43:39.7&+46:21:20.4&4.95&01/28/06&02/24/06&04/23/06&\\
RXJ0741&0.158&07:41:26.7&+25:58:30.6&1.00&01/29/06&02/25/06&04/24/06&\\
A620&0.135&08:05:43.2&+45:41:02.3&1.47&01/29/06&02/25/06&12/15/06\tablenotemark{$\dagger$}&\\
A665&0.182&08:30:57.4&+65:50:28.3&15.17&01/29/06&02/25/06&&\\
A743&0.133&09:06:44.9&+10:19:03.6&2.64&04/24/06&12/15/06&&\\
RXJ0958&0.147&09:58:12.5&+23:46:44.3&1.96&01/29/06&02/25/06&04/24/06&\\
RXJ1000&0.153&10:00:31.0&+44:08:44.6&2.75&02/25/06&04/24/06&&\\
A868&0.153&09:45:26.5&-08:39:06.7&4.60&01/29/06&02/25/06&&\\
A901&0.170&09:55:57.4&-09:59:01.8&10.73&01/29/06&02/25/06&&\\
A1045&0.138&10:35:0.4&+30:41:40.3&3.53&01/29/06&02/25/06&04/24/06&\\
A1068&0.138&10:40:44.8&+39:57:10.1&5.94&01/29/06&02/25/06&04/24/06&12/15/06\\
A990&0.144&10:23:39.9&+49:08:40.1&6.71&01/29/06&02/25/06&04/25/06&\\
A1264&0.127&11:27:01.8&+17:07:47.7&2.07&01/29/06&02/25/06&&\\
A1246&0.190&11:23:58.8&+21:28:49.0&6.31&01/29/06&02/24/06&12/15/06\tablenotemark{$\dagger$}&\\
A1553&0.165&12:30:48.9&+10:32:48.4&7.19&01/29/06&02/24/06&04/23/06&\\
A1204&0.171&11:13:20.5&+17:35:41.0&6.50&01/29/06&02/25/06&04/24/06&\\
Z1432&0.189&07:51:25.0&+17:30:50.4&5.18&01/30/06&04/25/06&&\\
A646&0.129&08:22:09.3&+47:05:49.3&4.94&01/30/06&02/26/06&12/15/06&\\
RXJ0821&0.110&08:21:02.0&+07:51:48.8&1.39&01/30/06&02/26/06&12/15/06&\\
Z1883&0.194&08:42:56.0&+29:27:26.9&6.21&01/30/06&02/26/06&&\\
A655&0.127&08:25:28.8&+47:08:04.1&4.90&01/30/06&02/26/06&&\\
RXJ0819&0.119&08:19:25.7&+63:37:26.1&2.80&01/30/06&02/26/06&&\\
A795&0.136&09:24:05.4&14:10:24.2&5.70&01/30/06&02/26/06&&\\
A923&0.116&10:06:38.9&+25:54:44.5&2.05&01/30/06&02/26/06&&\\
A1033&0.126&10:31:44.3&+35:02:28.3&5.12&01/30/06\tablenotemark{$\dagger$}&04/25/06\tablenotemark{$\dagger$}&12/15/06\tablenotemark{$\dagger$}&\\
A1132&0.136&10:58:23.7&+56:47:42.7&6.76&01/30/06&04/25/06&&\\
A1111&0.165&10:50:46.7&-02:21:24.1&4.12&01/30/06&02/25/06&&\\
Z3179&0.143&10:25:58.1&+12:41:10.6&4.71&01/30/06&04/25/06&&\\
RXJ1156&0.142&11:56:55.8&+24:15:37.4&3.10&01/30/06&04/25/06&&\\
A1235&0.104&11:22:16.0&+19:35:49.3&1.70&01/30/06&02/25/06&04/25/06&12/15/06\tablenotemark{$\dagger$}\\
A1366&0.116&11:44:36.9&+67:24:23.5&3.91&01/30/06&02/25/06&04/25/06&\\
A1902&0.160&14:21:40.5&+37:17:32.6&4.55&04/23/06&06/19/06&&\\
A1677&0.183&13:05:50.8&+30:54:13.7&5.37&04/24/06&06/19/06&&\\
A1914&0.171&14:25:56.7&+37:48:59.0&17.30&04/24/06&06/19/06&&\\
A1930&0.131&14:32:38.0&+31:38:50.5&3.99&04/24/06&06/19/06&&\\
A2259&0.164&17:20:09.9&+27:40:08.9&5.95&04/25/06&06/19/06&&\\
A2218&0.176&16:35:49.6&+66:12:44.8&9.55&04/25/06&06/19/06&&\\
RXJ1750&0.171&17:50:17.0&+35:04:59.7&5.34&04/25/06&06/19/06&&\\
\enddata
\tablenotetext{$\star$}{Denotes dates when 90Prime readout noise was anomalously high.  These epochs should not be considered as part of the main sample, but are included for those clusters for which they are the only other imaging epoch.}
\tablenotetext{$\dagger$}{Chip 3 data for this epoch was not used either due to the presence of a vary bright field star in the field, poor focus, or an automated data reduction error}
\end{deluxetable}

\clearpage

\begin{deluxetable}{lccccccc}
\tablecolumns{8}
\tablecaption{Summary of Intracluster SNe Candidates\label{table:ICsumm}}
\tablehead{
\colhead{Cluster} &
\colhead{$\alpha$} & 
\colhead{$\delta$} &
\colhead{$g$} & 
\colhead{$R_{BCG}$} &
\colhead{$R_{BCG}/r_{200}$} & 
\colhead{$m_{lim,g}$} &
\colhead{$M_{lim,g}$} \\
& 
\colhead{(J2000.0)} & 
\colhead{(J2000.0)} & 
\colhead{[mag]} & 
\colhead{[kpc]} &
\colhead{}&
\colhead{}&
\colhead{}
}
\startdata
A516   & 04:49:42.5 & $-$08:51:42.2 & 23.83 $\pm$ 0.90 & 1595 & 2.0 & 25.1 & $-$14.0\\ 
A1201 & 11:13:12.6 & +13:12:10.3 & 22.52 $\pm$ 0.09 & 2537 & 1.4 & 25.3 & $-$14.3\\
A1413 & 11:56:9.63 & +23:32:29.4 & 22.12 $\pm$ 0.05 & 2172 & 1.0 & 26.1 & $-$13.1\\ 
A2259 & 17:23:33.4 & +27:49:31.6 & 22.54 $\pm$ 0.08 & 7764 & 4.2 & 25.3 & $-$14.2\\ 
\enddata
\end{deluxetable}

\clearpage

\begin{deluxetable}{lcccccl}
\tablecolumns{7}
\tablecaption{List of Identified Variables -- Cluster Centered Chip 1\label{table:vartable}}
\tablehead{
\colhead{Cluster} & \colhead{Epoch} & \colhead{$\alpha$} & \colhead{$\delta$} & \colhead{$g$} & \colhead{Dist from BCG} &\colhead{Notes}\\
\colhead{} & \colhead{} & \colhead{(J2000.0)} & \colhead{J2000.0} & \colhead{mag} & \colhead{kpc}& \colhead{}}
\startdata
A136 & 2006-12-16 02:43:01 & 1:3:19.9 & 25:4:28.4 & 24.51 $\pm$ 0.60 & 2012 & z=0.172 galaxy; BCS \\ 
A136 & 2006-12-16 02:43:01 & 1:4:2.14 & 25:7:22.8 & 22.91 $\pm$ 0.12 & 589 &  \\ 
A136 & 2006-12-16 02:43:01 & 1:3:55.5 & 24:50:20.5 & 21.75 $\pm$ 0.06 & 2498 & z=0.246 galaxy; BCS \\ 
A351 & 2006-12-16 03:57:01 & 2:26:21.3 & -8:54:19.8 & 22.71 $\pm$ 0.17 & 1961 & SN 2006ub; SNIa; BCS \\ 
A351 & 2006-12-16 03:57:01 & 2:25:56.9 & -8:45:38.3 & 21.71 $\pm$ 0.07 & 740 & SN 2006ua; SNIa; BCS \\ 
Z0256 & 2006-12-16 05:11:56 & 2:59:35.2 & 0:18:38.6 & 22.87 $\pm$ 0.14 & 919 &  \\ 
Z0256 & 2006-12-16 05:11:56 & 2:59:7.80 & 0:22:33.9 & 22.81 $\pm$ 0.14 & 2054 & z=0.199 CG; BCS \\ 
Z0256 & 2006-12-16 05:11:56 & 3:0:18.8 & 0:19:11.6 & 23.59 $\pm$ 0.31 & 2455 &  \\ 
Z0256 & 2006-12-16 05:11:56 & 3:0:15.3 & 0:13:28.4 & 22.57 $\pm$ 0.14 & 2069 &  \\ 
A516 & 2006-12-16 04:34:40 & 4:49:42.5 & -8:51:42.2 & 23.83 $\pm$ 0.90 & 1595 & ?; BCS; IC Cand \\ 
A516 & 2006-12-16 04:34:40 & 4:50:52.5 & -8:45:3.96 & 22.09 $\pm$ 0.07 & 1206 & Stellar spec; BCS \\ 
A516 & 2006-12-16 04:34:40 & 4:50:29.1 & -8:50:2.11 & 22.93 $\pm$ 0.19 & 183 & Stellar spec; BCS \\ 
A743 & 2006-04-25 05:12:57 & 9:7:25.7 & 10:24:30.9 & 21.54 $\pm$ 0.06 & 1620 &  \\ 
A743 & 2006-12-16 08:53:47 & 9:6:55.2 & 10:14:13.6 & 22.00 $\pm$ 0.09 & 775 & z=0.267 Galaxy; BCS \\ 
A743 & 2006-12-16 08:53:47 & 9:6:32.4 & 10:9:29.1 & 21.32 $\pm$ 0.05 & 1425 & SN 2006ud; SNIa; BCS \\ 
A743 & 2006-12-16 08:53:47 & 9:7:12.5 & 10:23:34.9 & 22.29 $\pm$ 0.12 & 1158 &  \\ 
A565 & 2006-01-28 02:42:49 & 7:7:26.8 & 71:44:42.2 & 22.14 $\pm$ 0.09 & 528 &  \\ 
A565 & 2006-02-24 02:13:37 & 7:9:33.9 & 71:34:49.1 & 21.52 $\pm$ 0.05 & 1622 &  \\ 
A565 & 2006-04-23 03:04:32 & 7:10:10.5 & 71:50:20.4 & 21.56 $\pm$ 0.07 & 2077 &  \\ 
A565 & 2006-04-23 03:04:32 & 7:7:49.4 & 71:52:31.6 & 21.93 $\pm$ 0.09 & 1429 &  \\ 
A572 & 2006-02-24 02:48:27 & 7:15:26.0 & 54:37:29.0 & 21.80 $\pm$ 0.04 & 1188 &  \\ 
A572 & 2006-02-24 02:48:27 & 7:14:11.5 & 54:26:52.5 & 23.12 $\pm$ 0.15 & 1628 &  \\ 
A572 & 2006-02-24 02:48:27 & 7:13:8.08 & 54:26:39.5 & 22.93 $\pm$ 0.14 & 2025 &  \\ 
A572 & 2006-02-24 02:48:27 & 7:12:56.3 & 54:28:31.2 & 21.03 $\pm$ 0.03 & 1984 &  \\ 
A572 & 2006-02-24 02:48:27 & 7:12:38.0 & 54:41:0.311 & 20.90 $\pm$ 0.02 & 1666 &  \\ 
A572 & 2006-04-25 03:15:23 & 7:13:46.5 & 54:38:22.6 & 22.15 $\pm$ 0.06 & 615 &  \\ 
A572 & 2006-04-25 03:15:23 & 7:13:25.1 & 54:42:15.0 & 23.23 $\pm$ 0.18 & 897 &  \\ 
A580 & 2006-02-24 03:49:52 & 7:26:15.1 & 41:37:34.2 & 21.71 $\pm$ 0.04 & 1897 & RD  \\ 
A580 & 2006-04-24 03:05:59 & 7:25:5.34 & 41:32:48.4 & 22.15 $\pm$ 0.07 & 1753 &  \\ 
A580 & 2006-04-24 03:05:59 & 7:26:22.4 & 41:24:43.2 & 21.87 $\pm$ 0.06 & 640 & z=1.803 QSO; SDSS \\ 
A580 & 2006-04-24 03:05:59 & 7:26:20.5 & 41:14:48.2 & 20.40 $\pm$ 0.02 & 1204 &  \\ 
A580 & 2006-04-24 03:05:59 & 7:26:15.1 & 41:37:34.2 & 20.76 $\pm$ 0.02 & 1897 & RD  \\ 
A580 & 2006-04-24 03:05:59 & 7:26:12.5 & 41:22:34.1 & 22.20 $\pm$ 0.08 & 375 & CG; SDSS \\ 
RXJ0759 & 2006-01-28 05:07:38 & 8:0:37.4 & 53:47:50.4 & 22.01 $\pm$ 0.06 & 1728 &  \\ 
RXJ0759 & 2006-01-28 05:07:38 & 7:59:17.3 & 54:2:32.0 & 22.19 $\pm$ 0.07 & 466 &  \\ 
RXJ0759 & 2006-01-28 05:07:38 & 7:58:34.4 & 54:10:30.8 & 21.93 $\pm$ 0.06 & 1607 &  \\ 
RXJ0759 & 2006-04-24 03:52:36 & 7:59:40.3 & 53:55:49.2 & 22.64 $\pm$ 0.10 & 528 &  \\ 
RXJ0759 & 2006-04-24 03:52:36 & 7:58:32.1 & 54:6:42.1 & 22.71 $\pm$ 0.10 & 1362 & RD  \\ 
RXJ0759 & 2006-12-16 09:36:16 & 8:0:46.0 & 53:47:15.5 & 22.59 $\pm$ 0.07 & 1866 &  \\ 
RXJ0759 & 2006-12-16 09:36:16 & 7:59:52.3 & 53:47:44.5 & 21.27 $\pm$ 0.02 & 1466 &  \\ 
RXJ0759 & 2006-12-16 09:36:16 & 7:59:13.0 & 54:4:42.2 & 22.20 $\pm$ 0.04 & 680 & z=2.93 QSO; BCS \\ 
RXJ0759 & 2006-12-16 09:36:16 & 7:58:45.9 & 54:1:17.4 & 22.09 $\pm$ 0.04 & 934 & SN 2006uc; SNIa; BCS \\ 
RXJ0759 & 2006-12-16 09:36:16 & 7:58:32.1 & 54:6:42.9 & 22.71 $\pm$ 0.06 & 1363 & RD  \\ 
RXJ0804 & 2006-01-28 06:15:30 & 8:5:10.2 & 46:43:32.8 & 21.98 $\pm$ 0.04 & 1716 &  \\ 
RXJ0804 & 2006-02-24 05:20:32 & 8:5:6.84 & 46:35:19.0 & 22.58 $\pm$ 0.07 & 2672 &  \\ 
RXJ0804 & 2006-02-24 05:20:32 & 8:4:28.0 & 46:48:15.8 & 22.40 $\pm$ 0.06 & 292 &  \\ 
RXJ0804 & 2006-02-24 05:20:32 & 8:4:24.6 & 46:56:55.1 & 22.08 $\pm$ 0.04 & 1825 &  \\ 
RXJ0804 & 2006-02-24 05:20:32 & 8:4:8.89 & 46:39:10.9 & 19.13 $\pm$ 0.00 & 1561 &  \\ 
RXJ0804 & 2006-04-24 04:33:50 & 8:2:51.8 & 46:48:42.9 & 23.72 $\pm$ 0.32 & 2890 &  \\ 
RXJ0804 & 2006-12-16 07:07:14 & 8:5:18.5 & 46:35:40.3 & 21.60 $\pm$ 0.03 & 2845 & z=1.775 (phot) QSO/1 \\ 
RXJ0804 & 2006-12-16 07:07:14 & 8:3:24.0 & 46:55:18.9 & 21.95 $\pm$ 0.06 & 2386 &  \\ 
RXJ0804 & 2006-12-16 07:07:14 & 8:3:12.9 & 46:58:12.1 & 21.90 $\pm$ 0.05 & 3010 & z=1.775 (phot) QSO/1 \\ 
RXJ0804 & 2006-12-16 07:07:14 & 8:4:50.2 & 47:1:45.2 & 21.70 $\pm$ 0.05 & 2881 &  \\ 
RXJ0804 & 2006-12-16 07:07:14 & 8:4:21.4 & 46:57:15.4 & 22.83 $\pm$ 0.11 & 1885 &  \\ 
RXJ0804 & 2006-12-16 07:07:14 & 8:3:57.0 & 46:40:37.7 & 22.32 $\pm$ 0.07 & 1465 &  \\ 
A667 & 2006-01-28 06:52:55 & 8:28:49.1 & 44:44:38.0 & 20.47 $\pm$ 0.01 & 1177 & RD  \\ 
A667 & 2006-01-28 06:52:55 & 8:28:31.9 & 44:34:3.13 & 22.24 $\pm$ 0.06 & 1958 &  \\ 
A667 & 2006-01-28 06:52:55 & 8:28:15.9 & 44:47:24.2 & 22.29 $\pm$ 0.07 & 329 &  \\ 
A667 & 2006-02-24 05:57:42 & 8:27:9.95 & 44:29:25.1 & 24.18 $\pm$ 0.47 & 2966 &  \\ 
A667 & 2006-04-24 05:11:06 & 8:28:49.1 & 44:44:38.0 & 20.73 $\pm$ 0.02 & 1177 & RD  \\ 
A667 & 2006-04-24 05:11:06 & 8:26:41.5 & 44:55:37.4 & 23.25 $\pm$ 0.22 & 2716 &  \\ 
A750 & 2006-04-24 05:49:13 & 9:9:47.8 & 11:6:36.6 & 22.34 $\pm$ 0.09 & 1991 &  \\ 
A853 & 2006-01-28 08:36:18 & 9:43:12.5 & 15:35:9.70 & 22.00 $\pm$ 0.13 & 3140 &  \\ 
A853 & 2006-01-28 08:36:18 & 9:42:53.6 & 15:26:4.16 & 22.00 $\pm$ 0.12 & 1672 &  \\ 
A961 & 2006-02-24 07:52:05 & 10:17:29.6 & 33:45:15.6 & 22.83 $\pm$ 0.12 & 2063 &  \\ 
A1201 & 2006-01-28 10:18:32 & 11:13:36.0 & 13:38:31.0 & 22.57 $\pm$ 0.10 & 2759 &  \\ 
A1201 & 2006-01-28 10:18:32 & 11:12:17.8 & 13:21:6.03 & 21.55 $\pm$ 0.04 & 1779 & z=0.231 QSO; SDSS \\ 
A1201 & 2006-04-24 07:09:10 & 11:13:29.0 & 13:34:42.0 & 22.87 $\pm$ 0.12 & 2070 &  \\ 
A1201 & 2006-04-24 07:09:10 & 11:13:12.6 & 13:12:10.3 & 22.52 $\pm$ 0.09 & 2537 & IC Cand \\ 
A1201 & 2006-04-24 07:09:10 & 11:12:47.5 & 13:14:4.21 & 21.95 $\pm$ 0.05 & 2114 & QSO z=1.524; SDSS \\ 
A1413 & 2006-01-28 11:09:51 & 11:55:50.3 & 23:23:58.5 & 22.49 $\pm$ 0.08 & 1112 &  \\ 
A1413 & 2006-01-28 11:09:51 & 11:55:4.32 & 23:31:18.7 & 22.19 $\pm$ 0.07 & 1173 &  \\ 
A1413 & 2006-02-24 09:06:07 & 11:56:9.63 & 23:32:29.4 & 22.12 $\pm$ 0.05 & 2171 & IC Cand \\ 
A1413 & 2006-02-24 09:06:07 & 11:56:5.11 & 23:11:16.5 & 23.64 $\pm$ 0.24 & 2537 &  \\ 
A1413 & 2006-02-24 09:06:07 & 11:55:46.6 & 23:24:47.8 & 21.59 $\pm$ 0.04 & 989 &  \\ 
A1413 & 2006-04-24 07:59:00 & 11:55:55.2 & 23:15:44.3 & 21.56 $\pm$ 0.03 & 1808 &  \\ 
A1413 & 2006-04-24 07:59:00 & 11:55:44.4 & 23:37:29.5 & 22.29 $\pm$ 0.07 & 2198 &  \\ 
A1413 & 2006-04-24 07:59:00 & 11:55:42.0 & 23:28:55.6 & 22.03 $\pm$ 0.06 & 1088 &  \\ 
A1689 & 2006-01-28 11:51:19 & 13:10:50.2 & -1:24:20.7 & 18.85 $\pm$ 0.00 & 1958 & RD QUEST 088049/3 \\ 
A1689 & 2006-01-28 11:51:19 & 13:12:6.64 & -1:31:30.2 & 21.35 $\pm$ 0.03 & 2637 & RD  z=0.845 QSO/4 \\ 
A1689 & 2006-01-28 11:51:19 & 13:11:55.8 & -1:25:56.6 & 22.30 $\pm$ 0.07 & 1555 & RD  z=0.962 QSO/4 \\ 
A1689 & 2006-01-28 11:51:19 & 13:11:52.6 & -1:23:4.82 & 21.62 $\pm$ 0.04 & 1150 & CG/5 \\ 
A1689 & 2006-01-28 11:51:19 & 13:11:35.6 & -1:9:26.1 & 22.09 $\pm$ 0.06 & 2069 &  \\ 
A1689 & 2006-01-28 11:51:19 & 13:11:29.6 & -1:16:2.60 & 22.43 $\pm$ 0.08 & 833 & z=1.648 QSO/4 \\ 
A1689 & 2006-02-24 09:46:20 & 13:12:6.64 & -1:31:30.2 & 21.82 $\pm$ 0.05 & 2637 & RD  z=0.845 QSO/4 \\ 
A1689 & 2006-02-24 09:46:20 & 13:12:5.77 & -1:30:8.38 & 22.75 $\pm$ 0.12 & 2422 &  \\ 
A1689 & 2006-02-24 09:46:20 & 13:11:59.8 & -1:28:5.43 & 22.36 $\pm$ 0.16 & 1958 &  \\ 
A1689 & 2006-02-24 09:46:20 & 13:11:55.8 & -1:25:56.1 & 22.64 $\pm$ 0.11 & 1554 & RD  z=0.962 QSO/4 \\ 
A1689 & 2006-04-24 08:39:34 & 13:11:19.3 & -1:20:30.2 & 22.11 $\pm$ 0.07 & 485 & z=2.584 QSO/6 \\ 
A1689 & 2006-04-24 08:39:34 & 13:11:18.3 & -1:14:29.4 & 21.44 $\pm$ 0.03 & 1239 & 2QZ J131118.2-011430/4 \\ 
A1689 & 2006-04-24 08:39:34 & 13:10:55.5 & -1:27:25.1 & 21.63 $\pm$ 0.04 & 2029 & z=1.003 QSO/6 \\ 
A1689 & 2006-06-19 04:03:19 & 13:12:28.0 & -1:21:37.9 & 21.50 $\pm$ 0.04 & 2689 &  \\ 
A1689 & 2006-06-19 04:03:19 & 13:10:57.6 & -1:20:42.4 & 22.16 $\pm$ 0.08 & 1488 &  \\ 
A1689 & 2006-06-19 04:03:19 & 13:10:50.2 & -1:24:20.7 & 18.64 $\pm$ 0.00 & 1958 & RD QUEST 088049/3 \\ 
A1361 & 2006-01-28 12:29:23 & 11:44:20.2 & 46:28:29.6 & 18.56 $\pm$ 0.00 & 1269 & RD  \\ 
A1361 & 2006-01-28 12:29:23 & 11:43:44.1 & 46:23:0.144 & 21.70 $\pm$ 0.03 & 232 &  \\ 
A1361 & 2006-04-24 09:20:33 & 11:44:21.4 & 46:23:4.06 & 22.77 $\pm$ 0.10 & 940 &  \\ 
A1361 & 2006-04-24 09:20:33 & 11:44:20.2 & 46:28:30.0 & 18.58 $\pm$ 0.00 & 1270 & RD  \\ 
A1361 & 2006-04-24 09:20:33 & 11:42:23.4 & 46:26:11.0 & 22.24 $\pm$ 0.08 & 1778 &  \\ 
RXJ0741 & 2006-01-29 02:11:06 & 7:42:18.1 & 25:53:45.7 & 21.85 $\pm$ 0.07 & 2048 & RD  \\ 
RXJ0741 & 2006-01-29 02:11:06 & 7:41:26.6 & 26:7:36.5 & 21.20 $\pm$ 0.05 & 1490 &  \\ 
RXJ0741 & 2006-02-25 02:26:30 & 7:42:18.1 & 25:53:45.7 & 21.75 $\pm$ 0.06 & 2048 & RD  \\ 
RXJ0741 & 2006-02-25 02:26:30 & 7:40:34.2 & 26:3:31.5 & 22.29 $\pm$ 0.10 & 2097 &  \\ 
RXJ0741 & 2006-02-25 02:26:30 & 7:41:18.4 & 25:44:14.1 & 21.97 $\pm$ 0.09 & 2357 &  \\ 
RXJ0741 & 2006-04-25 03:53:15 & 7:40:52.7 & 25:51:51.3 & 22.97 $\pm$ 0.17 & 1658 &  \\ 
A620 & 2006-01-29 02:55:40 & 8:6:14.7 & 45:35:45.0 & 21.95 $\pm$ 0.05 & 1098 & RD Stellar spec; BCS \\ 
A620 & 2006-02-25 03:05:49 & 8:4:35.6 & 45:31:0.928 & 22.23 $\pm$ 0.08 & 2227 & CG; SDSS \\ 
A620 & 2006-02-25 03:05:49 & 8:5:6.89 & 45:32:37.7 & 21.20 $\pm$ 0.03 & 1513 & RD Stellar spec; BCS \\ 
A620 & 2006-12-16 05:52:00 & 8:5:6.89 & 45:32:37.7 & 21.70 $\pm$ 0.06 & 1513 & RD Stellar spec; BCS \\ 
A620 & 2006-12-16 05:52:00 & 8:6:14.7 & 45:35:45.5 & 21.19 $\pm$ 0.04 & 1097 & RD Stellar spec; BCS \\ 
A665 & 2006-01-29 04:31:29 & 8:30:1.46 & 65:45:2.30 & 21.92 $\pm$ 0.05 & 1453 & CG/7 \\ 
RXJ0958 & 2006-01-29 05:54:09 & 9:57:15.9 & 23:41:16.2 & 23.36 $\pm$ 0.37 & 2166 & RD  \\ 
RXJ0958 & 2006-02-25 05:03:22 & 9:57:15.9 & 23:41:16.7 & 22.29 $\pm$ 0.13 & 2167 & RD  \\ 
RXJ1000 & 2006-02-25 05:39:58 & 10:1:31.3 & 44:1:45.4 & 22.23 $\pm$ 0.06 & 2057 &  \\ 
RXJ1000 & 2006-04-25 07:19:59 & 9:59:47.7 & 44:15:18.8 & 22.89 $\pm$ 0.09 & 1618 &  \\ 
RXJ1000 & 2006-04-25 07:19:59 & 9:59:24.1 & 43:54:57.9 & 22.70 $\pm$ 0.11 & 2917 &  \\ 
A868 & 2006-02-25 06:16:49 & 9:46:19.4 & -8:49:27.9 & 22.08 $\pm$ 0.07 & 2661 &  \\ 
A901 & 2006-01-29 07:46:57 & 9:57:0.0908 & -9:53:9.53 & 20.76 $\pm$ 0.02 & 2872 &  \\ 
A901 & 2006-01-29 07:46:57 & 9:56:33.7 & -10:8:24.7 & 21.66 $\pm$ 0.04 & 2253 &  \\ 
A901 & 2006-02-25 06:53:39 & 9:55:34.2 & -9:51:15.2 & 21.31 $\pm$ 0.04 & 1677 &  \\ 
A1045 & 2006-04-25 07:58:26 & 10:34:41.5 & 30:47:3.26 & 21.92 $\pm$ 0.06 & 985 &  \\ 
A1045 & 2006-04-25 07:58:26 & 10:34:39.0 & 30:53:36.8 & 21.22 $\pm$ 0.03 & 1871 &  \\ 
A1068 & 2006-02-25 08:14:53 & 10:41:46.4 & 39:47:24.7 & 20.65 $\pm$ 0.06 & 2244 &  \\ 
A1068 & 2006-02-25 08:14:53 & 10:40:34.4 & 39:45:9.89 & 21.61 $\pm$ 0.16 & 1780 &  \\ 
A1068 & 2006-04-25 08:35:17 & 10:39:37.3 & 39:56:40.8 & 22.06 $\pm$ 0.14 & 1894 &  \\ 
A1068 & 2006-12-16 11:35:39 & 10:39:27.2 & 40:8:58.5 & 23.80 $\pm$ 0.26 & 2772 &  \\ 
A1068 & 2006-12-16 11:35:39 & 10:41:8.21 & 39:49:24.3 & 22.27 $\pm$ 0.05 & 1312 &  \\ 
A1068 & 2006-12-16 11:35:39 & 10:39:42.1 & 39:47:22.3 & 22.81 $\pm$ 0.10 & 2270 &  \\ 
A990 & 2006-01-29 09:51:56 & 10:22:15.2 & 49:2:7.73 & 23.03 $\pm$ 0.25 & 2326 &  \\ 
A990 & 2006-02-25 08:52:41 & 10:23:44.4 & 48:57:16.6 & 22.27 $\pm$ 0.07 & 1731 &  \\ 
A990 & 2006-04-26 05:39:01 & 10:23:45.3 & 49:16:47.7 & 23.57 $\pm$ 0.25 & 1239 &  \\ 
A990 & 2006-04-26 05:39:01 & 10:23:34.2 & 49:11:15.5 & 21.93 $\pm$ 0.04 & 417 &  \\ 
A990 & 2006-04-26 05:39:01 & 10:22:27.3 & 49:22:52.9 & 22.64 $\pm$ 0.16 & 2802 &  \\ 
A1264 & 2006-01-29 10:29:19 & 11:27:24.3 & 17:11:23.7 & 22.08 $\pm$ 0.11 & 884 &  \\ 
A1264 & 2006-01-29 10:29:19 & 11:26:41.2 & 17:3:10.9 & 21.97 $\pm$ 0.09 & 919 &  \\ 
A1264 & 2006-02-25 09:33:14 & 11:26:38.3 & 17:9:53.2 & 22.10 $\pm$ 0.14 & 814 &  \\ 
A1264 & 2006-02-25 09:33:14 & 11:26:34.6 & 17:3:51.1 & 22.11 $\pm$ 0.12 & 1035 &  \\ 
A1246 & 2006-12-16 12:11:23 & 11:24:48.9 & 21:36:5.22 & 21.45 $\pm$ 0.02 & 2614 & z=0.92 QSO; BCS \\ 
A1246 & 2006-12-16 12:11:23 & 11:24:41.5 & 21:12:9.94 & 24.46 $\pm$ 0.46 & 3689 &  \\ 
A1246 & 2006-12-16 12:11:23 & 11:24:2.51 & 21:36:58.1 & 22.54 $\pm$ 0.06 & 1559 & Stellar spec; BCS \\ 
A1246 & 2006-12-16 12:11:23 & 11:23:49.5 & 21:28:40.3 & 20.63 $\pm$ 0.01 & 411 & SN 2006ue; Cluster SNIa; BCS \\ 
A1553 & 2006-01-29 11:52:37 & 12:31:16.9 & 10:48:35.9 & 22.62 $\pm$ 0.17 & 2924 &  \\ 
A1553 & 2006-01-29 11:52:37 & 12:30:50.5 & 10:39:0.640 & 22.58 $\pm$ 0.13 & 1055 & z=0.213 QSO; SDSS \\ 
A1553 & 2006-02-24 11:45:49 & 12:30:10.6 & 10:23:11.6 & 22.69 $\pm$ 0.20 & 2283 &  \\ 
A1553 & 2006-04-24 10:03:28 & 12:30:45.8 & 10:32:50.4 & 22.34 $\pm$ 0.11 & 126 &  \\ 
A1553 & 2006-04-24 10:03:28 & 12:31:22.2 & 10:44:12.8 & 21.75 $\pm$ 0.06 & 2383 & z=0.092 Galaxy; SDSS \\ 
A1204 & 2006-01-29 12:31:42 & 11:12:55.8 & 17:37:36.7 & 22.14 $\pm$ 0.08 & 1078 &  \\ 
A1204 & 2006-02-25 12:13:56 & 11:14:16.6 & 17:40:57.0 & 22.26 $\pm$ 0.09 & 2513 &  \\ 
A1204 & 2006-04-25 09:11:22 & 11:13:59.2 & 17:39:2.83 & 22.06 $\pm$ 0.08 & 1718 &  \\ 
A1204 & 2006-04-25 09:11:22 & 11:13:53.7 & 17:30:14.0 & 21.93 $\pm$ 0.07 & 1681 &  \\ 
Z1432 & 2006-01-30 02:02:53 & 7:52:22.8 & 17:16:36.9 & 21.43 $\pm$ 0.10 & 3754 &  \\ 
Z1432 & 2006-02-26 02:29:16 & 7:51:54.2 & 17:21:43.8 & 21.83 $\pm$ 0.06 & 2173 &  \\ 
Z1432 & 2006-02-26 02:29:16 & 7:51:20.7 & 17:38:58.1 & 22.52 $\pm$ 0.38 & 1551 &  \\ 
A646 & 2006-12-16 10:22:04 & 8:22:7.01 & 46:56:0.475 & 21.03 $\pm$ 0.07 & 1356 & z=2.325 (phot) QSO/1 \\ 
A646 & 2006-12-16 10:22:04 & 8:21:5.55 & 47:16:33.4 & 21.40 $\pm$ 0.10 & 2104 & z=2.22 QSO; BCS \\ 
A646 & 2006-12-16 10:22:04 & 8:22:32.4 & 47:14:45.3 & 21.25 $\pm$ 0.09 & 1348 &  \\ 
RXJ0821 & 2006-01-30 03:29:06 & 8:21:52.2 & 7:39:4.20 & 22.57 $\pm$ 0.13 & 2145 &  \\ 
RXJ0821 & 2006-01-30 03:29:06 & 8:20:43.0 & 7:41:24.6 & 21.22 $\pm$ 0.04 & 1374 & RD  \\ 
RXJ0821 & 2006-01-30 03:29:06 & 8:20:42.6 & 8:3:56.1 & 22.32 $\pm$ 0.10 & 1569 &  \\ 
RXJ0821 & 2006-01-30 03:29:06 & 8:20:29.5 & 7:54:27.1 & 22.21 $\pm$ 0.09 & 1017 &  \\ 
RXJ0821 & 2006-01-30 03:29:06 & 8:20:28.5 & 7:41:6.98 & 22.95 $\pm$ 0.19 & 1629 & RD  \\ 
RXJ0821 & 2006-01-30 03:29:06 & 8:21:21.2 & 7:42:1.31 & 22.13 $\pm$ 0.08 & 1311 & RD  \\ 
RXJ0821 & 2006-01-30 03:29:06 & 8:21:14.9 & 7:40:31.0 & 22.62 $\pm$ 0.13 & 1413 &  \\ 
RXJ0821 & 2006-01-30 03:29:06 & 8:20:52.3 & 7:55:35.4 & 22.06 $\pm$ 0.07 & 537 & RD z=1.13 QSO; SDSS \\ 
RXJ0821 & 2006-01-30 03:29:06 & 8:20:45.6 & 8:3:25.8 & 21.77 $\pm$ 0.07 & 1480 & z=2.91 QSO; SDSS \\ 
RXJ0821 & 2006-02-26 04:02:42 & 8:21:21.2 & 7:42:1.31 & 22.31 $\pm$ 0.10 & 1311 & RD  \\ 
RXJ0821 & 2006-02-26 04:02:42 & 8:21:18.0 & 7:53:59.2 & 22.60 $\pm$ 0.13 & 544 & X-ray pt src/2 \\ 
RXJ0821 & 2006-02-26 04:02:42 & 8:20:52.4 & 7:55:34.9 & 22.24 $\pm$ 0.09 & 536 & RD z=1.13 QSO; SDSS \\ 
RXJ0821 & 2006-02-26 04:02:42 & 8:20:43.0 & 7:41:24.6 & 21.40 $\pm$ 0.04 & 1374 & RD  \\ 
RXJ0821 & 2006-02-26 04:02:42 & 8:20:28.5 & 7:41:6.08 & 22.59 $\pm$ 0.14 & 1630 & RD  \\ 
RXJ0821 & 2006-02-26 04:02:42 & 8:21:30.6 & 7:50:56.8 & 21.96 $\pm$ 0.08 & 860 &  \\ 
RXJ0821 & 2006-02-26 04:02:42 & 8:21:21.6 & 7:46:25.9 & 22.99 $\pm$ 0.21 & 874 &  \\ 
RXJ0821 & 2006-12-16 08:11:33 & 8:20:34.2 & 7:39:37.6 & 21.73 $\pm$ 0.07 & 1684 & Galaxy z=0.22; BCS \\ 
RXJ0821 & 2006-12-16 08:11:33 & 8:20:22.1 & 7:38:12.1 & 22.41 $\pm$ 0.15 & 2023 &  \\ 
RXJ0821 & 2006-12-16 08:11:33 & 8:20:19.2 & 7:53:24.7 & 21.84 $\pm$ 0.08 & 1288 &  \\ 
RXJ0821 & 2006-12-16 08:11:33 & 8:21:23.0 & 7:39:41.3 & 22.20 $\pm$ 0.11 & 1588 & z=0.90 QSO; BCS \\ 
RXJ0821 & 2006-12-16 08:11:33 & 8:21:16.3 & 8:3:38.1 & 22.48 $\pm$ 0.14 & 1485 &  \\ 
RXJ0821 & 2006-12-16 08:11:33 & 8:20:52.7 & 7:52:38.8 & 21.96 $\pm$ 0.08 & 294 & z=0.795 QSO; BCS \\ 
RXJ0821 & 2006-12-16 08:11:33 & 8:20:44.5 & 7:50:7.56 & 22.43 $\pm$ 0.13 & 558 & z=2.17 QSO; BCS \\ 
Z1883 & 2006-01-30 04:05:53 & 8:43:19.7 & 29:23:48.4 & 22.03 $\pm$ 0.06 & 1221 &  \\ 
A655 & 2006-01-30 04:42:47 & 8:25:49.6 & 47:10:28.4 & 17.45 $\pm$ 0.01 & 583 & CG AGN; SDSS/8 \\ 
RXJ0819 & 2006-01-30 05:19:40 & 8:17:52.7 & 63:42:0.576 & 22.99 $\pm$ 0.36 & 1451 &  \\ 
RXJ0819 & 2006-01-30 05:19:40 & 8:17:48.2 & 63:33:58.4 & 23.42 $\pm$ 0.56 & 1467 &  \\ 
RXJ0819 & 2006-02-26 06:25:49 & 8:19:39.7 & 63:45:46.4 & 20.36 $\pm$ 0.02 & 1093 &  \\ 
RXJ0819 & 2006-02-26 06:25:49 & 8:18:11.7 & 63:45:21.8 & 22.36 $\pm$ 0.13 & 1467 &  \\ 
A795 & 2006-01-30 05:56:39 & 9:24:58.8 & 13:56:55.7 & 22.01 $\pm$ 0.23 & 2703 &  \\ 
A795 & 2006-01-30 05:56:39 & 9:24:40.2 & 14:20:6.92 & 20.88 $\pm$ 0.09 & 1858 &  \\ 
A795 & 2006-01-30 05:56:39 & 9:24:4.61 & 13:56:16.7 & 18.99 $\pm$ 0.01 & 2041 &  \\ 
A795 & 2006-01-30 05:56:39 & 9:23:49.6 & 14:6:25.6 & 22.08 $\pm$ 0.24 & 797 &  \\ 
A1033 & 2006-01-30 07:14:07 & 10:31:58.6 & 35:1:17.9 & 21.72 $\pm$ 0.04 & 427 &  \\ 
A1033 & 2006-04-26 06:14:44 & 10:32:35.3 & 34:57:8.35 & 21.78 $\pm$ 0.04 & 1589 & RD  \\ 
A1033 & 2006-04-26 06:14:44 & 10:32:28.9 & 35:2:7.13 & 20.33 $\pm$ 0.01 & 1238 & z=0.128 QSO; CG; SDSS \\ 
A1033 & 2006-12-16 10:58:20 & 10:32:40.5 & 34:58:35.8 & 22.64 $\pm$ 0.10 & 1646 & z=1.56 BAL QSO; BCS \\ 
A1033 & 2006-12-16 10:58:20 & 10:32:35.2 & 34:57:8.41 & 22.62 $\pm$ 0.09 & 1587 & RD  \\ 
A1033 & 2006-12-16 10:58:20 & 10:31:31.0 & 35:16:23.7 & 22.77 $\pm$ 0.10 & 1920 &  \\ 
A1033 & 2006-12-16 10:58:20 & 10:30:31.9 & 34:56:47.9 & 23.16 $\pm$ 0.19 & 2149 & SNIb/c?; BCS \\ 
A1132 & 2006-01-30 07:50:49 & 10:59:6.18 & 56:59:46.8 & 22.30 $\pm$ 0.08 & 1934 &  \\ 
A1132 & 2006-01-30 07:50:49 & 10:58:48.6 & 56:48:0.285 & 22.03 $\pm$ 0.06 & 496 &  \\ 
A1132 & 2006-01-30 07:50:49 & 10:57:35.2 & 56:53:23.7 & 22.49 $\pm$ 0.11 & 1260 &  \\ 
Z3179 & 2006-01-30 09:50:19 & 10:25:20.4 & 12:34:45.2 & 22.21 $\pm$ 0.08 & 1689 & z=1.489 QSO; SDSS \\ 
RXJ1156 & 2006-01-30 10:29:24 & 11:57:30.6 & 24:27:43.2 & 21.26 $\pm$ 0.03 & 2168 &  \\ 
RXJ1156 & 2006-04-26 08:16:34 & 11:57:37.9 & 24:14:43.4 & 22.42 $\pm$ 0.08 & 1447 &  \\ 
RXJ1156 & 2006-04-26 08:16:34 & 11:56:27.6 & 24:12:26.9 & 21.68 $\pm$ 0.04 & 1072 &  \\ 
RXJ1156 & 2006-04-26 08:16:34 & 11:56:25.0 & 24:8:37.4 & 21.99 $\pm$ 0.05 & 1484 &  \\ 
RXJ1156 & 2006-04-26 08:16:34 & 11:56:20.4 & 24:18:52.6 & 22.63 $\pm$ 0.09 & 1301 &  \\ 
RXJ1156 & 2006-04-26 08:16:34 & 11:56:9.60 & 24:9:43.5 & 22.31 $\pm$ 0.08 & 1809 &  \\ 
RXJ1156 & 2006-04-26 08:16:34 & 11:56:7.56 & 24:8:39.3 & 23.11 $\pm$ 0.16 & 1951 &  \\ 
A1235 & 2006-01-30 11:09:02 & 11:22:51.3 & 19:35:51.5 & 22.43 $\pm$ 0.10 & 664 &  \\ 
A1235 & 2006-01-30 11:09:02 & 11:22:26.1 & 19:23:0.141 & 20.69 $\pm$ 0.03 & 1993 &  \\ 
A1235 & 2006-12-16 12:47:19 & 11:22:19.7 & 19:47:5.33 & 21.91 $\pm$ 0.11 & 1991 &  \\ 
A1235 & 2006-12-16 12:47:19 & 11:22:12.7 & 19:35:57.9 & 20.73 $\pm$ 0.04 & 1706 &  \\ 
A1235 & 2006-12-16 12:47:19 & 11:23:18.8 & 19:44:17.4 & 22.00 $\pm$ 0.12 & 973 &  \\ 
A1235 & 2006-12-16 12:47:19 & 11:22:51.3 & 19:28:34.6 & 20.97 $\pm$ 0.05 & 1065 &  \\ 
A1235 & 2006-12-16 12:47:19 & 11:22:44.9 & 19:48:37.7 & 21.47 $\pm$ 0.07 & 1689 &  \\ 
A1366 & 2006-04-26 09:30:39 & 11:46:48.9 & 67:37:6.08 & 21.81 $\pm$ 0.05 & 2253 & z=1.539 QSO; SDSS \\ 
A1366 & 2006-04-26 09:30:39 & 11:43:10.3 & 67:32:5.47 & 22.73 $\pm$ 0.11 & 1424 &  \\ 
A1902 & 2006-04-24 10:42:43 & 14:21:51.9 & 37:5:57.8 & 22.40 $\pm$ 0.08 & 1953 &  \\ 
A1902 & 2006-04-24 10:42:43 & 14:21:29.6 & 37:11:20.7 & 21.69 $\pm$ 0.04 & 1085 &  \\ 
A1902 & 2006-04-24 10:42:43 & 14:20:59.2 & 37:10:57.7 & 22.46 $\pm$ 0.09 & 1741 &  \\ 
A1902 & 2006-06-19 05:40:24 & 14:20:55.3 & 37:9:23.6 & 20.29 $\pm$ 0.01 & 2009 &  \\ 
A1677 & 2006-04-25 09:50:29 & 13:5:58.3 & 30:56:29.6 & 21.10 $\pm$ 0.03 & 513 &  \\ 
A1677 & 2006-06-19 04:59:20 & 13:6:6.00 & 30:47:28.2 & 22.70 $\pm$ 0.13 & 1385 &  \\ 
A1914 & 2006-04-25 10:31:46 & 14:26:20.3 & 37:45:0.308 & 22.11 $\pm$ 0.05 & 1074 &  \\ 
A1914 & 2006-06-19 06:24:29 & 14:25:28.4 & 37:45:39.9 & 22.31 $\pm$ 0.06 & 1133 &  \\ 
A1930 & 2006-06-19 07:06:52 & 14:32:16.4 & 31:40:39.0 & 21.92 $\pm$ 0.05 & 690 & z=0.129 QSO; CG; SDSS \\ 
A1930 & 2006-06-19 07:06:52 & 14:31:44.4 & 31:39:46.9 & 21.89 $\pm$ 0.04 & 1601 &  \\ 
A2259 & 2006-04-26 10:10:35 & 17:20:56.0 & 27:38:12.5 & 18.53 $\pm$ 0.00 & 1757 &  \\ 
A2259 & 2006-04-26 10:10:35 & 17:19:24.2 & 27:51:53.8 & 20.85 $\pm$ 0.02 & 2614 &  \\ 
A2259 & 2006-06-19 08:31:34 & 17:19:27.1 & 27:32:45.7 & 21.52 $\pm$ 0.05 & 2028 & z=1.447 QSO; SDSS \\ 
A2218 & 2006-06-19 07:49:38 & 16:34:2.04 & 66:15:47.7 & 21.96 $\pm$ 0.05 & 2011 &  \\ 
RXJ1750 & 2006-04-26 11:26:39 & 17:49:36.4 & 35:9:55.6 & 18.85 $\pm$ 0.01 & 1685 &  \\ 
RXJ1750 & 2006-04-26 11:26:39 & 17:50:57.1 & 35:14:47.0 & 19.85 $\pm$ 0.01 & 2231 &  \\ 
RXJ1750 & 2006-06-19 09:09:59 & 17:50:57.2 & 35:12:10.3 & 22.76 $\pm$ 0.22 & 1906 &  \\ 
RXJ1750 & 2006-06-19 09:09:59 & 17:49:23.7 & 35:7:15.8 & 21.93 $\pm$ 0.10 & 1942 &  \\ 
\enddata
\tablecomments{The following abbreviations were used under the 'Notes' column.  BCS -- Indicates that this transient was followed up spectroscopically by the Blue Channel Spectrograph on the MMT -- see \S~\ref{sec:spec}. CG -- Indicates that the transient is associated with a cluster galaxy.  IC Cand -- Intracluster SN candidate. RD -- This transient was detected as a variable in more than one epoch }
\tablerefs{(1) \citet{Richards05}; (2) \citet{Gandhi04}; (3) \citet{Rengstorf04}; (4) \citet{Croom04}; (5) \citet{Balogh02}; (6) \citet{Hewett95}; (7) \citet{Oegerle91}; (8) \citet{VandenBerk06} }
\end{deluxetable}

\clearpage

\begin{deluxetable}{lcccccl}
\tablecolumns{7}
\tablecaption{List of Identified Variables  -- Background Chip 3\label{table:vartable3}}
\tablehead{
\colhead{Cluster} & \colhead{Epoch} & \colhead{$\alpha$} & \colhead{$\delta$} & \colhead{g} & \colhead{Dist from BCG} &\colhead{Notes}\\
\colhead{} & \colhead{} & \colhead{(J2000.0)} & \colhead{J2000.0} & \colhead{mag} & \colhead{kpc}& \colhead{}}
\startdata
RXJ1750 & 2006-04-26 11:26:39 & 17:53:2.73 & 35:5:57.7 & 20.95 $\pm$ 0.04 & 5925 &  \\ 
RXJ1750 & 2006-04-26 11:26:39 & 17:52:46.3 & 35:7:9.76 & 21.72 $\pm$ 0.08 & 5349 &  \\ 
A2218 & 2006-04-26 10:46:38 & 16:40:49.7 & 66:22:48.2 & 21.35 $\pm$ 0.03 & 5670 &  \\ 
A2218 & 2006-04-26 10:46:38 & 16:40:42.8 & 66:11:10.9 & 22.32 $\pm$ 0.07 & 5302 &  \\ 
A2259 & 2006-04-26 10:10:35 & 17:23:33.4 & 27:49:31.6 & 22.54 $\pm$ 0.08 & 7764 & IC Cand \\ 
A2259 & 2006-04-26 10:10:35 & 17:22:7.20 & 27:26:29.7 & 21.87 $\pm$ 0.10 & 4963 &  \\ 
A1677 & 2006-06-19 04:59:20 & 13:9:27.7 & 30:44:19.4 & 20.54 $\pm$ 0.02 & 8794 &  \\ 
A1677 & 2006-06-19 04:59:20 & 13:9:3.27 & 31:6:26.6 & 22.18 $\pm$ 0.07 & 7930 &  \\ 
A1689 & 2006-01-28 11:51:19 & 13:14:46.1 & -1:8:52.2 & 21.92 $\pm$ 0.06 & 9305 &  \\ 
A1689 & 2006-01-28 11:51:19 & 13:13:46.3 & -1:13:19.2 & 22.07 $\pm$ 0.06 & 6431 &  \\ 
A1689 & 2006-02-24 09:46:20 & 13:13:34.8 & -1:18:3.00 & 22.17 $\pm$ 0.07 & 5781 & z=1.675 (phot) QSO/1 \\ 
A1689 & 2006-06-19 04:03:19 & 13:14:30.8 & -1:29:45.9 & 20.82 $\pm$ 0.03 & 8514 &  \\ 
A1689 & 2006-06-19 04:03:19 & 13:14:17.9 & -1:35:12.1 & 22.13 $\pm$ 0.10 & 8208 &  \\ 
A1689 & 2006-06-19 04:03:19 & 13:13:48.7 & -1:25:39.3 & 22.12 $\pm$ 0.09 & 6472 &  \\ 
A1902 & 2006-04-24 10:42:43 & 14:24:37.2 & 37:20:33.4 & 22.76 $\pm$ 0.16 & 5836 &  \\ 
A1902 & 2006-04-24 10:42:43 & 14:26:2.16 & 37:3:4.55 & 21.92 $\pm$ 0.07 & 8965 &  \\ 
A1914 & 2006-04-25 10:31:46 & 14:29:27.7 & 37:45:21.0 & 21.98 $\pm$ 0.05 & 7316 & z=1.69 QSO; SDSS \\ 
A1914 & 2006-04-25 10:31:46 & 14:28:22.1 & 37:58:14.5 & 22.96 $\pm$ 0.10 & 5264 &  \\ 
A1914 & 2006-06-19 06:24:29 & 14:30:9.63 & 38:0:41.9 & 22.19 $\pm$ 0.05 & 8942 &  \\ 
A1930 & 2006-06-19 07:06:52 & 14:35:9.50 & 31:31:48.3 & 21.86 $\pm$ 0.04 & 4624 & z=0.50 QSO; SDSS \\ 
A1930 & 2006-06-19 07:06:52 & 14:35:0.177 & 31:41:35.7 & 22.97 $\pm$ 0.17 & 4250 &  \\ 
A1930 & 2006-06-19 07:06:52 & 14:34:55.9 & 31:39:24.5 & 21.93 $\pm$ 0.05 & 4108 &  \\ 
A1553 & 2006-04-24 10:03:28 & 12:34:12.5 & 10:26:23.9 & 22.71 $\pm$ 0.16 & 8565 &  \\ 
RXJ0958 & 2006-01-29 05:54:09 & 10:1:32.4 & 23:40:0.701 & 22.28 $\pm$ 0.10 & 7137 &  \\ 
RXJ0958 & 2006-01-29 05:54:09 & 10:1:25.3 & 23:56:20.2 & 21.05 $\pm$ 0.03 & 6955 &  \\ 
RXJ0958 & 2006-01-29 05:54:09 & 10:0:49.4 & 23:54:33.7 & 21.82 $\pm$ 0.06 & 5664 & RD  \\ 
RXJ0958 & 2006-01-29 05:54:09 & 10:0:7.77 & 23:44:56.0 & 22.60 $\pm$ 0.12 & 4078 &  \\ 
RXJ0958 & 2006-02-25 05:03:22 & 10:0:49.4 & 23:54:33.7 & 21.62 $\pm$ 0.05 & 5664 & RD  \\ 
A1361 & 2006-01-28 12:29:23 & 11:47:27.2 & 46:24:37.6 & 22.30 $\pm$ 0.06 & 4998 &  \\ 
A1361 & 2006-02-24 10:30:22 & 11:48:37.0 & 46:34:16.5 & 22.58 $\pm$ 0.08 & 6694 &  \\ 
A1361 & 2006-02-24 10:30:22 & 11:47:7.16 & 46:34:29.2 & 22.24 $\pm$ 0.05 & 4825 &  \\ 
A1361 & 2006-04-24 09:20:33 & 11:47:7.34 & 46:7:12.0 & 23.04 $\pm$ 0.11 & 4909 &  \\ 
A1361 & 2006-04-24 09:20:33 & 11:47:4.65 & 46:21:42.6 & 22.42 $\pm$ 0.05 & 4490 &  \\ 
A1413 & 2006-01-28 11:09:51 & 11:58:15.1 & 23:14:42.1 & 22.74 $\pm$ 0.12 & 6291 &  \\ 
A1413 & 2006-04-24 07:59:00 & 11:58:19.8 & 23:31:38.1 & 21.48 $\pm$ 0.03 & 6374 &  \\ 
A1413 & 2006-04-24 07:59:00 & 11:57:48.9 & 23:13:47.5 & 21.12 $\pm$ 0.02 & 5450 &  \\ 
A961 & 2006-02-24 07:52:05 & 10:18:58.7 & 33:47:15.8 & 22.41 $\pm$ 0.08 & 4477 &  \\ 
A961 & 2006-02-24 07:52:05 & 10:18:51.2 & 33:45:37.6 & 21.05 $\pm$ 0.02 & 4222 &  \\ 
A961 & 2006-04-24 06:26:29 & 10:19:55.8 & 33:45:5.33 & 22.02 $\pm$ 0.05 & 5971 &  \\ 
A961 & 2006-04-24 06:26:29 & 10:19:48.4 & 33:26:2.09 & 22.98 $\pm$ 0.12 & 5948 &  \\ 
A961 & 2006-04-24 06:26:29 & 10:19:23.1 & 33:32:10.4 & 22.59 $\pm$ 0.08 & 5075 &  \\ 
A853 & 2006-01-28 08:36:18 & 9:45:19.1 & 15:14:37.7 & 21.71 $\pm$ 0.09 & 7665 &  \\ 
A853 & 2006-01-28 08:36:18 & 9:45:4.39 & 15:12:47.9 & 22.22 $\pm$ 0.14 & 7143 &  \\ 
A853 & 2006-01-28 08:36:18 & 9:44:50.7 & 15:12:37.1 & 19.44 $\pm$ 0.01 & 6611 &  \\ 
A853 & 2006-01-28 08:36:18 & 9:44:36.3 & 15:22:24.2 & 22.29 $\pm$ 0.29 & 5782 &  \\ 
A853 & 2006-01-28 08:36:18 & 9:44:10.3 & 15:23:54.8 & 21.17 $\pm$ 0.06 & 4720 &  \\ 
A853 & 2006-04-25 05:53:14 & 9:44:35.5 & 15:9:31.2 & 21.93 $\pm$ 0.08 & 6186 &  \\ 
A667 & 2006-04-25 05:53:14 & 8:31:29.9 & 44:32:30.7 & 22.24 $\pm$ 0.06 & 5904 & RD  \\ 
A667 & 2006-04-24 05:11:06 & 8:32:24.1 & 44:53:53.6 & 22.66 $\pm$ 0.11 & 7063 & z=1.525 (phot) QSO/1 \\ 
A667 & 2006-04-24 05:11:06 & 8:31:29.9 & 44:32:30.7 & 22.25 $\pm$ 0.08 & 5905 & RD  \\ 
A580 & 2006-01-28 04:25:28 & 7:29:51.0 & 41:32:55.8 & 21.38 $\pm$ 0.03 & 5735 & RD  \\ 
A580 & 2006-01-28 04:25:28 & 7:28:53.6 & 41:34:36.7 & 22.52 $\pm$ 0.07 & 4470 &  \\ 
A580 & 2006-04-24 03:05:59 & 7:29:51.1 & 41:32:56.3 & 19.74 $\pm$ 0.01 & 5737 & RD  \\ 
A580 & 2006-04-24 03:05:59 & 7:29:17.8 & 41:38:2.40 & 21.29 $\pm$ 0.03 & 5161 & NVSS source/2 \\ 
A580 & 2006-04-24 03:05:59 & 7:28:46.9 & 41:29:54.6 & 21.52 $\pm$ 0.04 & 4156 &  \\ 
A646 & 2006-01-30 02:51:19 & 8:25:49.5 & 47:10:27.5 & 18.60 $\pm$ 0.01 & 5210 & z=0.13 CG AGN/3 \\ 
A646 & 2006-12-16 10:22:04 & 8:25:10.5 & 47:16:34.3 & 21.86 $\pm$ 0.07 & 4498 & z=1.425 (phot) QSO/1 \\ 
A743 & 2006-12-16 10:22:04 & 9:10:13.1 & 10:4:23.3 & 21.97 $\pm$ 0.12 & 7557 &  \\ 
A743 & 2006-12-16 10:22:04 & 9:9:27.2 & 10:5:8.28 & 21.67 $\pm$ 0.07 & 5998 &  \\ 
A743 & 2006-12-16 08:53:47 & 9:8:58.4 & 10:25:32.0 & 21.05 $\pm$ 0.04 & 4743 &  \\ 
RXJ1000 & 2006-02-25 05:39:58 & 10:5:7.00 & 44:0:39.8 & 21.08 $\pm$ 0.03 & 8018 &  \\ 
RXJ1000 & 2006-04-25 07:19:59 & 10:5:3.80 & 44:17:23.7 & 21.85 $\pm$ 0.06 & 7906 & z=0.957 QSO; SDSS \\ 
RXJ1000 & 2006-04-25 07:19:59 & 10:3:17.2 & 44:1:52.6 & 21.75 $\pm$ 0.05 & 4890 &  \\ 
RXJ0804 & 2006-01-28 06:15:30 & 8:9:14.8 & 46:39:38.9 & 20.86 $\pm$ 0.02 & 9560 & z=0.975 (phot) QSO/1 \\ 
RXJ0804 & 2006-04-24 04:33:50 & 8:7:44.3 & 46:38:44.0 & 20.56 $\pm$ 0.02 & 6729 &  \\ 
RXJ0804 & 2006-04-24 04:33:50 & 8:7:18.7 & 47:1:18.1 & 22.82 $\pm$ 0.11 & 6259 &  \\ 
A750 & 2006-01-28 07:59:36 & 9:11:3.81 & 11:4:28.3 & 21.65 $\pm$ 0.05 & 4689 & z=0.928 QSO; SDSS \\ 
A750 & 2006-02-24 06:35:19 & 9:11:56.1 & 11:9:16.5 & 22.53 $\pm$ 0.07 & 6974 &  \\ 
A750 & 2006-04-24 05:49:13 & 9:11:36.7 & 11:11:11.7 & 22.71 $\pm$ 0.12 & 6308 &  \\ 
A750 & 2006-04-24 05:49:13 & 9:11:33.0 & 10:57:3.77 & 21.90 $\pm$ 0.05 & 5794 &  \\ 
A750 & 2006-04-24 05:49:13 & 9:11:22.5 & 11:8:58.3 & 22.91 $\pm$ 0.14 & 5632 &  \\ 
A750 & 2006-04-24 05:49:13 & 9:11:5.82 & 10:58:12.3 & 22.24 $\pm$ 0.07 & 4665 &  \\ 
RXJ0741 & 2006-04-25 03:53:15 & 7:44:23.5 & 26:8:3.07 & 21.35 $\pm$ 0.05 & 6684 & RD  \\ 
RXJ0741 & 2006-01-29 02:11:06 & 7:43:56.4 & 26:9:21.5 & 21.94 $\pm$ 0.06 & 5784 &  \\ 
RXJ0741 & 2006-01-29 02:11:06 & 7:44:46.9 & 25:48:10.3 & 20.83 $\pm$ 0.02 & 7574 & z=0.141 CG QSO; SDSS/4 \\ 
RXJ0741 & 2006-02-25 02:26:30 & 7:44:49.8 & 25:46:11.8 & 20.87 $\pm$ 0.02 & 7756 &  \\ 
RXJ0741 & 2006-02-25 02:26:30 & 7:44:23.5 & 26:8:3.07 & 22.22 $\pm$ 0.08 & 6685 & RD  \\ 
A620 & 2006-01-29 02:55:40 & 8:9:43.3 & 45:32:35.2 & 20.21 $\pm$ 0.01 & 6157 &  \\ 
A620 & 2006-02-25 03:05:49 & 8:8:20.2 & 45:37:37.0 & 22.00 $\pm$ 0.05 & 3971 &  \\ 
A1045 & 2006-02-25 07:36:44 & 10:38:30.8 & 30:54:31.3 & 20.53 $\pm$ 0.08 & 6866 &  \\ 
A990 & 2006-02-25 07:36:44 & 10:27:46.8 & 49:19:41.3 & 22.44 $\pm$ 0.10 & 6325 &  \\ 
A990 & 2006-02-25 08:52:41 & 10:26:21.7 & 48:57:20.9 & 23.04 $\pm$ 0.09 & 4379 &  \\ 
A1264 & 2006-01-29 10:29:19 & 11:30:21.0 & 17:17:44.7 & 22.08 $\pm$ 0.05 & 6622 &  \\ 
A1204 & 2006-01-29 12:31:42 & 11:15:59.4 & 17:40:15.1 & 22.32 $\pm$ 0.11 & 6665 &  \\ 
RXJ0821 & 2006-12-16 08:11:33 & 8:23:5.83 & 7:41:52.1 & 21.75 $\pm$ 0.05 & 3881 &  \\ 
Z1883 & 2006-01-30 04:05:53 & 8:45:48.2 & 29:20:54.8 & 20.29 $\pm$ 0.01 & 7363 &  \\ 
A655 & 2006-02-26 05:47:17 & 8:30:3.34 & 47:16:21.7 & 22.13 $\pm$ 0.09 & 6446 &  \\ 
A1132 & 2006-01-30 07:50:49 & 11:4:26.6 & 56:34:7.51 & 21.06 $\pm$ 0.04 & 7484 &  \\ 
A1132 & 2006-01-30 07:50:49 & 11:3:16.7 & 56:54:40.8 & 21.81 $\pm$ 0.06 & 5865 &  \\ 
A1132 & 2006-04-26 08:52:28 & 11:2:56.0 & 56:51:34.7 & 22.15 $\pm$ 0.07 & 5406 &  \\ 
A1132 & 2006-04-26 08:52:28 & 11:2:4.42 & 56:48:44.3 & 22.06 $\pm$ 0.07 & 4366 &  \\ 
Z3179 & 2006-01-30 09:50:19 & 10:29:31.1 & 12:28:4.00 & 21.90 $\pm$ 0.18 & 8082 &  \\ 
Z3179 & 2006-04-26 06:59:56 & 10:28:26.5 & 12:44:23.6 & 22.08 $\pm$ 0.06 & 5476 &  \\ 
RXJ1156 & 2006-04-26 08:16:34 & 12:0:27.6 & 24:5:57.1 & -10.83 $\pm$ 0.02 & 7388 &  \\ 
RXJ1156 & 2006-04-26 08:16:34 & 11:59:25.3 & 24:24:16.7 & -9.60 $\pm$ 0.07 & 5264 &  \\ 
A1235 & 2006-01-30 11:09:02 & 11:26:30.8 & 19:22:58.9 & 22.44 $\pm$ 0.11 & 5465 &  \\ 
A1235 & 2006-04-26 07:39:05 & 11:25:22.5 & 19:23:30.3 & 21.88 $\pm$ 0.05 & 3698 &  \\ 
A1235 & 2006-04-26 07:39:05 & 11:25:42.7 & 19:28:35.9 & 21.16 $\pm$ 0.03 & 4046 &  \\ 
A1235 & 2006-04-26 07:39:05 & 11:24:53.6 & 19:33:1.66 & 22.16 $\pm$ 0.06 & 2655 &  \\ 
A1366 & 2006-01-30 12:26:14 & 11:49:46.3 & 67:36:15.9 & 23.07 $\pm$ 0.09 & 4006 &  \\ 
A1366 & 2006-02-25 11:33:20 & 11:52:49.5 & 67:37:42.0 & 22.74 $\pm$ 0.07 & 6142 &  \\ 
A1366 & 2006-02-25 11:33:20 & 11:50:5.13 & 67:36:14.5 & 22.80 $\pm$ 0.07 & 4214 &  \\ 
A565 & 2006-04-23 03:04:32 & 7:15:39.7 & 71:52:47.9 & 21.89 $\pm$ 0.09 & 4917 &  \\ 
A136 & 2006-12-16 02:43:01 & 1:6:55.7 & 25:5:40.9 & 22.57 $\pm$ 0.15 & 5952 &  \\ 
A136 & 2006-12-16 02:43:01 & 1:6:47.5 & 24:58:14.5 & 22.56 $\pm$ 0.15 & 5764 &  \\ 
Z0256 & 2006-12-16 05:11:56 & 3:2:55.7 & 0:2:32.8 & 23.68 $\pm$ 0.53 & 10061 &  \\ 
\enddata
\tablecomments{The following abbreviations were used under the 'Notes' column. IC Cand -- Intracluster SN candidate. RD -- This transient was detected as a variable in more than one epoch }
\tablerefs{(1) \citet{Richards05}; (2) \citet{McMahon02}; (3) \citet{Hao05}; (4) \citet{VandenBerk06}}

\end{deluxetable}

\clearpage

\begin{deluxetable}{lccccccc}
\tablecolumns{8}
\tablecaption{Blue Channel Followup of Transients \label{table:bcs}}

\tablehead{
\colhead{Cluster/ID} & \colhead{$\alpha$} & \colhead{$\delta$} &\colhead{UT Date} & \colhead{Classification} & \colhead{$z$} & \colhead{Discovery} & \colhead{Exp.}\\
\colhead{} & \colhead{(J2000.0)} & \colhead{J2000.0} & \colhead{2007 Dec 20} & \colhead{} & \colhead{} & \colhead{(mag)} & \colhead{(sec)}}
\startdata
Abell 1033/ID0&10:32:40.6&+34:58:35.8&0.54&BAL QSO&1.56&22.64$\pm$0.10&600\\
Abell 1033/ID6&10:30:31.9&+34:56:47.0&0.53&SNIb/c?&0.1?&23.16$\pm$0.19&600\\
Abell 1246/ID2&11:24:49.9&+21:36:05.2&0.51&QSO&0.92&21.45$\pm$0.02&600\\
Abell 1246/ID9&11:24:02.5&+21:36:58.1&0.52&Stellar spec&0.0&22.54$\pm$0.06&600\\
Abell 1246/ID10&11:23:49.6&+21:28:40.4&0.50&Cluster SNIa&0.19&20.63$\pm$0.01&900\\
Abell 136/ID3&01:03:55.5&+24:50:20.6&0.08&Galaxy&0.246&21.75$\pm$0.06&1800\\
Abell 136/ID7&01:03:19.9&+25:04:28.4&0.10&Galaxy&0.172&24.5$\pm$0.6&1800\\
Abell 351/ID1&02:26:21.3&-08:54:19.8&0.15&SNIa &0.33&22.71$\pm$0.17&1800\\
Abell 351/ID5&02:25:57.0&-08:45:38.4&0.13&SNIa &0.27&21.71$\pm$0.07&900\\
Abell 516/ID3&04:50:52.5&-08:45:03.5&0.23&Stellar spec&0.0&22.09$\pm$0.07&1800\\
Abell 516/ID8&04:50:29.2&-08:50:01.7&0.26&Stellar spec&0.0&22.93$\pm$0.19&1800\\
Abell 516/ID11&04:49:42.6&-08:51:43.1&0.21&?; IC Candidate&?&23.83$\pm$0.90&1800\\
Abell 620/ID5&08:06:14.8&+45:35:45.5&0.29&Stellar spec&0.0&21.19$\pm$0.04&600\\
Abell 620/ID12&08:05:06.9&+45:32:37.3&0.30&Stellar spec&0.0&21.70$\pm$0.06&1800\\
Abell 646/ID13&08:22:07.0&+46:56:00.5&0.32&QSO&2.36&21.03$\pm$0.07&1800\\
Abell 646/ID19&08:21:05.6&+47:16:33.0&0.35&QSO&2.22&21.40$\pm$0.10&900\\
RXJ0759/ID2&07:59:13.0&+54:04:42.2&0.36&QSO&2.93&22.20$\pm$0.04&900\\
RXJ0759/ID3&07:58:45.9&+54:01:17.5&0.38&Galaxy&0.21&22.09$\pm$0.04&1350\\
RXJ0821/ID4&08:21:23.0&+07:39:41.8&0.45&QSO&0.90&22.20$\pm$0.11&900\\
RXJ0821/ID8&08:20:52.8&+07:52:38.4&0.40&QSO&0.795&21.96$\pm$0.08&900\\
RXJ0821/ID9&08:20:44.6&+07:50:07.6&0.43&QSO&2.17&22.43$\pm$0.13&1350\\
RXJ0821/ID15&08:20:34.3&+07:39:38.1&0.42&Galaxy&0.22&21.73$\pm$0.07&1350\\
Z0256/ID19&02:59:07.8&+00:22:33.0&0.17&Cluster Galaxy&0.199&22.81$\pm$0.14&1800\\
Abell 743/ID10&09:06:55.3&+10:14:13.7&0.48&AGN&0.267&22.00$\pm$0.09&900\\
Abell 743/ID12&09:06:32.5&+10:09:28.7&0.46&SNIa &0.175&21.32$\pm$0.05&1350\\
\enddata
\end{deluxetable}


\end{document}